\documentclass[runningheads,a4paper]{llncs}

\usepackage[utf8x]{inputenc}
\usepackage{amssymb,amsmath}
\usepackage{bm,bbm}
\usepackage{subfigure,tikz}
\usepackage{booktabs}
\usepackage{multirow}
\usepackage[ugly]{units}
\usepackage{url}
\usepackage{rotating}

\urldef{\mails}\path|{ljmtorres, sidnei, corina}@dpi.inpe.br, acfrery@gmail.com|    
\newcommand{\keywords}[1]{\par\addvspace\baselineskip
\noindent\keywordname\enspace\ignorespaces#1}

\begin{document}

\title{Speckle Reduction in Polarimetric SAR Imagery with Stochastic Distances and Nonlocal Means}

\titlerunning{Speckle Reduction in Polarimetric SAR Imagery with SD and NL-means}

\author{${ }^{1}$Leonardo Torres\and ${ }^{1}$Sidnei J.\ S.\ Sant'Anna \and ${ }^{1}$Corina C.\ Freitas \and \\${ }^{2}$Alejandro C.\ Frery\thanks{The authors are grateful to CNPq, CAPES, FAPEAL and FAPESP for supporting this research.}}

\authorrunning{Torres, Sant'Anna, Freitas \& Frery}

\institute{
${ }^{1}$Instituto Nacional de Pesquisas Espaciais -- INPE\\
Divis\~ao de Processamento de Imagens -- DPI\\
Av. dos Astronautas, 1758, 12227-010\\
S\~ao Jos\'e dos Campos -- SP, Brazil\\
\vspace*{.3cm}
${ }^{2}$Universidade Federal de Alagoas -- UFAL\\ 
Laborat\'orio de Computa\c c\~ao Cient\'ifica e An\'alise Num\'erica -- LaCCAN\\
Av. Lourival Melo Mota, s/n, Tabuleiro dos Martins, 57072-900\\
Macei\'o -- AL, Brazil\\
\vspace*{.3cm}
\mails
}

\toctitle{Speckle Reduction in Polarimetric SAR Imagery with SD and NL-means}
\tocauthor{Torres, Medeiros \& Frery}
\maketitle

\begin{abstract}
This paper presents a technique for reducing speckle in Polarimetric Synthetic Aperture Radar (PolSAR) imagery using Nonlocal Means and a statistical test based on stochastic divergences.
The main objective is to select ho\-mo\-ge\-ne\-ous pixels in the filtering area through statistical tests between distributions.
This pro\-po\-sal uses the complex Wishart model to describe PolSAR data, but the technique can be extended to other models.
The weights of the location-variant linear filter are function of the $p$-values of tests which verify the hypothesis that two samples come from the same distribution and, therefore, can be used to compute a local mean.
The test stems from the family of ($h$-$\phi$) divergences which originated in Information Theory.
This novel technique was compared with the Boxcar, Refined Lee and IDAN filters.
Image quality assessment methods on simulated and real data are employed to validate the performance of this approach.
We show that the proposed filter also enhances the polarimetric entropy and preserves the scattering information of the targets.
\end{abstract}

\section{Introduction}\label{sec:intro}

Among the remote sensing technologies, Polarimetric Synthetic Aperture Radar (PolSAR) has achieved a prominent position. 
PolSAR imaging is a well-developed coherent microwave remote sensing technique for providing large-scale two-di\-men\-sional (2-D) high spatial resolution images of the Earth’s surface dielectric properties~\cite{LeePottier2009Book}.

In SAR systems, the value at each pixel is a complex number: the amplitude and phase information of the returned signal.
Full PolSAR data is comprised of four complex channels which result from the combination of the horizontal and vertical transmission modes, and horizontal and vertical reception modes.

The speckle phenomenon in SAR data hinders the interpretation these data and reduces the accuracy of segmentation, classification and analysis of objects contained within the image.
Therefore, reducing the noise effect is an important task, and multilook processing is often used for this purpose in single- and full-channel data.
In the latter, such processing yields a covariance matrix in each pixel, but further noise reduction is frequently needed.

According to Lee and Pottier~\cite{LeePottier2009Book}, Polarimetric SAR image smoothing requires preserving the target polarimetric signature.
Such requirement can be posed as:
(i)~each element of the image should be filtered in a similar way to multilook processing by averaging the covariance matrix of neighboring pixels; and 
(ii)~homogeneous regions in the neighborhood should be adaptively selected to preserve resolution, edges and the image quality.
The second requirement, i.e. selecting homogeneous areas given similarity criterion, is a common problem in pattern recognition.
It boils down to identifying observations from different stationary stochastic processes.

Usually, the Boxcar filter is the standard choice because of its simple design.
However, it has poor performance since it does not discriminate different targets.
Lee et al.~\cite{Lee1999PolSARspeckleFiltering,Lee1991MultiPol_SARimagery} propose techniques for speckle reduction based on the multiplicative noise model using the minimum mean-square error (MMSE) criterion.
Lee et al.~\cite{Lee2006ScatteringModelBased_SpeckleFiltering} proposed a methodology for selecting neighboring pixels with similar scattering characteristics, known as Refined Lee filter.
Other techniques use the local linear minimum mean-squared error (LLMMSE) criterion proposed by Vasile et al.~\cite{Vasile2006PolSAR}, in a similar adaptive technique, but the decision to select homogeneous areas is based on the intensity information of the polarimetric coherency matrices, namely intensity-driven adaptive-neighborhood (IDAN).

{\c C}etin and Karl~\cite{Cetin2001FeatureEnhancedSAR} presented a technique for image formation based on regularized image reconstruction.
This approach employs a tomographic model which allows the incorporation of prior information about, among other features, the sensor.
The resulting images have many desirable properties, reduced speckled among them.
Our approach deals with data already produced and, thus, does not require interfering in the processing protocol of the data.

Osher et al.~\cite{Osher2005IteratedRegularizationMethod} presented a novel iterative regularization method for inverse problems based on the use of Bregman distances using a total variation denoising technique tailored to additive noise.
The authors also propose a generalization for multiplicative noise, but no results with this kind of contamination are show. 
The main contributions were the rigorous convergence results and effective stopping criteria for the general procedure, that provides information on how to obtain an approximation of the noise-free image intensity.
Goldstein and Osher~\cite{Goldstein2009SplitBregmanMethod} presented an improvement of this work using the class of $L_1$-regularized optimization problems, that originated in functional analysis for finding extrema of convex functionals.
The authors apply this technique to the Rudin-Osher-Fatemi model for image denoising and to a compressed sensing problem that arises in magnetic resonance imaging.
Our work deals with full polarimetric data, for which, to the best of our knowledge, there are no similar results that take into account its particular nature: the pixels values are definite positive Hermitian complex matrices.

Soccorsi et al.~\cite{Soccorsi2010HMMSARimagesRestoration} presented a despeckling technique for single-look complex SAR image using nonquadratic regularization.
They use an image model, a gradient, and a prior model, to compute the objective function.
We employ the full polarimetric information provided by the multilook scaled complex Wishart distribution.

Chambolle~\cite{Chambolle2004TVminimization} proposed a Total Variation approach for a number of problems in image restoration (denoising, zooming and mean curvature motion), but under the Gaussian additive noise assumption.

Li et al.~\cite{Li2011FeatureEnhancementCurveletTransform} propose the use of a particle swarm optimization algorithm and an extension of the curvelet transform for speckle reduction.
They employ the homomorophic transformation, so their technique can be used either in amplitude or intensity data, but not in complex-valued imagery, as is the case we present here.

Wong and Fieguth~\cite{Wong2010BayesianApproachSARdata} presented a novel approach for performing blind decorrelation of SAR data. 
They use a similarity technique between patches of the point-spread function using a Bayesian least squares estimation approach based on a Fisher-Tippett log-scatter model.
In a similar way, S{\o}lbo and Eltoft~\cite{Solbo2004HomomorphicWavelet} assume a Gamma distribution in a wavelet-based speckle reduction procedure, and they estimate all the parameters locally without imposing a fixed number of looks (which they call ``degree of heterogeneity'') for the whole image.

Buades et al.~\cite{Buades2005DenoisingAlgorithms} proposed a methodology, termed Nonlocal Means (NL-means), which consists in using similarities between patches as the weights of a mean filter; it is known to be well suited for combating additive Gaussian noise.
Deledalle et al.~\cite{Deledalle2010PolSAREstimationNLM} applied this methodology to PolSAR data using the Kullback-Leibler distance between two zero-mean complex circular Gaussian laws.
Following the same strategy, Chen et al.~\cite{Chen2011NonLocalFilteringPolSAR} used the test for equality between two complex Wishart matrices proposed by Conradsen et al.~\cite{Conradsen2003testStatisticWishart}.

This paper proposes a new approach for speckle noise filtering in PolSAR imagery: an adaptive nonlinear extension of the NL-means algorithm.
This is an extension of previous works~\cite{SpeckleReductionStochasticDistancesCIARP2012,PolarimetricSARImageSmoothingStochasticDistancesCIARP2012}, where we used an  approach similar to that of Nagao and Matsuyma~\cite{NagaoMatsuyama}.
Overlapping samples are compared based on stochastic distances between distributions, and the $p$-values resulting from such comparisons are used to build the weights of an adaptive linear filter.
The goodness-of-fit tests are derived from the divergences discussed by Frery et al.~\cite{Frery2011InformationTheoryPolSAR} and Nascimento et al.~\cite{Nascimento2010}.
The new proposal is called \textit{Stochastic Distances Nonlocal Means} (SDNLM), and amounts to using those observations which are not rejected by a test seeking for a strong stationary process.

This paper is organized as follows.
First, we summarize the basic principles that lead to the complex Wishart model for full polarimetric data.
In Section~\ref{sec:nonlocal} we recall the Nonlocal Means method.
Our approach for reducing speckle in PolSAR data using stochastic distances between two complex Wishart distributions is proposed in Section~\ref{sec:distances}.
Image Quality Assessment is briefly discussed in Section~\ref{sec:assessment}.
Results are presented in Section~\ref{sec:results}, while Section~\ref{sec:conclu} concludes the paper.

\section{The Complex Wishart Distribution}\label{sec:model}

PolSAR imaging results in a complex scattering matrix, which includes intensity and relative phase data~\cite{Frery2011InformationTheoryPolSAR}.
Such matrices have usually four distinct complex elements, namely $S_{V\!V}$, $S_{V\!H}$, $S_{H\!V}$, and $S_{H\!H}$, where $H$ and $V$ refer to the horizontal and vertical wave polarization states, respectively.
In a reciprocal medium, which is most common situation in remote sensing, $S_{V\!H}=S_{H\!V}$ so the complex signal backscattered from each resolution cell can be characterized by a scattering vector $\bm{Y}$ with three complex elements \cite{Ulaby1990RadarPolarimetriy}. 

Thus, we have a scattering complex random vector
$
\bm{Y}=[S_{H\!H},S_{V\!H},S_{V\!V}]^t,
$
where $[\cdot]^t$ indicates vector transposition.
In general, PolSAR data are locally modeled by a multivariate zero-mean complex circular Gaussian distribution that characterize the scene reflectivity~\cite{Touzi2004ReviewPolarimetry,Ulaby1990RadarPolarimetriy}, whose probability density function is
$$
f(\bm Y;\bm{\Sigma})=\frac{1}{\pi^3 \vert\bm{\Sigma}\vert}\exp\bigl\{-\bm Y^{*t}\bm{\Sigma}^{-1}\bm Y\bigr\},
$$
where $\vert \cdot \vert$ is the determinant, and the superscript `$*$' denotes the complex conjugate of a vector; $\bm{\Sigma}$ is the covariance matrix of $\bm{Y}$.
This distribution is defined on $\mathbb C^3$.
The covariance matrix $\bm{\Sigma}$, besides being Hermitian and positive definite, has all the information which characterizes the scene under analysis.

Multilook processing enhances the signal-to-noise ratio.
It is performed averaging over $L$ ideally independent looks of the same scene, and it yields the sample covariance matrix $\bm{Z}$ given, in each pixel, by
$
\bm{Z}=L^{-1}\sum_{\iota=1}^{L}\bm{Y}_{\iota} \bm{Y}_{\iota}^{*},
$
where $L$ is the number of looks.

Goodman~\cite{Goodman1963ComplexWishartDistributed} proved that $\bm{Z}$ follows a scaled multilook complex Wishart distribution, denoted by $\bm{Z}\thicksim\mathcal{W}(\bm{\Sigma},L)$, and characterized by the following probability density function:
\begin{equation}
 f_{\bm{Z}}(\bm{Z}';\bm{\Sigma},L) = \frac{L^{3L}\vert \bm{Z}'\vert^{L-3}}{\vert\bm{\Sigma}\vert^L \Gamma_3(L)} \exp\bigl\{-L\ \mathrm{tr}\bigl(\bm{\Sigma}^{-1} \bm{Z}'\bigr)\bigr\},
 \label{eq:denswishart}
\end{equation}
where, for $L\geq 3$, $\Gamma_3(L)=\pi^3 \prod^2_{i=0} \Gamma(L-i)$, $\Gamma(\cdot)$ is the gamma function, $\mathrm{tr}(\cdot)$ is the trace operator, and the covariance matrix $\bm{Z}$ is given by
$$
\bm{\Sigma} = E\{\bm{Y}\bm{Y}^{*t}\} = \left[\begin{array}{ccc}
                                    E\{S_{H\!H} S_{H\!H}^*\}\; & E\{S_{H\!H} S_{V\!H}^*\}\; & E\{S_{H\!H} S_{V\!V}^*\} \\
                                    E\{S_{V\!H} S_{H\!H}^*\}\; & E\{S_{V\!H} S_{V\!H}^*\}\; & E\{S_{V\!H} S_{V\!V}^*\} \\
                                    E\{S_{V\!V} S_{H\!H}^*\}\; & E\{S_{V\!V} S_{V\!H}^*\}\; & E\{S_{V\!V} S_{V\!V}^*\} \\
                                    \end{array} \right],
$$
where $E\{\cdot\}$ denote expectation.
Anfinsen et al.~\cite{EstimationEquivalentNumberLooksSAR} removed the restriction $L\geq 3$.
The resulting distribution has the same form as in~\eqref{eq:denswishart} and is termed the ``relaxed'' Wishart. 
We assume this last model, and we allow variations of $L$ along the image.

The support of this distribution is the cone of positive definite Hermitian complex matrices~\cite{EntropyBasedStatisticalAnalysisPolSAR}.

The parameters are usually estimated by maximum likelihood (ML) due to its statistical properties.
Let $\bm{Z}_r$ be a random matrix which follows a $\mathcal{W}(\bm{\Sigma},L)$ law.
Its log-likelihood function is given by
\begin{align*}
 \ell_r(\bm{\Sigma},L) = &{}\  3L\log L + (L-3)\log\vert\bm{Z}_r\vert - L\log\vert\bm{\Sigma}\vert - 3\log\pi - \nonumber\\ 
 & \sum_{q=0}^{2}\log\Gamma(L-q) - L\mathrm{tr}(\bm{\Sigma}^{-1}\bm{Z}_r),\label{eq:logLikelihood}
\end{align*}
resulting in the following score function: 
\begin{equation*}
\nabla \ell_r = 
\left(
\begin{array}{c}
L\mathrm{vec}(\bm{\Sigma}^{-1}\bm{Z}_r\bm{\Sigma}^{-1} - \bm{\Sigma}^{-1})\\
3(\log L+1) + \log\vert\bm{Z}_r\vert - \log\vert\bm{\Sigma}\vert - \sum_{j=0}^{2}\psi^0(L-j) - \mathrm{tr}(\bm{\Sigma}^{-1}\bm{Z}_r)
\end{array}
\right),
\end{equation*}
where $\psi^0$ is the digamma function.
Let $\{\bm{Z}_1, \bm{Z}_2, \dots, \bm{Z}_N \}$ be an i.i.d.\ random sample of size $N$ from the $\mathcal{W}(\bm{\Sigma},L)$ law.
The ML estimator $(\widehat{\bm{\Sigma}},\widehat{L})$ of its parameters is 
$
\widehat{\bm{\Sigma}} = \overline{\bm{Z}} = N^{-1}\Sigma_{r=1}^{N}\bm{Z}_r
$
and the solution of 
\begin{equation}
 3\log\widehat{L} + \frac1N\sum_{r=1}^{N}\log\vert\bm{Z}_r\vert - \log\vert\bm{\overline{Z}}\vert - \sum_{q=0}^{2}\psi^0(\widehat{L}-q)=0.\label{eq:MLENL}
\end{equation}
The case $L<q$ was also treated by Anfinsen et al.~\cite{EstimationEquivalentNumberLooksSAR}.

\section{Nonlocal Means}\label{sec:nonlocal}

The NL-means method was proposed by Buades et al.~\cite{Buades2005DenoisingAlgorithms} based on the redundancy of neighboring patches in images.
In this method, the noise-free estimated value of a pixel is defined as a weighted mean of pixels in a certain region.
Under the additive white Gaussian noise assumption, these weights are calculated based on Euclidean distances which are used to measure the similarity between a central region patch and neighboring patches in a search window.
The filtered pixel is computed as:
\begin{equation}
  g(x,y) = \frac{\sum_{u,v \in W} f(x+u,y+v)\,w(u,v)}{\sum_{u,v \in W} w(u,v)},
  \label{eq:NLMeansMethod}
\end{equation}
where $w(u,v)$ are the weights defined on the search window $W$.
The resulting image $g$ is, thus, the convolution of the input image $f$ with the mask $w=w'/\sum w'(u,v)$.
The factor $w'(u,v)$ is inversely proportional to a distance between the patches, and is given by
\begin{equation*}
  w'(u,v) = \exp\Bigl\{ -\frac{1}{h} \sum_{k \in P}\lvert f(\rho_u(k))-f(\rho_v(k)) \rvert^2 \Bigr\},
  \label{eq:NLMeansWeight}
\end{equation*}
where $h>0$ controls the intensity, in a similar way the temperature controls the Simulated Annealing algorithm, $f(\rho_u(k))$ and $f(\rho_v(k))$ are the observations in the $k$-th pixels of the patches centered in $u$ and $v$, respectively.
When $h\to\infty$ the weights tend to be equal, while when $h\to0$ they tend to zero unless  $f(\rho_u(k))=f(\rho_v(k))$.
In the first case the filter becomes a mean over the search window; in the last case, the filtered value will remain unaltered.

The NL-Means proposal and its extensions rely on computing the weights of a convolution mask as functions of similarity measures: the closer (in some sense) two patches are, the heavier the contribution of the central pixel to the filter.

Deledalle et al.~\cite{CompareNoisyPatchesbeyondGaussian} analyze several similarity criteria for data which depart from the Gaussian assumption.
In particular, the authors consider the Gamma and Poisson noises, because they are good image models.
Deledalle et al.~\cite{Deledalle2009IterativeDenoising} extended the NL-means method to speckled imagery using statistical inference in an iterative procedure.
The authors derive the weights using the likelihood function of Gaussian and square root of Gamma (termed ``Nakagami-Rayleigh'') noises.
This idea is extended by Deledalle et al.~\cite{Deledalle2010PolSAREstimationNLM} to PolSAR data under the complex Gaussian distribution, and by Su et al.~\cite{TwoStepMultitemporalNLMSAR} to multitemporal PolSAR data.

Chen et al.~\cite{Chen2011NonLocalFilteringPolSAR} presented a NL-Means filter for PolSAR data under the complex Wishart distribution.
The authors employ the likelihood ratio test of equality between two $\mathcal{W}(\bm{\Sigma},L)$ laws with the same number of looks $L$, as presented by Conradsen et al.~\cite{Conradsen2003testStatisticWishart}, to calculate the weights between the patches.

Our proposal addresses the problem in a more general fashion: we use the $p$-value of a goodness-of-fit test between two samples.
The tests are derived using an Information Theory approach to compute stochastic divergences, which are turned into distances and then scaled to exhibit good asymptotic properties: they obey a $\chi^2$ distribution.
The weights are computed with a soft threshold which incorporates all the observations which were not rejected by the test, and some of the others.
As presented in the next section, all these elements can be easily generalized to other models, tests and weight functions.

\section{Stochastic Distances Filter}\label{sec:distances}

\subsection{The weights}\label{sec:weights}

\begin{figure}[hbt]
\centering
\begin{tikzpicture}
\node () at (1.75,1.75)[fill=blue!06,shape=rectangle,draw=black,text width=2.25cm,text height=2.25cm]{};
\node () at (1.75,1.75)[fill=blue!25,shape=rectangle,draw=black,text width=1.25cm,text height=1.25cm]{};
\node () at (1.75,1.75)[]{\footnotesize$\bm{Z}_1$};

\node () at (0.75,2.75)[fill=blue!15,shape=rectangle,draw=black,text width=1.25cm,text height=1.25cm]{};
\node () at (0.75,2.75)[]{\footnotesize$\bm{Z}_2$};

\node () at (1.25,2.25)[fill=blue!35,shape=rectangle,draw=black,text width=0.25cm,text height=0.25cm]{};
\node () at (1.75,1.75)[shape=rectangle,draw=black,line width=.4mm,text width=2.25cm,text height=2.25cm]{};

  \draw[very thick,dashed,line width=.23mm] (0.0,-0.5) -- (0.0,4.0);
  \draw[very thick,dashed,line width=.23mm] (0.5,-0.5) -- (0.5,4.0);
  \draw[very thick,dashed,line width=.23mm] (1.0,-0.5) -- (1.0,4.0);
  \draw[very thick,dashed,line width=.23mm] (1.5,-0.5) -- (1.5,4.0);
  \draw[very thick,dashed,line width=.23mm] (2.0,-0.5) -- (2.0,4.0);
  \draw[very thick,dashed,line width=.23mm] (2.5,-0.5) -- (2.5,4.0);
  \draw[very thick,dashed,line width=.23mm] (3.0,-0.5) -- (3.0,4.0);
  \draw[very thick,dashed,line width=.23mm] (3.5,-0.5) -- (3.5,4.0);
  
  \draw[very thick,dashed,line width=.23mm] (-0.5,0.0) -- (4.0,0.0);
  \draw[very thick,dashed,line width=.23mm] (-0.5,0.5) -- (4.0,0.5);
  \draw[very thick,dashed,line width=.23mm] (-0.5,1.0) -- (4.0,1.0);
  \draw[very thick,dashed,line width=.23mm] (-0.5,1.5) -- (4.0,1.5);
  \draw[very thick,dashed,line width=.23mm] (-0.5,2.0) -- (4.0,2.0);
  \draw[very thick,dashed,line width=.23mm] (-0.5,2.5) -- (4.0,2.5);
  \draw[very thick,dashed,line width=.23mm] (-0.5,3.0) -- (4.0,3.0);
  \draw[very thick,dashed,line width=.23mm] (-0.5,3.5) -- (4.0,3.5);
  
  \draw[-, red,very thick,line width=.23mm] (0.93,2.75) -- (2.75,2.75) -- (0.75,2.25) -- (2.75,2.25) -- (0.75,1.75) -- (1.25,1.75);
  \draw[->,red,very thick,line width=.23mm] (2.25,1.75) -- (2.75,1.75) -- (0.75,1.25) -- (2.75,1.25) -- (0.75,0.75) -- (2.75,0.75);
\end{tikzpicture} 
\vspace*{-.5cm}
\caption{Central pixel $\bm{Z}_1$ and its neighboring $\bm{Z}_i$, $i=\{2,\dots,25\}$ with patches $3\times3$ pixels and filtering window $5\times5$ pixels.}\label{fig:patches}
\end{figure}
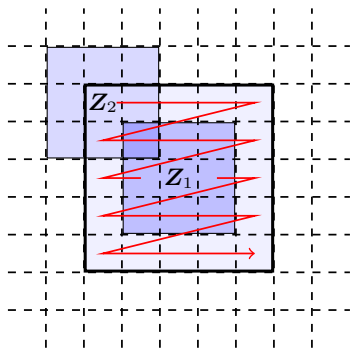

In this paper we use a $7\times7$ search window.
The shape and size of the neighboring patches and the central patch are the same: squares of $3\times3$ pixels. 
The central patch, with center pixel $\bm{Z}_1$, is thus compared with $24$ neighboring patches, whose center pixels are $\bm{Z}_i$, $i=2,\dots,25$, as illustrated in Figure~\ref{fig:patches}.

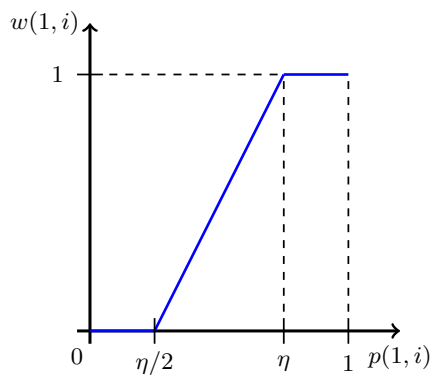
\begin{figure}[hbt]
\centering
\begin{tikzpicture}[scale=1.7]
\node () at (-0.1,-0.2)[]{\footnotesize$0$};
\node () at (-0.25,2.0)[]{\footnotesize$1$};
\node () at (-0.35,2.4)[]{\footnotesize$w(1,i)$};
\node () at (0.5,-0.25)[]{\footnotesize$\eta/2$};
\node () at (1.5,-0.25)[]{\footnotesize$\eta$};
\node () at (2.0,-0.25)[]{\footnotesize$1$};
\node () at (2.4,-0.20)[]{\footnotesize$p(1,i)$};
  \draw[->,very thick,line width=.4mm] (-0.1,0) -- (2.4,0);
  \draw[->,very thick,line width=.4mm] (0,-0.1) -- (0,2.4);
  \draw[blue,very thick,line width=.35mm] (0,0) -- (.5,0);
  \draw[blue,very thick,line width=.35mm] (.5,0) -- (1.5,2);
  \draw[very thick,line width=.23mm] (-0.1,2) -- (0.1,2);
  \draw[very thick,line width=.23mm] (0.5,-0.1) -- (0.5,0.1);
  
  \draw[blue,very thick,line width=.35mm] (1.5,2) -- (2,2);
  \draw[very thick,line width=.23mm] (1.5,-0.1) -- (1.5,0.1);
  \draw[very thick,dashed,line width=.23mm] (1.5,0.15) -- (1.5,2);
  
  \draw[very thick,line width=.23mm] (2,-0.1) -- (2,0.1);
  \draw[very thick,dashed,line width=.23mm] (2,0.15) -- (2,2);
  \draw[very thick,dashed,line width=.23mm] (0.15,2) -- (1.5,2);
\end{tikzpicture} 
\vspace*{-.5cm}
\caption{Weight function.}\label{fig:WeightFunction}
\end{figure}

The estimate of the noise-free observation at $\bm{Z}_1$ is a weighted sum of the observations at $\bm{Z}_2,\dots,\bm{Z}_{25}$, being each weight a function of the $p$-value ($p({1,i})$) observed in a test of same distribution between two complex Wishart laws: 
\begin{equation}
 w(1,i) = 
\left\{
\begin{array}{ll} 
1  &\text{if }\; p({1,i}) \geq \eta, \\ 
\frac2\eta p({1,i})-1 & \text{if }\; \frac{\eta}{2} < p({1,i}) < \eta, \\
0 & \text{otherwise},
\end{array}
\right.
\label{eq:weightNLMeansTorres}
\end{equation}
where $\eta$ is the significance of the test, specified by the user.
By definition, $p(1,1)=1$.
This function is illustrated in Figure~\ref{fig:WeightFunction}.

In this way we employ a soft threshold instead of an accept-reject decision.
This allows the use of more evidence than with a binary decision, as the one used in~\cite{Torres2012IGARSS,PolarimetricSARImageSmoothingStochasticDistancesCIARP2012} which was $1$ if the sample was not rejected and $0$ otherwise.

When all weights are computed, the mask of convolution coefficients is scaled to add one.

This setup can be generalized in many ways, among them:
\begin{itemize}
\item The support, i.e., the set of positions $i$ do not need to be local; it can extend arbitrarily.
\item The shape and size of the local windows.
\item The weight function; we opted for the piecewise linear function presented in equation~\eqref{eq:weightNLMeansTorres} as a good compromise between generality and computational cost.
\item The test which produces each $p$-value.
\end{itemize}

The rationale for using the weight function specified in equation~\eqref{eq:weightNLMeansTorres} is the following.
If the $p$-values were used instead, the presence of a single sample with excellent match to the central sample would dominate the weights in the mask, forcing other samples that were not rejected by the test to be practically discarded.
For instance, consider the case $p(1,i_1)=0.89$, $p(1,i_2)=0.05$ and all other $p$-values close to zero.
Without the weight function, the nonzero weights would be, approximately, $0.46$ and $0.03$, so the second observation would have a negligible influence on the filtered value while the first, along with the central value, dominate the result.
Using the aforementioned expression, the three weights would equal $1/3$.
This increases the smoothing effect without loosing the discriminatory ability.

\subsection{The statistical test}

As defined in the previous section, filtering each pixel requires computing a number of goodness-of-fit tests between the patch around the central pixel $\bm{Z}_1$ and the patches surrounding pixels $\bm{Z}_i$, $2\leq i\leq 25$.
This will be performed using tests derived from stochastic distances between samples.

Denote $\bm{\widehat{\theta}}_1$ the estimated parameter in the central region $\bm{Z}_1$, and $\big(\bm{\widehat{\theta}}_2,\ldots,\bm{\widehat{\theta}}_{25}\big)$ the estimated parameters in the remaining areas.
To account for possible de\-par\-tures from the homogeneous model, we estimate $\bm{\widehat{\theta}}=(\widehat{\bm{\Sigma}},\widehat{L})$ by maximum like\-li\-hood.

The proposal is based on the use of stochastic distances between the patches.
Consider that $\bm{Z}_1$ and $\bm{Z}_i$, $2\leq i\leq 25$, are random matrices defined on the same probability space, whose distributions are characterized by the den\-si\-ties $f_{\bm{Z}_1}(\bm{Z}';\bm{\theta}_1)$ and $f_{\bm{Z}_i}(\bm{Z}';\bm{\theta}_i)$, res\-pec\-ti\-ve\-ly, where $\bm{\theta}_1$ and $\bm{\theta}_i$ are parameters.
Assuming that the densities have the same support given by the cone of Hermitian positive definite matrices $\bm{\mathcal{A}}$, the $h$-$\phi$ divergence between $f_{\bm{Z}_1}$ and $f_{\bm{Z}_i}$ is given by
\begin{equation*}
 D_{\phi}^{h}(\bm{Z}_1,\bm{Z}_i) = h \Big( \int_{\bm{\mathcal{A}}}\phi \Big( \frac{f_{\bm{Z}_1}(\bm{Z}';\bm{\theta}_1)}{f_{\bm{Z}_i}(\bm{Z}';\bm{\theta}_i)} \Big) \;f_{\bm{Z}_i}(\bm{Z}';\bm{\theta}_i)\;\mathrm{d}\bm{Z}' \Big),
\end{equation*}
where $h\colon(0,\infty)\rightarrow[0,\infty)$ is a strictly increasing function with $h(0)=0$ and $h'(x)>0$ for every $x \in \mathbbm{R}$, and $\phi\colon (0,\infty)\rightarrow[0,\infty)$ is a convex function~\cite{Salicru1994}.
Choices of functions $h$ and $\phi$ result in several divergences.

Divergences sometimes are not symmetric.
A simple solution, described in Frery et al.~\cite{Frery2011InformationTheoryPolSAR,Nascimento2010,AnalyticExpressionsStochasticDistancesBetweenRelaxedComplexWishartDistributions}, is to define a new measure $d_{\phi}^{h}$ given by
\begin{equation*}
 d_{\phi}^{h}(\bm{Z}_1,\bm{Z}_i) = \frac{D_{\phi}^{h}(\bm{Z}_1,\bm{Z}_i)+D_{\phi}^{h}(\bm{Z}_i,\bm{Z}_1)}{2}.
\end{equation*}
Distances, in turn, can be conveniently scaled to present good statistical prop\-er\-ties that make them suitable as test statistics~\cite{Salicru1994}:
\begin{equation}
 \mathcal{S}_{\phi}^{h}(\bm{\widehat{\theta}}_1,\bm{\widehat{\theta}}_i) = 
 \frac{2mn}{(m+n)h'(0)\phi''(1)}\;d^{h}_{\phi}(\bm{\widehat{\theta}}_1,\bm{\widehat{\theta}}_i),
\label{eq:GeneralTestStatistic}
\end{equation}
where $\bm{\widehat{\theta}}_1$ and $\bm{\widehat{\theta}}_i$ are maximum likelihood estimators based on samples size $m$ and $n$, respectively.
When $\bm\theta_1=\bm\theta_i$, under mild conditions $\mathcal{S}_{\phi}^{h}(\bm{\widehat{\theta}}_1,\bm{\widehat{\theta}}_i)$ is asymptotically $\chi^2_M$ distributed, being $M$ the dimension of $\bm{\theta}_1$.
Ob\-serv\-ing $\mathcal{S}_{\phi}^{h}(\bm{\widehat{\theta}}_1,\bm{\widehat{\theta}}_i)$, the (asymptotic) $p$-value of the test is $p=\Pr( \chi^2_{M}>\mathcal{S}_{\phi}^{h}(\bm{\widehat{\theta}}_1,\bm{\widehat{\theta}}_i))$, and the null hypothesis $\bm{\theta}_1=\bm{\theta}_i$ can be rejected at significance level $\eta$ if $p\leq \eta$;
details can be seen in the work by Salicr{\'u} et al.~\cite{Salicru1994}.
Since we are using the same central sample for $24$ nontrivial tests (notice that $\mathcal{S}_{\phi}^{h}(\bm{\widehat{\theta}}_1,\bm{\widehat{\theta}}_1)=0$).

Frery et al.~\cite{Frery2011InformationTheoryPolSAR,AnalyticExpressionsStochasticDistancesBetweenRelaxedComplexWishartDistributions} obtained several distances between $\mathcal{W}(\bm{\Sigma},L)$ distributions.
The test used in this paper was derived from the Hellinger distance, yielding:
\begin{align}
 \mathcal{S}_{H}(\bm{\widehat{\theta}}_1,\bm{\widehat{\theta}}_i) =&{} \frac{8mn}{m+n}\;\Bigg[ 1- \frac{ 
 \Big\vert\big( \frac{ \widehat{L}_1\widehat{\bm{\Sigma}}_1^{-1} + \widehat{L}_i\widehat{\bm{\Sigma}}_i^{-1}}{2} \big)^{-1} \Big\vert^{(\widehat{L}_1+\widehat{L}_i)/2}}
 {\vert\widehat{\bm{\Sigma}}_1\vert^{\widehat{L}_1/2}\; \vert\widehat{\bm{\Sigma}}_i\vert^{\widehat{L}_i/2}}\;
 \sqrt{\widehat{L}_1^{3\widehat{L}_1}\; \widehat{L}_i^{3\widehat{L}_i}}\nonumber\\
& \prod_{q=0}^{2}\frac{\Gamma(\frac{\widehat{L}_1+\widehat{L}_i}{2}-q)}{\sqrt{\Gamma(\widehat{L}_1-q)\; \Gamma(\widehat{L}_i-q)}} \Bigg],\label{eq:HellingerTestStatistic}
\end{align}
provided $L\geq 3$.
If $L<3$, the following expression can be used
\begin{equation*}
\mathcal{S}_{H} (\bm{\widehat{\theta}}_1,\bm{\widehat{\theta}}_i) = \frac{8mn}{m+n}\Bigg[ 1-\bigg( \frac{\big\vert \big( \frac{\widehat{\bm{\Sigma}}_1^{-1} + \widehat{\bm{\Sigma}}_i^{-1}}{2} \big)^{-1} \big\vert}{\sqrt{\vert \widehat{\bm{\Sigma}}_1\vert\;\vert \widehat{\bm{\Sigma}}_i\vert}} \bigg)^{(\widehat{L}_1+\widehat{L}_i)/2} \Bigg];
\end{equation*}
it was also derived by Frery et al.~\cite{Frery2011InformationTheoryPolSAR,AnalyticExpressionsStochasticDistancesBetweenRelaxedComplexWishartDistributions} for the case when $L_1=L_2=L$; in this last case, the equivalent number of looks $\widehat L$ should be estimated using all the available data, for instance as presented by Anfinsen et al.~\cite{EstimationEquivalentNumberLooksSAR}.
The setup presented in Section~\ref{sec:weights} leads to $m=n=9$ since squared windows of side $3$ are used.

The tests based on the Kullback-Leibler, Bhattacharyya and R\'enyi of order $\beta\in\{.1, .5, .9\}$ were also used.
They produced almost exactly the same results as of the Hellinger distance, at the expense of more computational load.

Although the distribution of the test statistics given in equation~\eqref{eq:GeneralTestStatistic} is only known in the limit when $m,n\to\infty$ such that $m/(m+n)\to\lambda\in(0,1)$, it has been observed that the difference between the asymptotic and empirical distributions is negligible in samples as small as the ones considered in this setup~\cite{AnalyticExpressionsStochasticDistancesBetweenRelaxedComplexWishartDistributions,Frery2011InformationTheoryPolSAR}.

The filter obtained using the $p$-values produced by the test statistic given in equation~\eqref{eq:HellingerTestStatistic} can be applied iteratively.
The complex Wishart distribution is preserved by convolutions, and since the number of looks is estimated in every pairwise comparison, the evolution of the filtered data is always controlled.
This also holds for any filter derived from $h$-$\phi$ divergences and the setup presented in Section~\ref{sec:weights}.

Computational information about our implementation is provided in~\ref{app:ComputationalInformation}.

\section{Image Quality Assessment}\label{sec:assessment}

According to Wang et al.~\cite{ImageAssessmentDifficult} image quality assessment in general, and filter performance evaluation in particular, are hard tasks and crucial for most image processing applications.

We assess the filters using simulated and real input data by means of visual inspection and quantitative measures.
Visual inspection includes edges, small features and color preservation.
The measures are computed on the $H\!H$, $H\!V$ and $V\!V$ intensity channels, and they amount to
(a)~the Equivalent Number of Looks (ENL) in homogeneous areas; and
(b)~the Structural Similarity Index Measure ($\operatorname{SSIM}$), proposed by Wang et al.~\cite{Wang2004SSIMindex} for measuring the similarity between two images.
(c)~the Blind/Referenceless Image Spatial QUality Evaluator ($\operatorname{BRISQUE}$), proposed by Mittal et al.~\cite{Mittal2012NoReferenceIQASpatialDomain} to a holistic measure of quality on no-reference images.

The equivalent number of looks $\widehat N$ can be estimated by maximum likelihood solving equation~\eqref{eq:MLENL}.
The ENL although it is a measure of the signal to noise ratio, and it must be considered carefully since it will repute highly blurred data as excellent: for this measure, the best result is a completely flat (constant) image, i.e., an image with no information whatsoever.
The Boxcar filter usually has a higher ENL than other proposals because it will always use all the pixels in the filtering window and the result is a large loss of spatial resolution.
Therefore, the equivalent number of looks should be used with care and only to assess noise reduction and not image quality in general.

The $\operatorname{SSIM}$ index is a structural information measure that represents the structure of objects in the scene, regardless the average luminance and contrast.
The $\operatorname{SSIM}$ index takes into account three factors:
(I)~correlation between edges;
(II)~brightness distortion; and
(III)~distortion contrast.
Let $f$ and $g$ be the original data and the filtered version, respectively, the $\operatorname{SSIM}$ index is expressed by 
\begin{equation*}
\operatorname{SSIM}(f,g)=
\frac{\operatorname{Cov}(f,g)+C_1}{\widehat{\sigma}_{f} \widehat{\sigma}_{g}+C_1} 
 \frac{2\overline{f}\overline{g}+C_2}{\overline{f}^2+\overline{g}^2+C_2} 
 \frac{2 \widehat{\sigma}_{f} \widehat{\sigma}_{g}+C_3}{\widehat{\sigma}_{f}^2+\widehat{\sigma}_{g}^2+C_3},
\end{equation*}
where $\overline{f}$ and $\overline{g}$ are sample means, $\widehat{\sigma}_{f}^2$ and $\widehat{\sigma}_{g}^2$ are the sample variances, $\operatorname{Cov}(f,g)$ is the sample covariance between $f$ and $g$, and the constants $C_1,C_2,C_3$ are intended to stabilize the index.
Following~\cite{Wang2004SSIMindex}, we used the following values: $C_1=(K_1\mathcal{L})^2$, $C_2=(K_2\mathcal{L})^2$ and $C_3=C_2/2$, where $\mathcal{L}$ is the observed dynamic range and both $K_1,K_2\ll 1$; we used $K_1=0.01$ and $K_2=0.03$.

The $\operatorname{SSIM}$ is defined for scalar-valued images and it ranges in the $[-1,1]$ interval, being the bigger value observed is the better result.
It was computed on each intensity channel as the means over squared windows of side $8$, and the reported value is the mean over the three channels.
This index is a particular case of the Universal Quality Index proposed by Wang and Bovik~\cite{WangBovik:02}.
We also applied this last measure to all the images here discussed, and the results were in full agreement with those reported by the $\operatorname{SSIM}$ index.

The $\operatorname{BRISQUE}$ is a model that operates in the spatial domain and requires no-reference image.
This image quality evaluator does not compute specific distortions such as ringing, blurring, blocking, or aliasing, but quantifies possible losses of ``naturalness'' in the image. 
This approach is based on the principle that natural images possess certain regular statistical properties that are measurably modified by the presence of distortions. No transformation to another coordinate frame (DFT, DCT, wavelets, etc) is required, distinguishing it from previous blind/no-reference approaches.
The $\operatorname{BRISQUE}$ is defined for scalar-valued images and it ranges in the $[0,100]$ interval, and smaller values indicate better results.

\section{Results}\label{sec:results}

The proposed filter, termed ``SDNLM (\textit{Stochastic Distances Nonlocal Means}) filter'' was compared with the Refined Lee (Scattering-model-based), IDAN (intensity-driven adaptive-neighborhood) and Boxcar filters.
These filters act on a well defined neighborhood (as our implementation of the SDNLM): a squared search window of side of $5\times5$ pixels, and the former is adaptive (as is the SDNLM).

\subsection{Simulated data}

\subsubsection{Sampling from the Wishart distribution}\label{sec:SamplingWishartDistribution}

The simulated data was obtained mimicking real data from six classes on a phantom,  a segmented image of size $496\times496$.
The original data were produced by a polarimetric sensor aboard the R99-B Brazilian Air Force aircraft in October 2005 over Campinas, S\~ao Paulo State, Brazil.
The sensor operates in the L-band, and produces imagery with four nominal looks.
The data were simulated in single look, i.e., in the lowest signal-to-noise possible configuration.
The covariance matrices are reported in~\ref{app:CovarianceMatrices}.

\begin{figure}[hbt]
\centering
  \subfigure[Phantom\label{fig:phantom}]{\includegraphics[width=.45\linewidth]{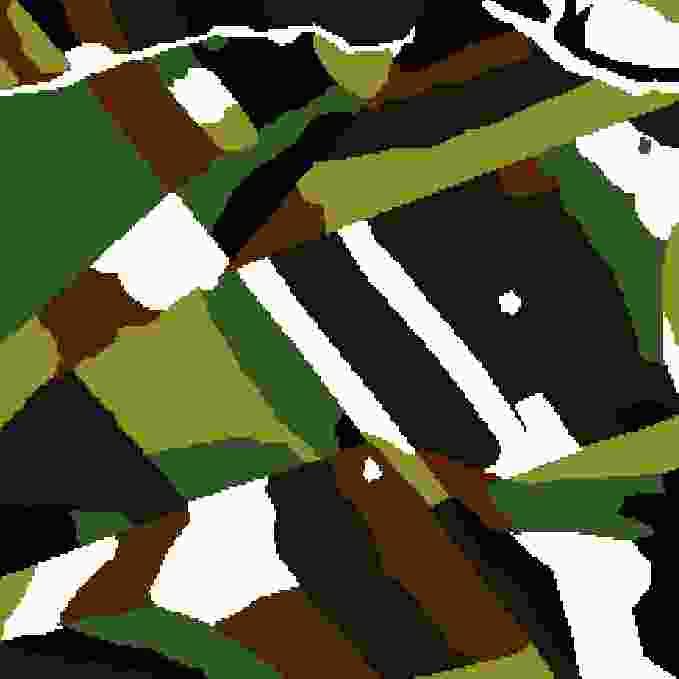}}
  \subfigure[Simulated single-look data\label{fig:R99B}]{\includegraphics[width=.45\linewidth]{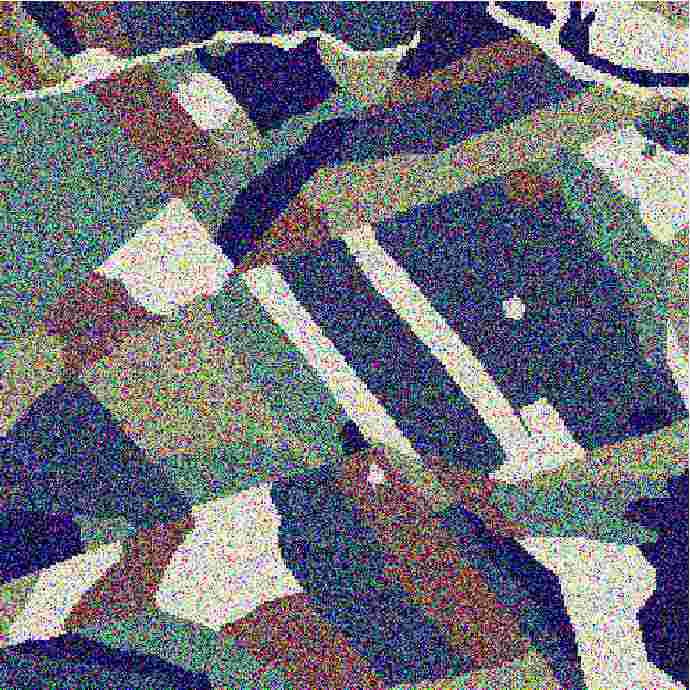}}
  \subfigure[Boxcar filter\label{fig:R99B_Boxcar}]{\includegraphics[width=.45\linewidth]{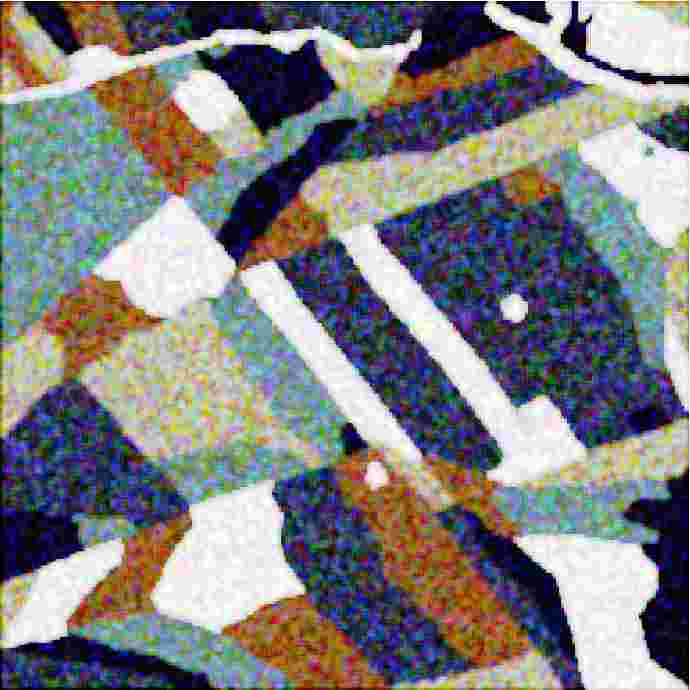}}
  \subfigure[Refined Lee filter\label{fig:R99B_Lee}]{\includegraphics[width=.45\linewidth]{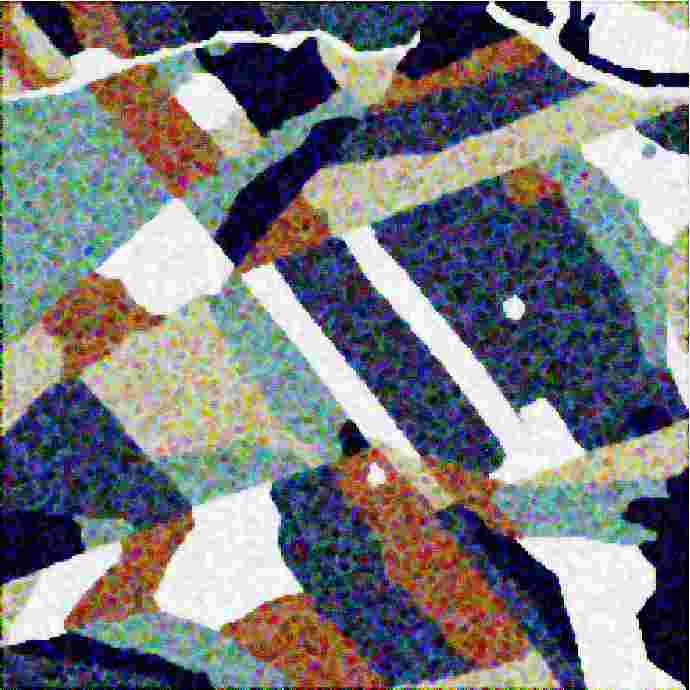}}
  \subfigure[IDAN filter\label{fig:R99B_IDAN}]{\includegraphics[width=.45\linewidth]{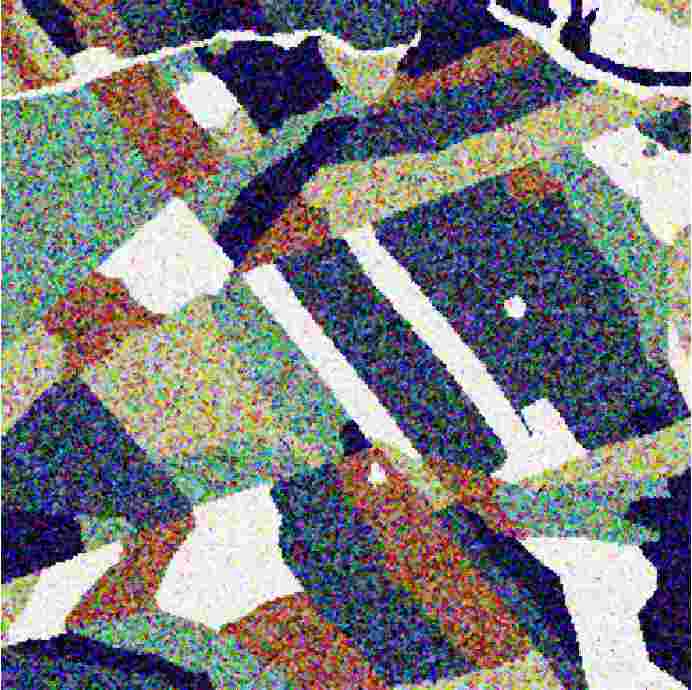}}
  \subfigure[SDNLM filter\label{fig:R99B_SDNLM}]{\includegraphics[width=.45\linewidth]{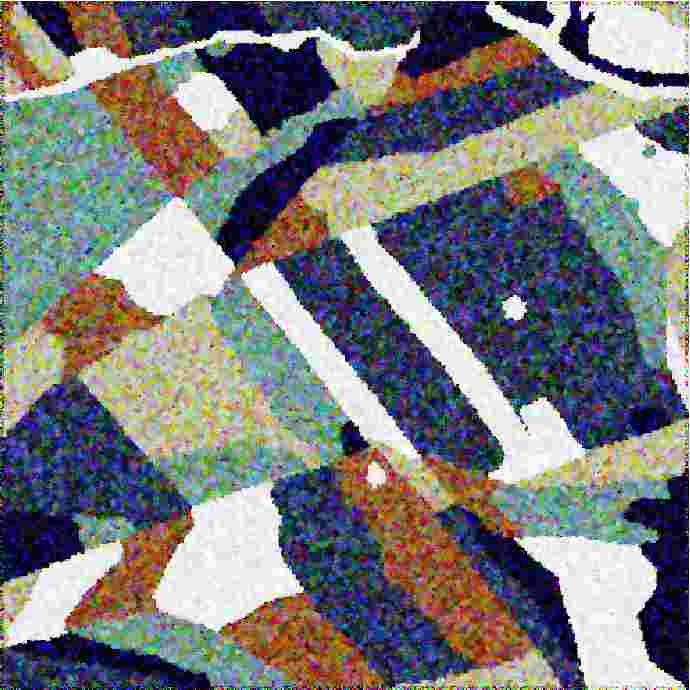}}
\caption{Original single-look PolSAR simulated data and filtered versions after one iteration.}
\label{fig:SimulImg_R99B}
\end{figure}

Figures~\ref{fig:phantom} and~\ref{fig:R99B} show, respectively, the phantom and simulated single-look PolSAR image with with false color using $\vert S_{H\!H} \vert^2$ in the Red channel, $\vert S_{H\!V} \vert^2$ in the Green channel, and $\vert S_{V\!V} \vert^2$ in the Blue channel.
The filtered versions with the Boxcar, Refined Lee, IDAN and SDNLM filters are shown in Figures~\ref{fig:R99B_Boxcar}, \ref{fig:R99B_Lee}, \ref{fig:R99B_IDAN} and~\ref{fig:R99B_SDNLM}, respectively.
The latter was obtained at the $\eta=90\%$ confidence level.

The noise reduction is noticeable in all the four filtered images.
The speckled effect was mitigated without loosing the color balance, but the Boxcar filter produces a blurred image.
The Refined Lee filter also introduces some blurring, but less intense that the Boxcar filter.
The IDAN filter also reduces noise, but less effectively that the Refined Lee.
The SDNLM filter is the one which best preserves edges.
This is particularly noticeable in the fine strip which appears light, almost white, to the upper left region of the image.
After applying the Boxcar filter it appears wider than it is.
The star-shaped bright spot in the middle of the blue field looses detail after Boxcar, Refined Lee and IDAN filters are applied, while the SDNLM noise reduction technique preserves its details well.
The Refined Lee filter introduces a noticeable pixellation effect which is evident mainly in the brown areas.

\begin{figure}[hbt]
\centering
  \subfigure[Phantom edges\label{fig:phantom_canny}]{\includegraphics[width=.45\linewidth,angle=270]{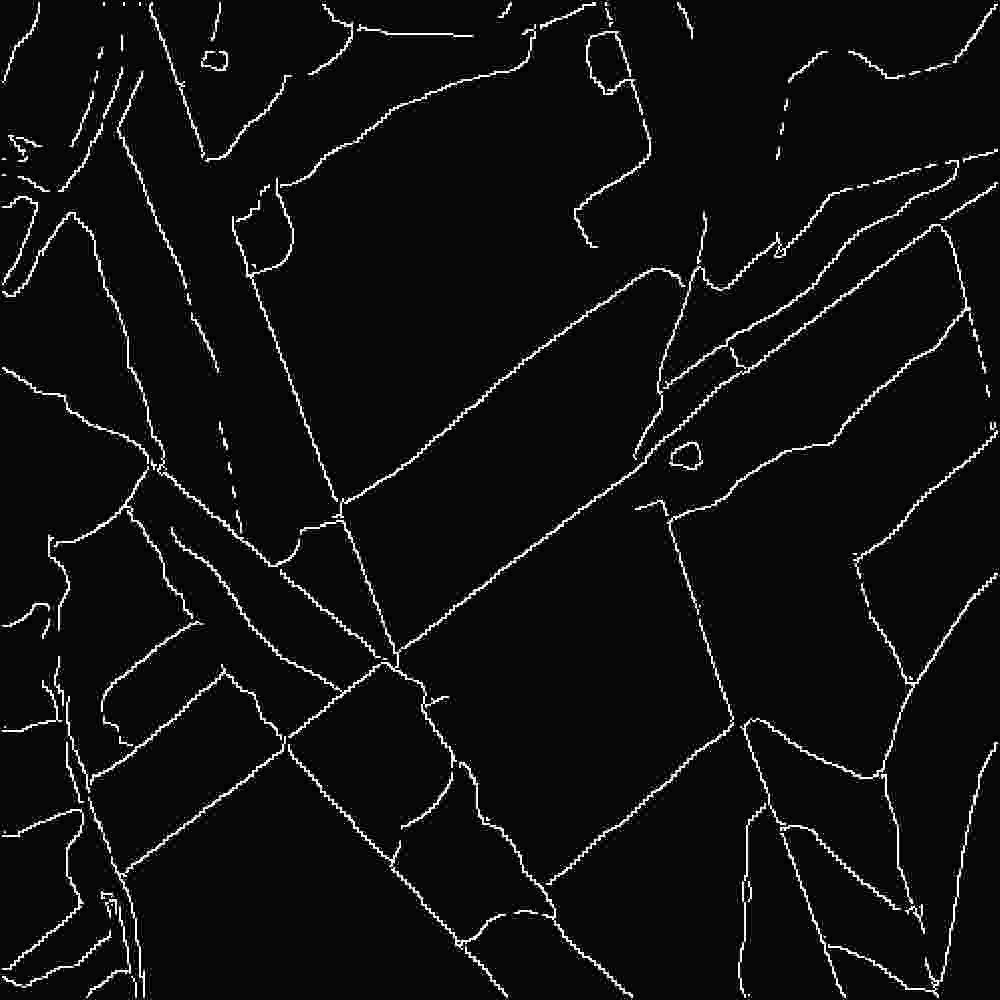}}
  \subfigure[Edges of the simulated $H\!H$ band \label{fig:R99B_canny}]{\includegraphics[width=.45\linewidth,angle=270]{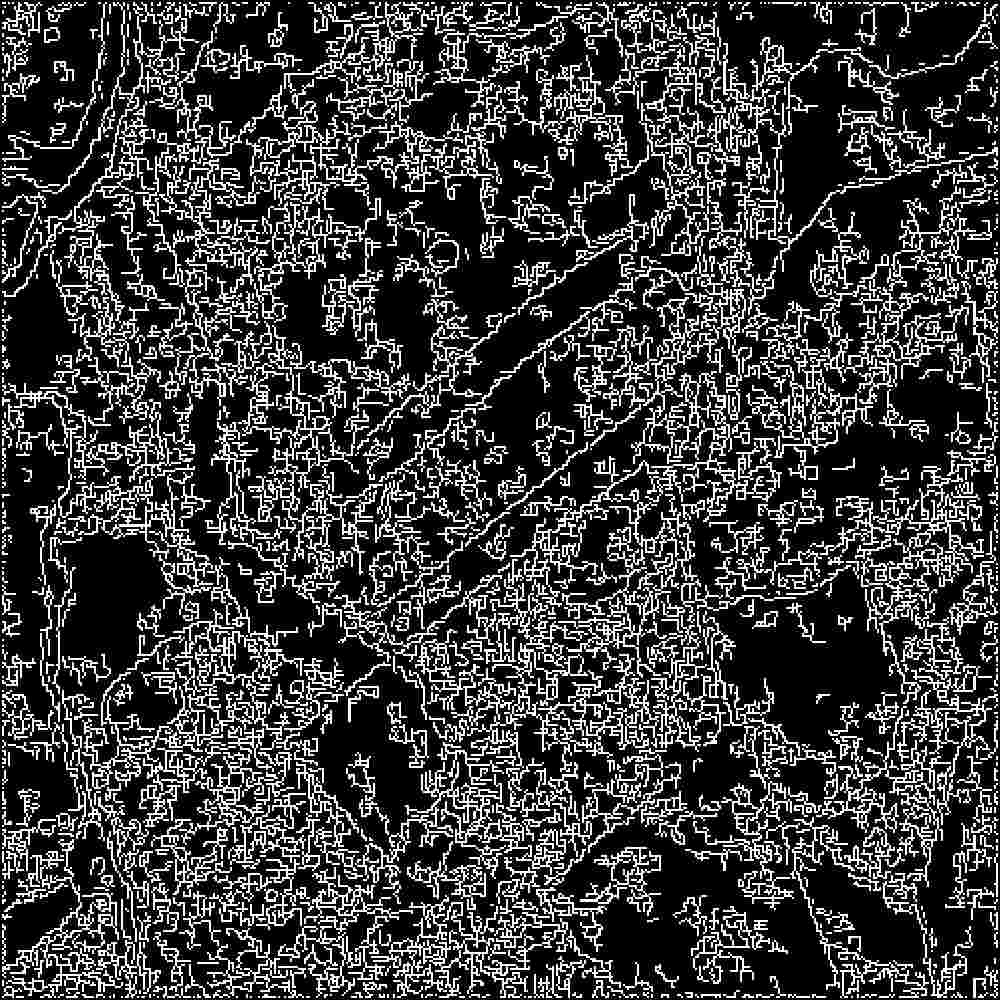}}  
  \subfigure[Boxcar filter edges\label{fig:R99B_Boxcar_canny}]{\includegraphics[width=.45\linewidth,angle=270]{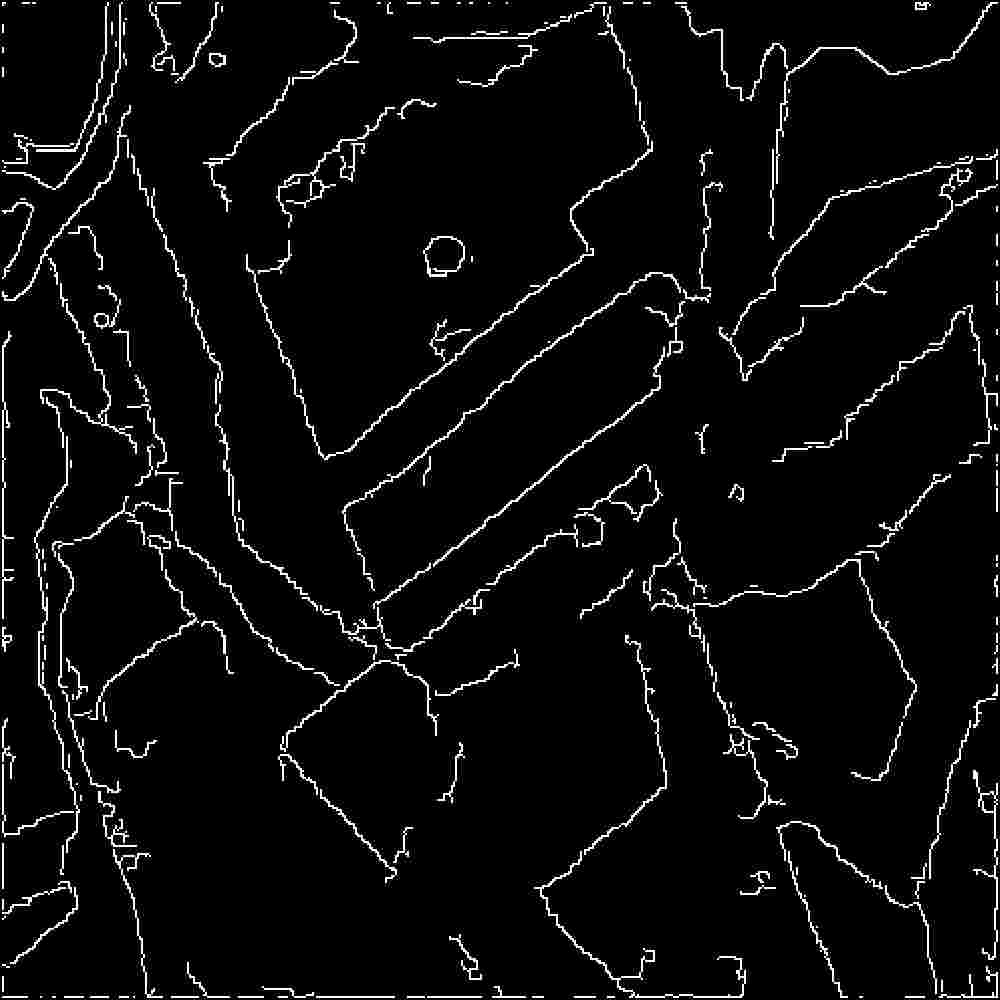}}
  \subfigure[Refined Lee filter edges\label{fig:R99B_Lee_canny}]{\includegraphics[width=.45\linewidth,angle=270]{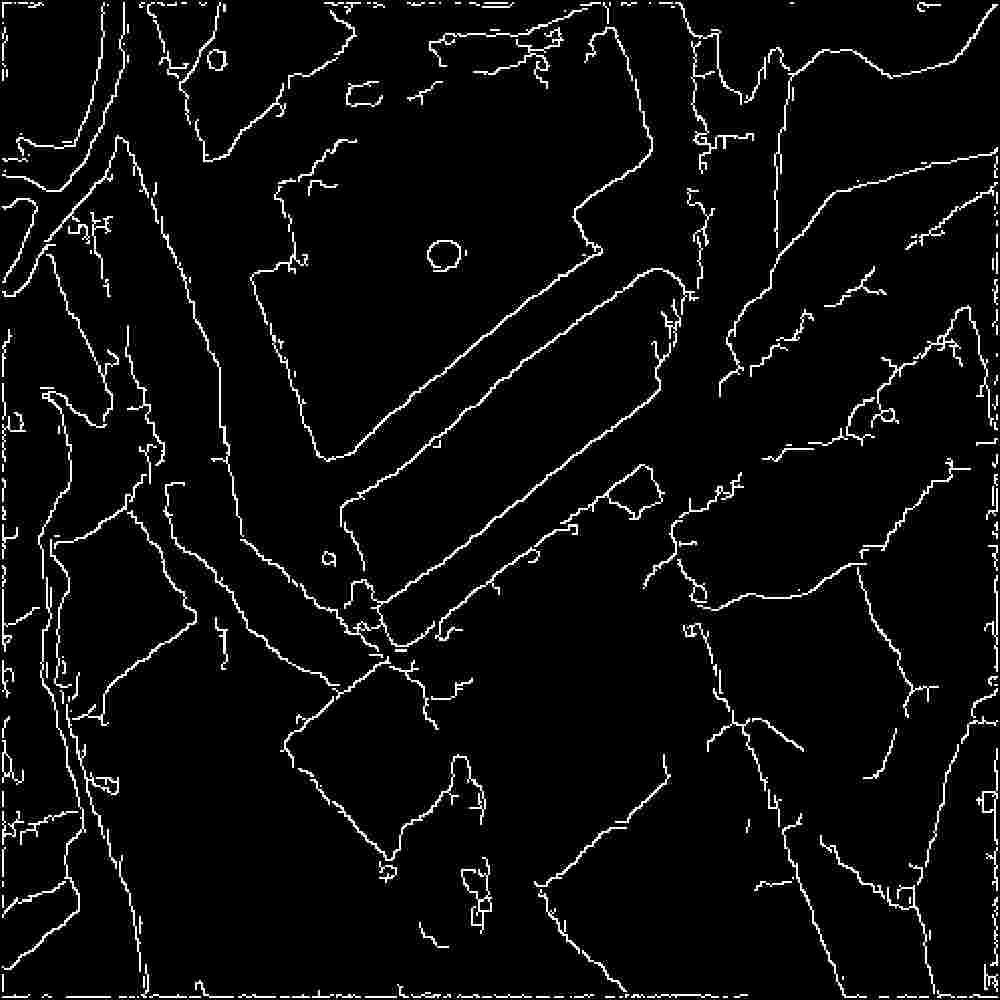}}
  \subfigure[IDAN filter edges\label{fig:R99B_IDAN_canny}]{\includegraphics[width=.45\linewidth,angle=270]{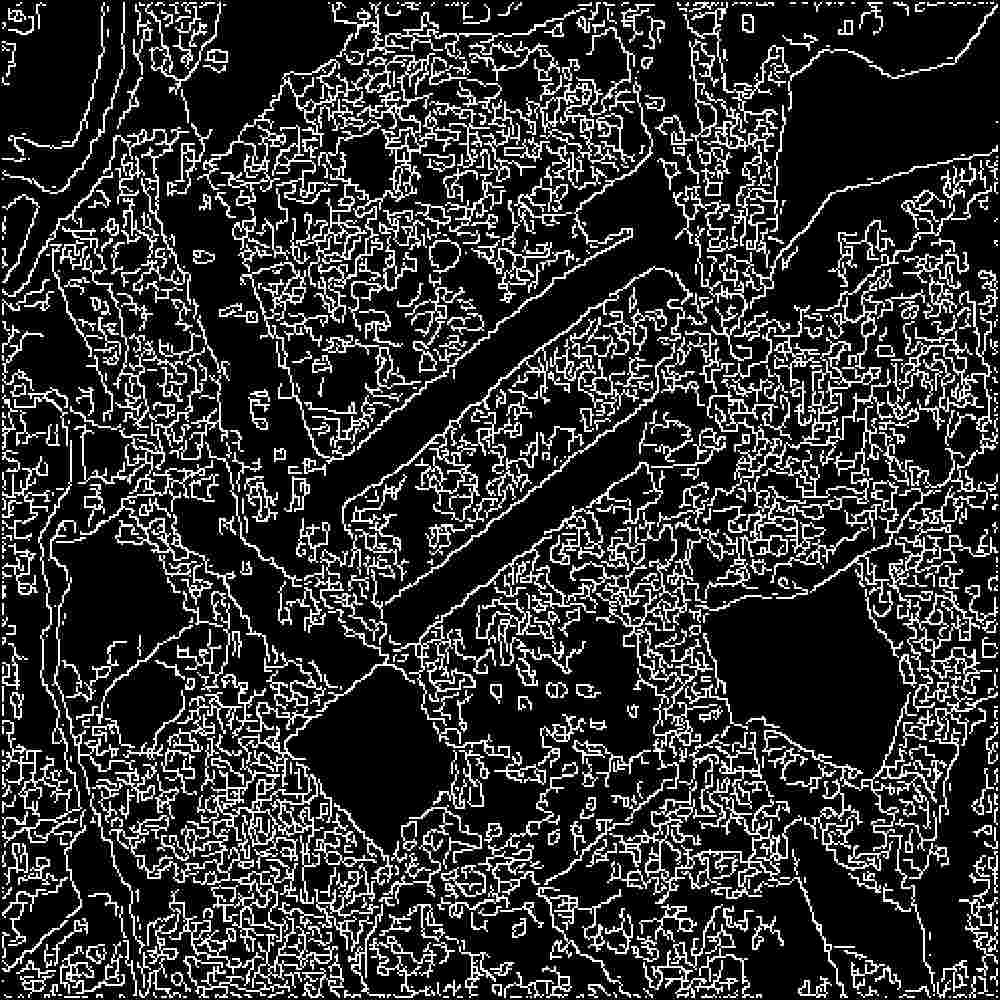}}
  \subfigure[SDNLM filter edges\label{fig:R99B_SDNLM_canny}]{\includegraphics[width=.45\linewidth,angle=270]{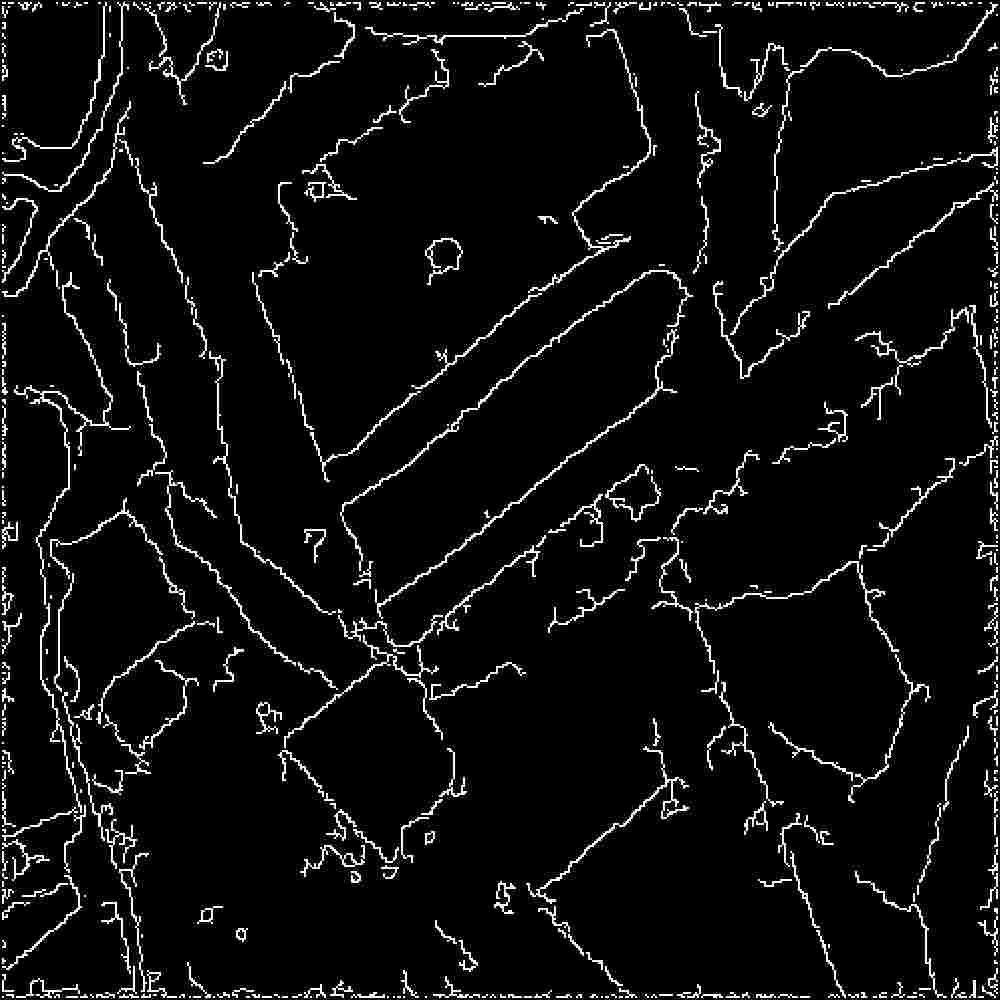}}
\caption{Edges detected by the Canny filter applied to the $H\!H$ polarization channel of the original and filtered images.}
\label{fig:SimulImg_R99B_canny}
\end{figure}

The Canny detector was applied to the $H\!H$ band of all the available data presented in Figure~\ref{fig:SimulImg_R99B}, and the results are presented in Figure~\ref{fig:SimulImg_R99B_canny}.
Even when applied to the phantom, the Canny detector fails to produce continuous edges in a few situations.
In particular, it completely misses the small light features and a few edges between regions; see Figure~\ref{fig:phantom_canny}.
Edge detection with this method is, in practice, impossible using the original single-look data, cf.\ Figure~\ref{fig:R99B_canny} with the other results, although a few large linear features are visible in the clutter of noisy edges.
As expected, the edges detected in the Boxcar filtered image are smooth, but they neither grant continuity nor identify fine details; this information is lost by the filter, see Figure~\ref{fig:R99B_Boxcar_canny}.
The edges detected in the image processed by the IDAN filter are only marginally better than those observed in the original, unfiltered, data; see Figure~\ref{fig:R99B_IDAN_canny}.
Figures~\ref{fig:R99B_Lee_canny} and~\ref{fig:R99B_SDNLM_canny} are the edges detected on the data filtered by the Refined Lee and SDNLM filters, respectively.
Although they look alike, it is noticeable that the latter preserves better the small details; see, for instance, the star-shaped object to the center-right of the image.
It appears round in the former, while in the latter it is possible to identify minute variations.

The phantom is available, so it is possible to make a quantitative assessment of the results.
Table~\ref{tab:QIndexesSimulImg_R99B} presents the assessment of the filters in the three intensity channels $H\!H$, $H\!V$ and $V\!V$ of the data presented in Figure~\ref{fig:SimulImg_R99B} using the Equivalent Number of Looks (ENL) and the $\operatorname{SSIM}$ index.

\begin{table}[hbt]
\centering
\caption{Image quality indexes in the images shown in Figure~\ref{fig:SimulImg_R99B}.}
\begin{tabular}{c r@{.}l r@{.}l r@{.}l r@{.}l r@{.}l r@{.}l}
\toprule
\multirow{2}{*}{\textbf{Filter}} & \multicolumn{6}{c}{\textbf{ENL}} & \multicolumn{6}{c}{\textbf{$\operatorname{SSIM}$ Index}} \\ \cmidrule(lr{.5em}){2-7} \cmidrule(lr{.5em}){8-13}
    & \multicolumn{2}{c}{$H\!H$} & \multicolumn{2}{c}{$H\!V$} & \multicolumn{2}{c}{$V\!V$} & \multicolumn{2}{c}{$H\!H$} & \multicolumn{2}{c}{$H\!V$} & \multicolumn{2}{c}{$V\!V$} \\
\midrule
Boxcar      & \textbf{15}&\textbf{696} & 5&768  & \textbf{25}&\textbf{111} &  0&083 & 0&038 & 0&083 \\
Refined Lee & 11&665 & \textbf{10}&\textbf{136} & 14&398 & 0&164 & 0&092 & 0&144 \\ 
IDAN        &  2&164 &  3&171 &  1&977 & 0&199  & 0&137 & 0&188 \\ 
SDNLM 80\%  &  7&269 &  5&999 & 11&217 & \textbf{0}&\textbf{234}& \textbf{0}&\textbf{150} & \textbf{0}&\textbf{230} \\  
SDNLM 90\%  &  8&786 &  6&578 & 13&559 & 0&181  & 0&101 & 0&177 \\ 
SDNLM 99\%  & 14&429 &  7&129 & 23&787 & 0&101  & 0&055 & 0&101 \\ 
\bottomrule 
\end{tabular}
\label{tab:QIndexesSimulImg_R99B}
\end{table}

The ENL was estimated on homogeneous areas far from edges, so no smudging from other areas contaminated these values.
As expected, the most intense blurring produces the best results with respect to this criterion: the Boxcar filter outperforms in two out of three bands and, when, it is not the best, the Refined Lee filter is.
Regarding the Equivalent Number of Looks, IDAN produces worse results, but a good performance in $\operatorname{SSIM}$ index.
The SDNLM filter improves with respect to the ENL criterion when the significance level increases.
In order to make it competitive with the Boxcar, Refined Lee and IDAN filters, the application to real data was done using $\eta=90\%$. 
Regarding the $\operatorname{SSIM}$ index, our proposal consistently outperforms the other three filters and, as expected, the smallest the significance of the test the better the performance is with respect to this criterion since the least the image is blurred.

\subsubsection{Physical-based simulation}\label{sec:PhysicalBasedSimulation}

Sant'Anna et al.~\cite{SantAnna2008Sensors} proposed a methodology for simulating PolSAR imagery taking into account the electromagnetic characteristics of the targets and of the sensing system. 
The simulated images are more realistic than the ones obtained by merely stipulating the posterior distribution given the classes, in particular spatial correlation among pixels emerges and mixture of classes in the borders as observed.

Each simulated pixel is a complex scattering matrix based on a phantom image (an idealized cartoon model) with five distinct regions.
Multifrequency sets of single-look PolSAR images have been generated in the \textsf{L}-, \textsf{C}- and \textsf{X}-bands, corresponding to \unit[1.25], \unit[5.3] and \unit[9.6]{GHz}, respectively.
The acquisition geometry is that of an airborne monostatic sensor flying at \unit[6,000]{m} of altitude and \unit[35]{\textdegree} grazing angle imaging a \unit[$290\times290$]{m$^2$} area terrain. 
The \unit[3.0]{m} spatial resolution and \unit[2.8]{m} pixel spacing were set in the range and the azimuth directions. 
The data have $128\times128$ pixels; details in~\cite{SantAnna2008Sensors}.

\begin{figure}[hbt]
\centering
  \subfigure[Simulated \textsf{L}-band\label{fig:Band_L}]{\includegraphics[width=.18\linewidth]{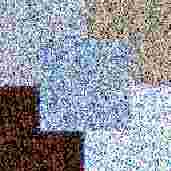}}  
  \subfigure[Boxcar filter\label{fig:Band_L_Boxcar}]{\includegraphics[width=.18\linewidth]{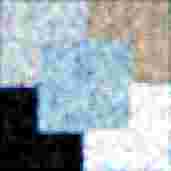}}   
  \subfigure[Refined Lee filter\label{fig:Band_L_Lee}]{\includegraphics[width=.18\linewidth]{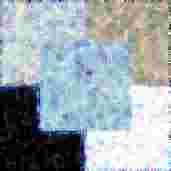}}
  \subfigure[IDAN filter\label{fig:Band_L_IDAN}]{\includegraphics[width=.18\linewidth]{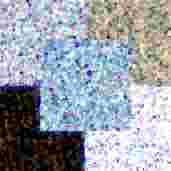}}
  \subfigure[SDNLM filter\label{fig:Band_L_SDNLM}]{\includegraphics[width=.18\linewidth]{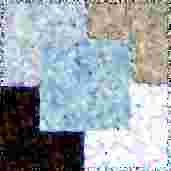}}\\
  \subfigure[Simulated \textsf{C}-band\label{fig:Band_C}]{\includegraphics[width=.18\linewidth]{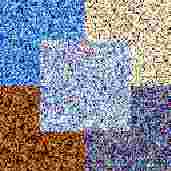}}   
  \subfigure[Boxcar filter\label{fig:Band_C_Boxcar}]{\includegraphics[width=.18\linewidth]{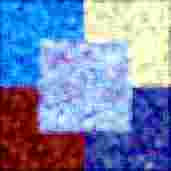}}   
  \subfigure[Refined Lee filter\label{fig:Band_C_Lee}]{\includegraphics[width=.18\linewidth]{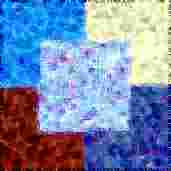}}
  \subfigure[IDAN filter\label{fig:Band_C_IDAN}]{\includegraphics[width=.18\linewidth]{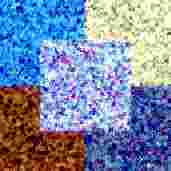}}
  \subfigure[SDNLM filter\label{fig:Band_C_SDNLM}]{\includegraphics[width=.18\linewidth]{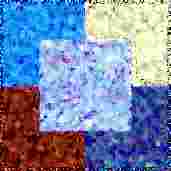}}\\
  \subfigure[Simulated \textsf{X}-band\label{fig:Band_X}]{\includegraphics[width=.18\linewidth]{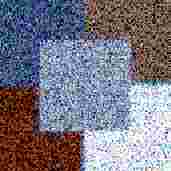}}   
  \subfigure[Boxcar filter\label{fig:Band_X_Boxcar}]{\includegraphics[width=.18\linewidth]{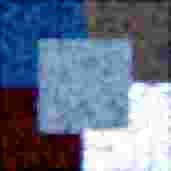}}  
  \subfigure[Refined Lee filter\label{fig:Band_X_Lee}]{\includegraphics[width=.18\linewidth]{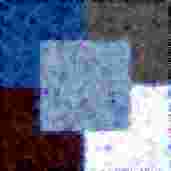}}
  \subfigure[IDAN filter\label{fig:Band_X_IDAN}]{\includegraphics[width=.18\linewidth]{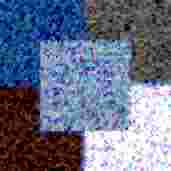}}
  \subfigure[SDNLM filter\label{fig:Band_X_SDNLM}]{\includegraphics[width=.18\linewidth]{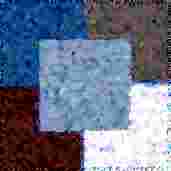}} 
\caption{Original physical-based single-look images and their filtered versions after one iteration.}
\label{fig:SimulImgBands}
\end{figure}

Figure~\ref{fig:SimulImgBands} shows the simulated images in the \textsf{L}-, \textsf{C}- and \textsf{X}-bands (Figures~\ref{fig:Band_L}, \ref{fig:Band_C} and~\ref{fig:Band_X}, resp.) and its filtered versions by the 
Boxcar (Figures~\ref{fig:Band_L_Boxcar}, \ref{fig:Band_C_Boxcar} and~\ref{fig:Band_X_Boxcar}, resp.), 
Refined Lee (Figures~\ref{fig:Band_L_Lee}, \ref{fig:Band_C_Lee} and~\ref{fig:Band_X_Lee}, resp.), IDAN (Figures~\ref{fig:Band_L_IDAN}, \ref{fig:Band_C_IDAN} and~\ref{fig:Band_X_IDAN}, resp.) and SDNLM filters (Figures~\ref{fig:Band_L_SDNLM}, \ref{fig:Band_C_SDNLM} and~\ref{fig:Band_X_SDNLM}, resp.) with $\eta=90\%$.

The three main visual drawbacks of the Boxcar, Refined Lee and IDAN filters are noticeable in these results, namely the excessive blurring introduced by the former, and the pixellate effect produced by the latter.
The SDNLM filter consistently presents a good compromise between smoothing and edge preservation.

Table~\ref{tab:QIndexesSimulatedImages} presents the assessment of the filters in three intensity channels by means of the ENL (computed in the central region, a homogeneous area), and by means of the  $\operatorname{SSIM}$ index. 
The best results are highlighted in boldface.
The results are consistent with those observed in Table~\ref{tab:QIndexesSimulImg_R99B}: the Boxcar filter is the best with respect to the ENL computed in homogeneous areas far from edges, but the SDNLM filter outperforms the other three filters when a more sophisticated metric is used: the $\operatorname{SSIM}$ index, which takes into account not only smoothness but structural information.
Regarding these data, the ideal significance level lies approximately between $\eta=80\%$ and $\eta=90\%$; these values provide a good smoothing without compromising the structural information.

\begin{table}[hbt]
\centering\footnotesize
\caption{Image quality indexes in the images shown in Figure~\ref{fig:SimulImgBands}.}\label{tab:QIndexesSimulatedImages}
\begin{tabular}{c c r@{.}l r@{.}l r@{.}l r@{.}l r@{.}l r@{.}l}
\toprule
\multirow{2}{*}{\rotatebox{90}{\textbf{band}}} & \multirow{2}{*}{\textbf{Filter}}  & \multicolumn{6}{c}{\textbf{ENL}} & \multicolumn{6}{c}{\textbf{$\operatorname{SSIM}$ Index}} \\ \cmidrule(lr{.5em}){3-8} \cmidrule(lr{.5em}){9-14}
  &  & \multicolumn{2}{c}{$H\!H$} & \multicolumn{2}{c}{$H\!V$} & \multicolumn{2}{c}{$V\!V$} & \multicolumn{2}{c}{$H\!H$} & \multicolumn{2}{c}{$H\!V$} & \multicolumn{2}{c}{$V\!V$} \\
\midrule
\multirow{6}{*}{\rotatebox{90}{\textsf{L}-}} 
  & Boxcar      & \textbf{19}&\textbf{294}& \textbf{21}&\textbf{952} & \textbf{23}&\textbf{072} & 0&067 & 0&077 & 0&115 \\ 
  & Refined Lee & 12&639 & 14&059 & 14&850 & 0&159 & 0&163 & 0&186 \\ 
  & IDAN        &  4&278 &  4&400 &  4&655 & 0&239 & 0&205 & 0&237 \\ 
  & SDNLM 80\%  &  8&683 &  9&619 &  9&588 & \textbf{0}&\textbf{246}& \textbf{0}&\textbf{231} & \textbf{0}&\textbf{251} \\ 
  & SDNLM 90\%  &  9&737 & 11&128 & 10&780 & 0&120 & 0&187 & 0&220 \\ 
  & SDNLM 99\%  & 17&559 & 20&359 & 21&125 & 0&100 & 0&168 & 0&140 \\ 
\midrule
\multirow{6}{*}{\rotatebox{90}{\textsf{C}-}} 
  & Boxcar      & \textbf{24}&\textbf{008}& \textbf{21}&\textbf{970} & \textbf{23}&\textbf{547} & 0&078 & 0&077 & 0&087 \\ 
  & Refined Lee & 14&305 & 12&947 & 12&434 & 0&177 & 0&165 & 0&144 \\
  & IDAN        &  2&626 &  3&079 &  4&492 & 0&268 & 0&253 & 0&243 \\ 
  & SDNLM 80\%  & 10&290 &  9&217 &  9&429 & \textbf{0}&\textbf{271}& \textbf{0}&\textbf{265} & \textbf{0}&\textbf{244} \\ 
  & SDNLM 90\%  & 11&840 & 10&599 & 10&801 & 0&212 & 0&211 & 0&194 \\ 
  & SDNLM 99\%  & 22&061 & 19&576 & 21&045 & 0&110 & 0&110 & 0&114 \\ 
\midrule
\multirow{6}{*}{\rotatebox{90}{\textsf{X}-}} 
  & Boxcar      & \textbf{18}&\textbf{553}& \textbf{23}&\textbf{125} & \textbf{23}&\textbf{747} & 0&080 & 0&102 & 0&154 \\ 
  & Refined Lee &  9&694 & 13&603 & 14&526 & 0&169 & 0&195 & 0&235 \\ 
  & IDAN        &  1&945 &  2&151 &  2&986 & 0&259 & 0&259 & 0&273 \\ 
  & SDNLM 80\%  &  8&348 &  9&463 &  9&574 & \textbf{0}&\textbf{267}& \textbf{0}&\textbf{270} & \textbf{0}&\textbf{300} \\ 
  & SDNLM 90\%  &  9&371 & 11&049 & 11&620 & 0&210 & 0&222 & 0&261 \\ 
  & SDNLM 99\%  & 17&023 & 21&082 & 21&547 & 0&121 & 0&131 & 0&190 \\ 
\bottomrule 
\end{tabular}
\end{table}

\subsection{Real data}\label{sec:RealData}

In the remainder of this section the data will be presented in false color using the Pauli decomposition \cite{LeePottier2009Book}.
This representation of PolSAR data has the advantage of being interpretable in terms of types of backscattering mechanisms.
It consists of assigning $\vert S_{H\!H} + S_{V\!V} \vert^2$ to the Red channel,  $\vert S_{H\!H} - S_{V\!V} \vert^2$ to the Green channel, and $2\vert S_{H\!V} \vert^2$ to the Blue channel.
This is one of many possible representations, and it is noteworthy that the data are filtered in their original domain, before being decomposed for visualization.

\begin{figure}[hbt]
\centering
  \subfigure[Map of the area\label{fig:FriscoGoogleMap}]{\includegraphics[width=.43\linewidth]{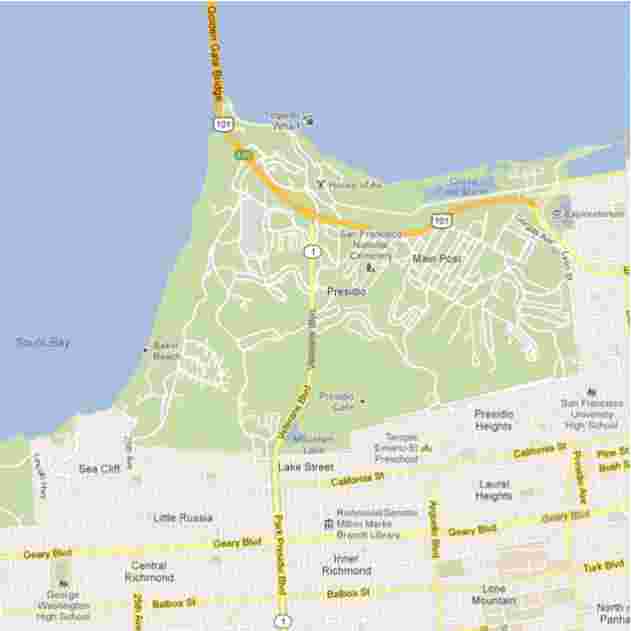}}
  \subfigure[AIRSAR \textsf{L}-band data\label{fig:Pauli_SanFran}]{\includegraphics[width=.43\linewidth,viewport= 335 335 695 695,clip=TRUE]{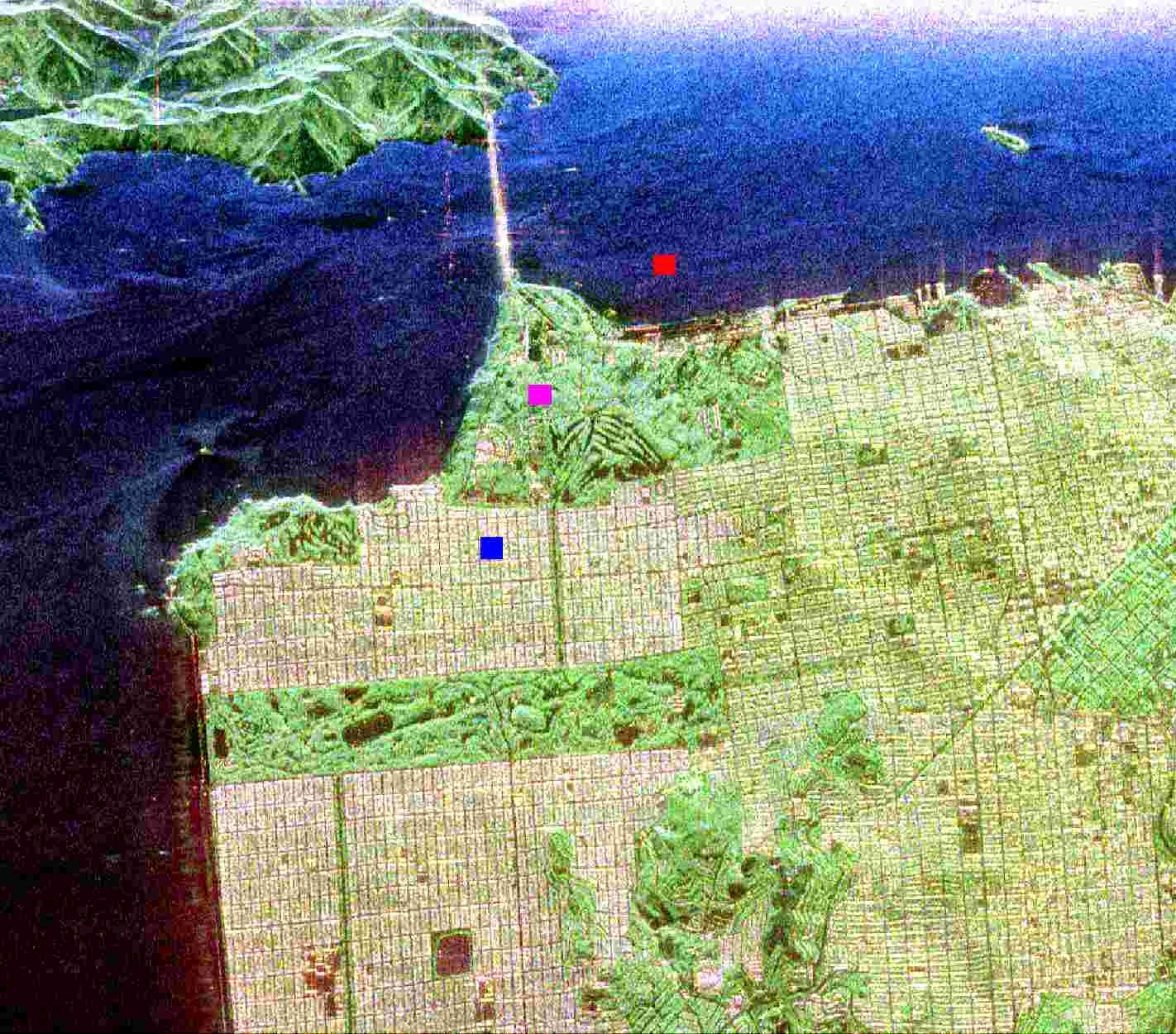}}
  \subfigure[Boxcar filter\label{fig:Pauli_SanFran_BOX}]{\includegraphics[width=.43\linewidth,viewport= 320 320 670 670,clip=TRUE]{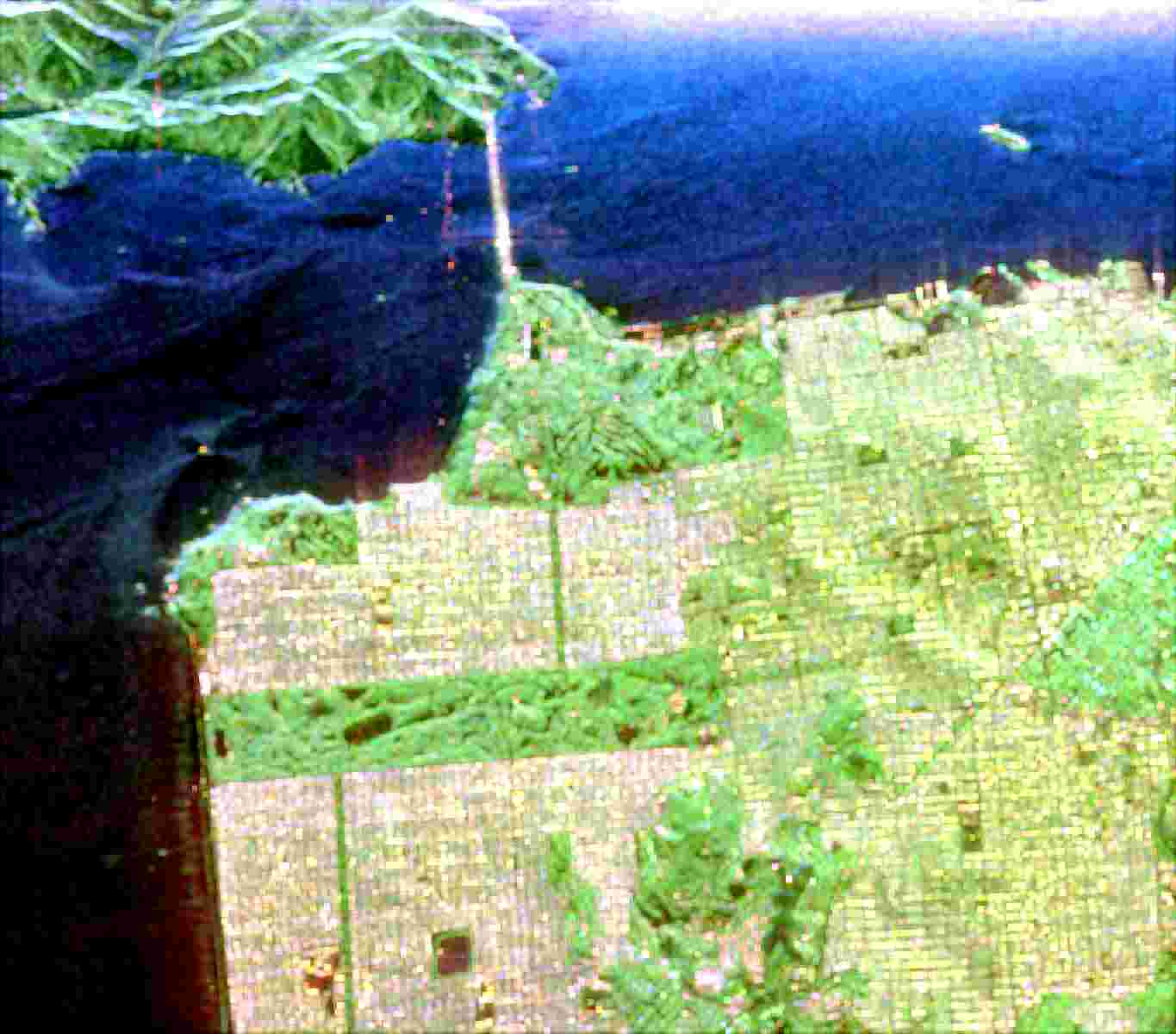}}
  \subfigure[Refined Lee filter\label{fig:Pauli_SanFran_LEE}]{\includegraphics[width=.43\linewidth,viewport= 320 320 670 670,clip=TRUE]{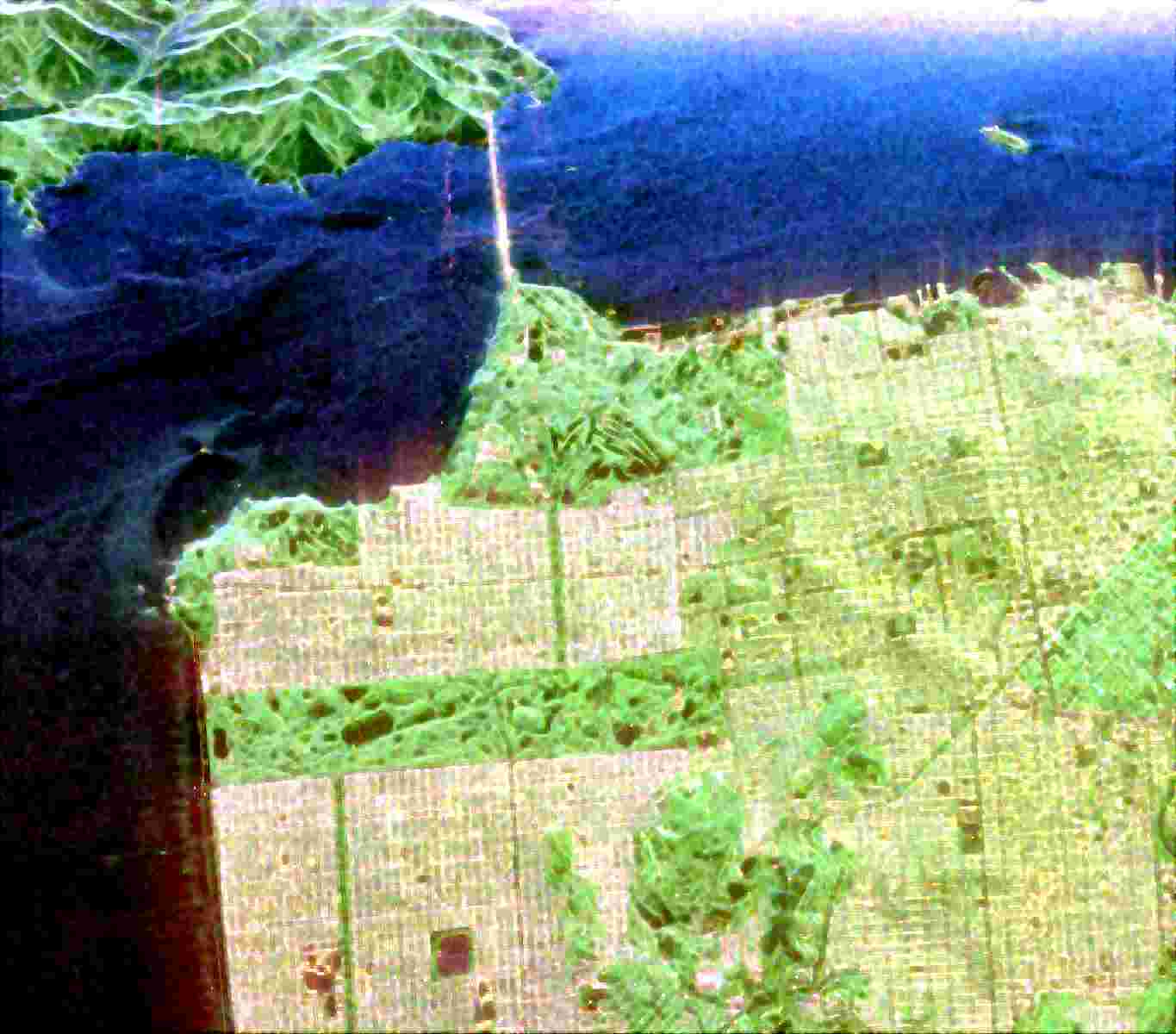}}
  \subfigure[IDAN filter\label{fig:Pauli_SanFran_IDAN}]{\includegraphics[width=.43\linewidth,viewport= 320 320 670 670,clip=TRUE]{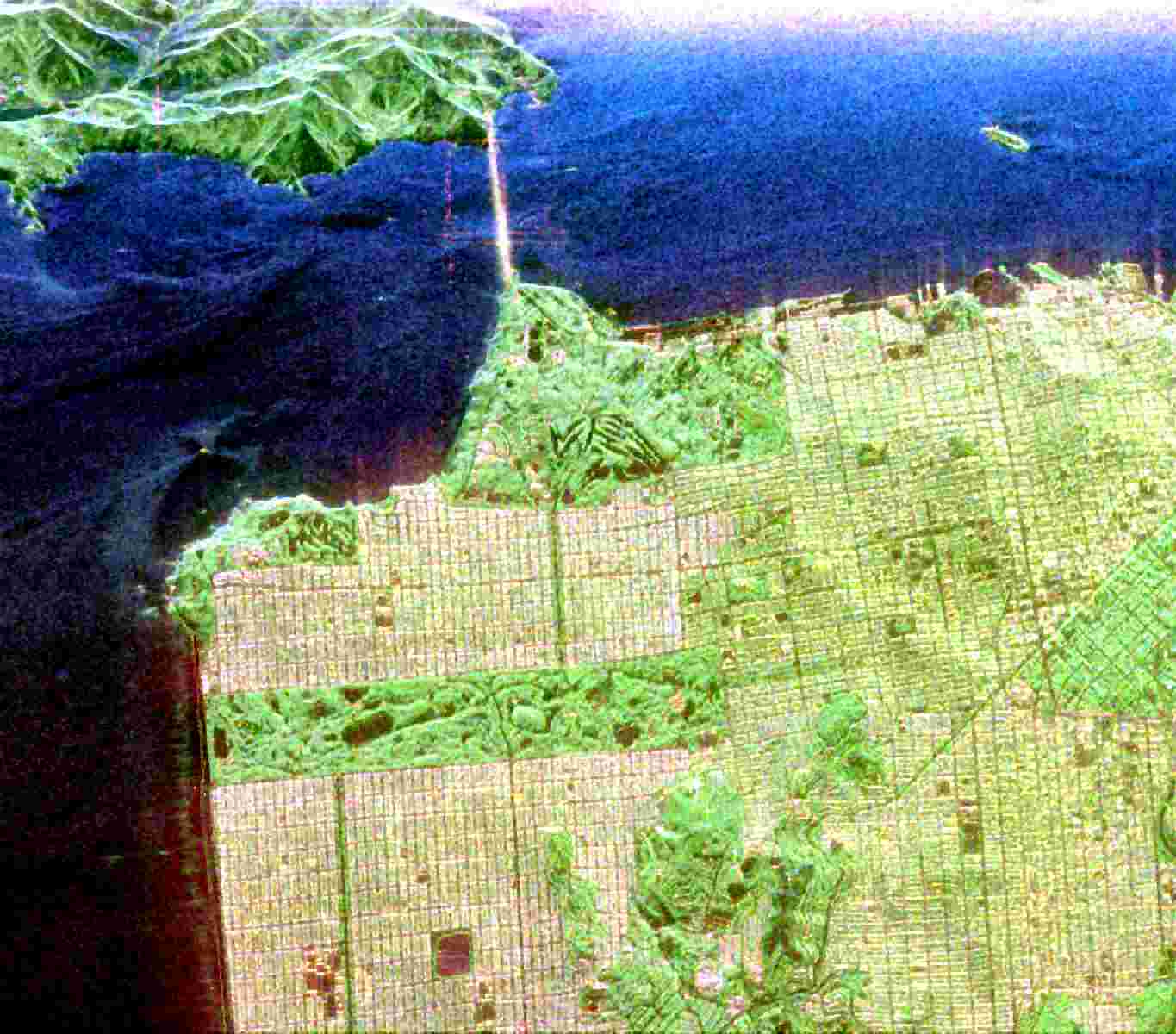}}  
  \subfigure[SDNLM filter\label{fig:Pauli_SanFran_SDNLM_90}]{\includegraphics[width=.43\linewidth,viewport= 320 320 670 670,clip=TRUE]{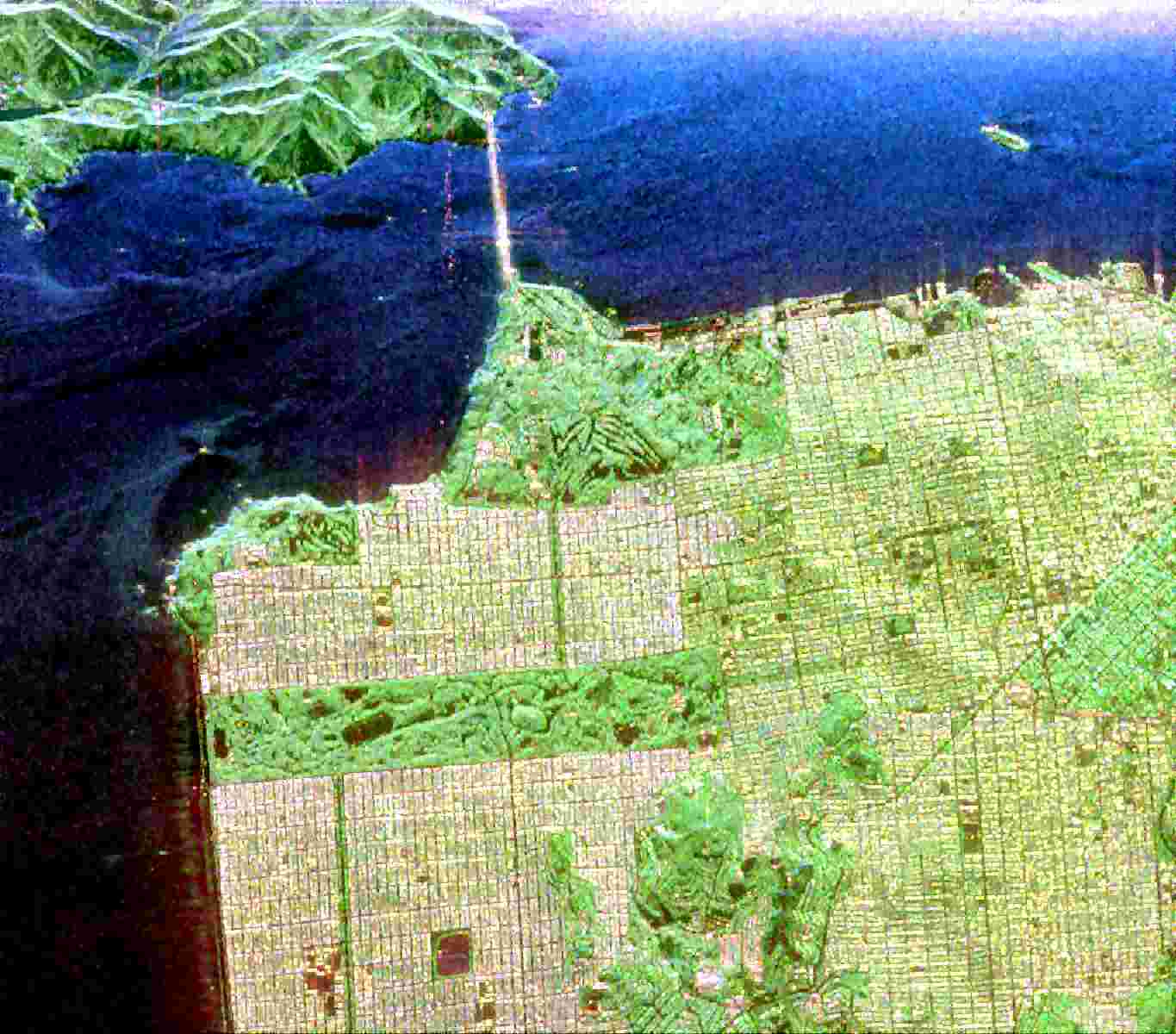}}
\caption{Pauli decomposition of the original AIRSAR image over San Francisco and its filtered versions.}
\label{fig:SanFran}
\end{figure}

A National Aeronautics and Space Administration Jet Propulsion Laboratory (NASA/JPL) Airborne SAR (AIRSAR) image of the San Francisco Bay was used for evaluating the proposed filter, see \url{http://earth.eo.esa.int/polsarpro/datasets.html}.
The original PolSAR image was generated in the \textsf{L}-band, four nominal looks, and \unit[$10\times10$]{m} spatial resolution.
The test region has $350\times350$ pixels, and is shown in Figure~\ref{fig:Pauli_SanFran}, along with a Google Map\textcopyright\ of the area (Figure~\ref{fig:FriscoGoogleMap}, see \url{http://goo.gl/maps/HJkPf}).

The four filters employ a kernel of $5\times5$ pixels, and the patches in our proposal are $3\times3$ windows.
Figure~\ref{fig:Pauli_SanFran_BOX} shows the effect of the Boxcar filter.
Albeit the noise reduction is evident, it is also clear that the blurring introduced eliminates useful information as, for instance, the Presidio Golf Course: the curvilinear dark features to the center of the image: a forested area.
Figure~\ref{fig:Pauli_SanFran_LEE} is the result of applying the Refined Lee filter, which shows a good performance, but some details in the edges are eliminated.
In particular, the Mountain Lake, the small brown spot to the center of the image, is blurred, as well as the blocks in the urban area.
The results of the IDAN filter and of our proposal with $\eta=90\%$ are shown in Figures~\ref{fig:Pauli_SanFran_IDAN} and~\ref{fig:Pauli_SanFran_SDNLM_90}, respectively.
Both filters are able to smooth the image in a selective way, but the SDNLM filter enhances more the signal-to-noise ratio while preserving fine details than the IDAN filter.

In image quality assessment, the $\operatorname{SSIM}$ requires a reference image, as was the case of sections~\ref{sec:SamplingWishartDistribution} and~\ref{sec:PhysicalBasedSimulation}, therefore this index is not applied with ease on real images. 
Table~\ref{tab:QIndexesRealImages} presents the result of assessing the filters in the three intensity channels by means of the ENL in the large forest area, a homogeneous area, and of the $\operatorname{BRISQUE}$ index.
Again, the results are consistent with what was observed before: the mere evaluation of the noise reduction by the ENL suggests the Boxcar filter as the best one, but the natural scene distortion-generic  $\operatorname{BRISQUE}$ index is better after applying the SDNLM filter in all intensity channels.

\begin{table}[hbt]
\centering\footnotesize
\caption{Image quality indexes in the real PolSAR image.}\label{tab:QIndexesRealImages}
\begin{tabular}{c r@{.}l r@{.}l r@{.}l r@{.}l r@{.}l r@{.}l}
\toprule
\multirow{2}{*}{\textbf{Filter}} & \multicolumn{6}{c}{\textbf{ENL}} & \multicolumn{6}{c}{\textbf{$\operatorname{BRISQUE}$ Index}} \\ \cmidrule(lr{.5em}){2-7} \cmidrule(lr{.5em}){8-13}
 & \multicolumn{2}{c}{$H\!H$} & \multicolumn{2}{c}{$H\!V$} & \multicolumn{2}{c}{$V\!V$} & \multicolumn{2}{c}{$H\!H$} & \multicolumn{2}{c}{$H\!V$} & \multicolumn{2}{c}{$V\!V$} \\
\midrule
Real data   &  3&867 &  4&227 &  4&494 & 58&258 & 70&845 & 61&593 \\ 
\midrule
Boxcar      & \textbf{14}&\textbf{564} & \textbf{25}&\textbf{611} & \textbf{18}&\textbf{946} & 36&498 & 37&714 & 36&792 \\ 
Refined Lee & 11&491 & 20&415 & 15&407 & 44&997 & 51&547 & 49&412 \\ 
IDAN        &  2&994 &  3&732 &  3&923 & 28&823 & 34&853 & 34&691 \\ 
SDNLM 80\%  &  7&263 & 11&532 &  8&299 & \textbf{27}&\textbf{841} & \textbf{27}&\textbf{256} & \textbf{33}&\textbf{541} \\ 
SDNLM 90\%  &  8&177 & 12&404 &  9&013 & 28&622 & 35&622 & 35&016 \\ 
SDNLM 99\%  & 10&828 & 18&379 & 13&075 & 31&026 & 33&881 & 36&881 \\ 
\bottomrule 
\end{tabular}
\end{table}

\subsection{Effect of the filters on the scattering characteristics}

Polarimetric target decompositions aim at expressing the physical properties of the scattering mechanisms.
Among them, the entropy-based decomposition proposed by Cloude and Pottier~\cite{Cloude1996TargetDecompositionTheorems} is widely used in PolSAR image classification.
It extracts two parameters from each observed covariance matrix: the scattering entropy $H\in[0,1]$, and $\overline{\alpha}\in[0,90]$, an indicator of the type of scattering.
The $(H,\overline{\alpha})$ plane is then divided into nine regions which provide both a classification rule and an interpretation for the observed data.

\begin{figure}[hbt]
\centering
  \subfigure[Entropy of the Original data\label{fig:entropy_SanFran}]{\includegraphics[width=.43\linewidth,viewport= 188 188 394 394,clip=TRUE]{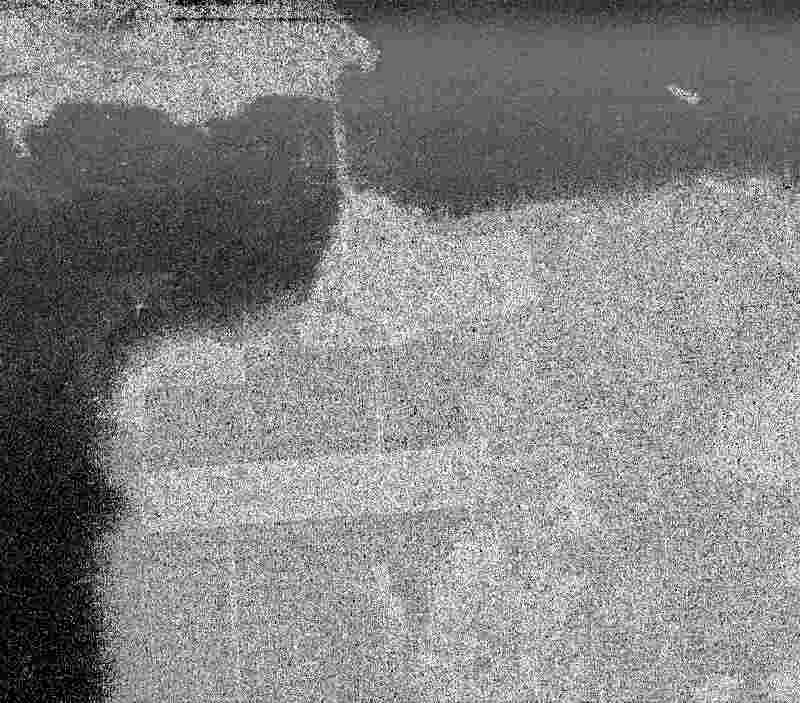}}\\
  \subfigure[Entropy of the Boxcar filter\label{fig:entropy_SanFran_BOX}]{\includegraphics[width=.43\linewidth,viewport= 188 188 394 394,clip=TRUE]{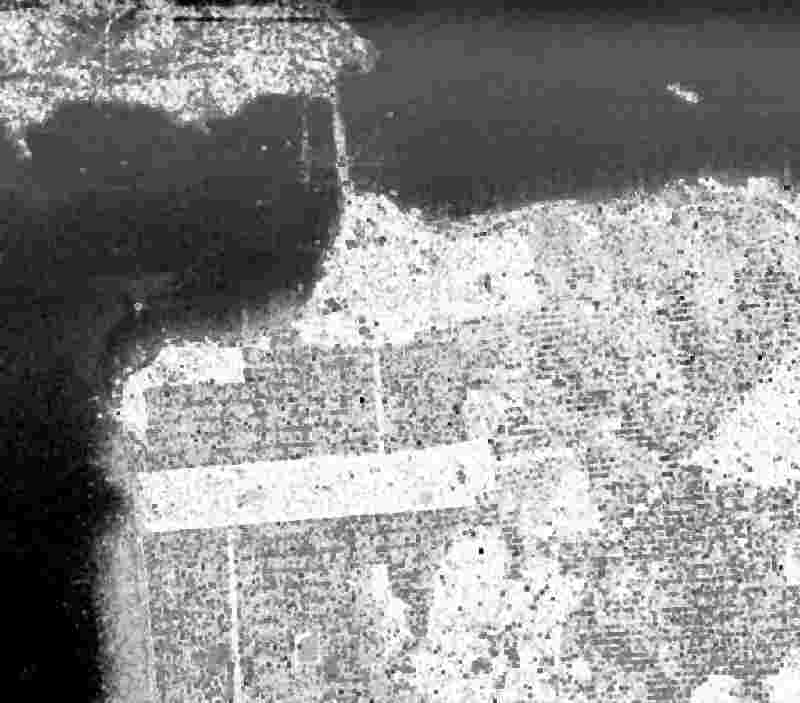}}
  \subfigure[Entropy of the Refined Lee filter\label{fig:entropy_SanFran_LEE_Refined}]{\includegraphics[width=.43\linewidth,viewport= 188 188 394 394,clip=TRUE]{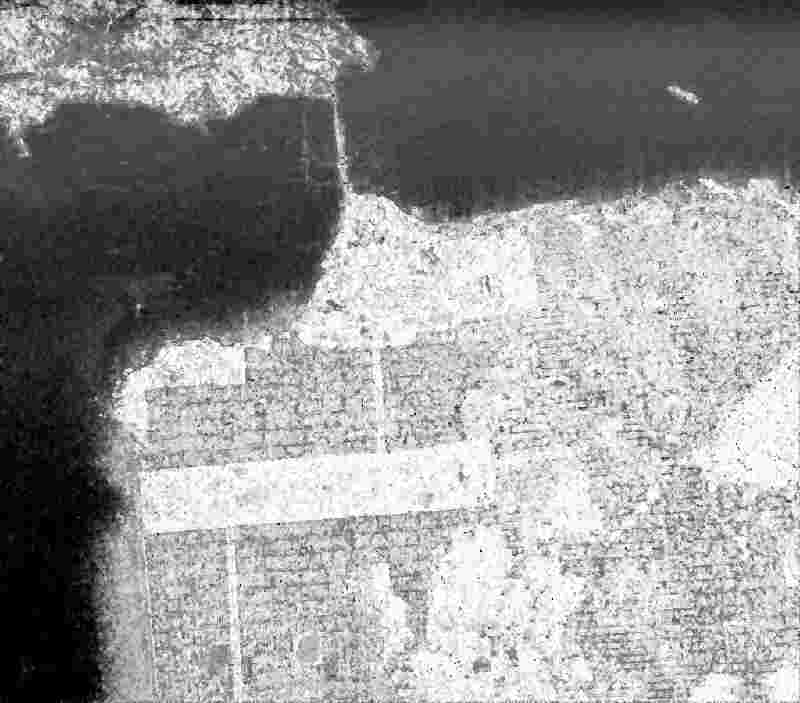}}
  \subfigure[Entropy of the IDAN filter\label{fig:entropy_SanFran_IDAN}]{\includegraphics[width=.43\linewidth,viewport= 188 188 394 394,clip=TRUE]{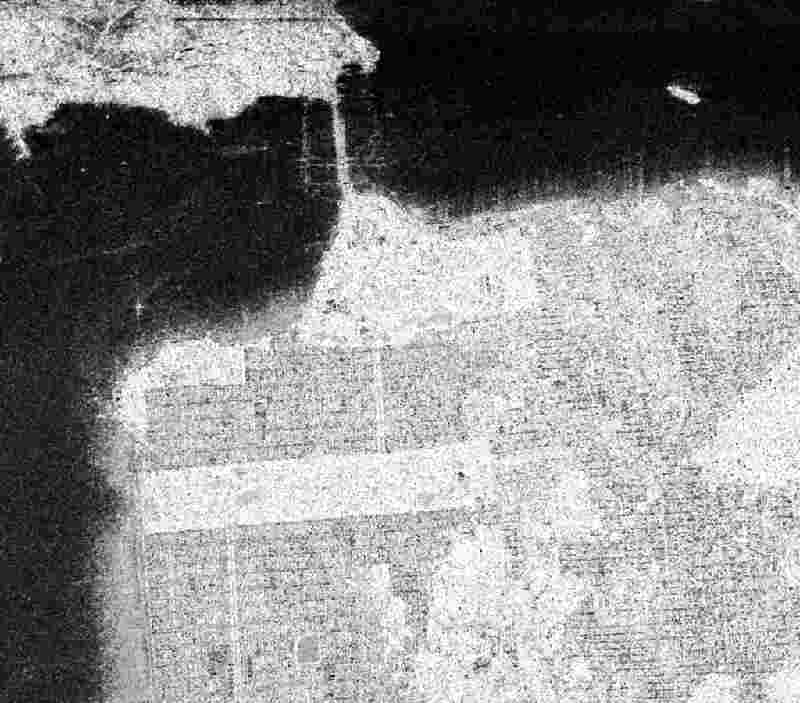}}
  \subfigure[Entropy of the SDNLM filter\label{fig:entropy_SanFran_SDNLM_90}]{\includegraphics[width=.43\linewidth,viewport= 188 188 394 394,clip=TRUE]{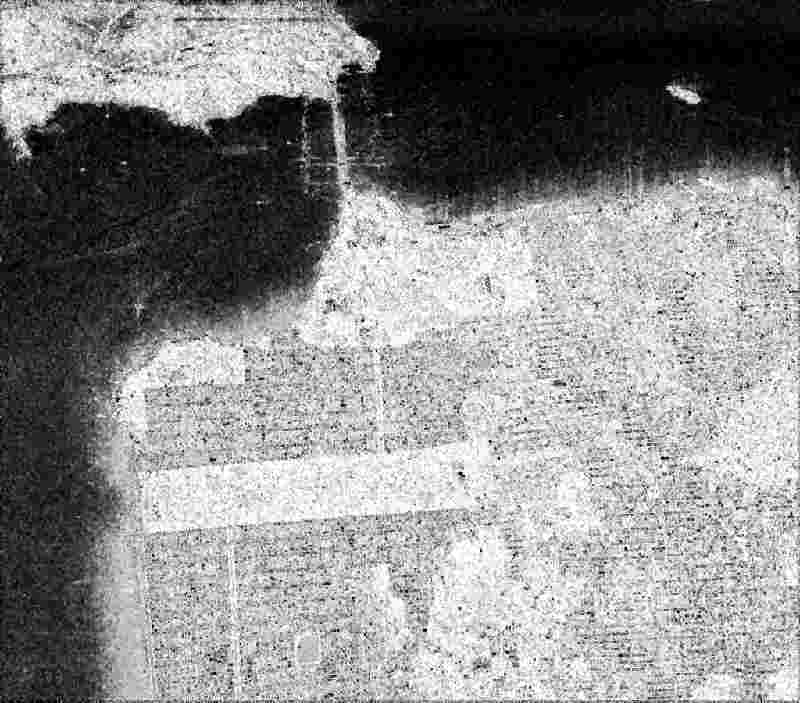}}
\caption{Entropy in the AIRSAR \textsf{L}-band data.}
\label{fig:EntropySanFran}
\end{figure}

Figure~\ref{fig:EntropySanFran} shows the effect of the filters on the entropy of the data.
Three classes of entropy are barely visible corresponding, in increasing brightness, to the sea, urban area an forest. 
The difference between the classes is small. 
All filters enhance the discrimination ability of the entropy, cf.\ figures Figures~\ref{fig:entropy_SanFran_BOX}, \ref{fig:entropy_SanFran_LEE_Refined}, \ref{fig:entropy_SanFran_IDAN} and~\ref{fig:entropy_SanFran_SDNLM_90} with Figure~\ref{fig:entropy_SanFran}.
Again, the best preservation of detail is obtained by the IDAN and SDNLM techniques that, among others, retain the information of the Presidio Golf Course.
Additionally, our proposal is the one that best preserves the low entropy small spots within the urban area, a feature typical of the high variability of these areas due to their heterogeneous composition.

\begin{figure}[hbt]
\centering
  \subfigure[Original data\label{fig:plotHalpha_SanFran}]{\includegraphics[width=.48\linewidth]{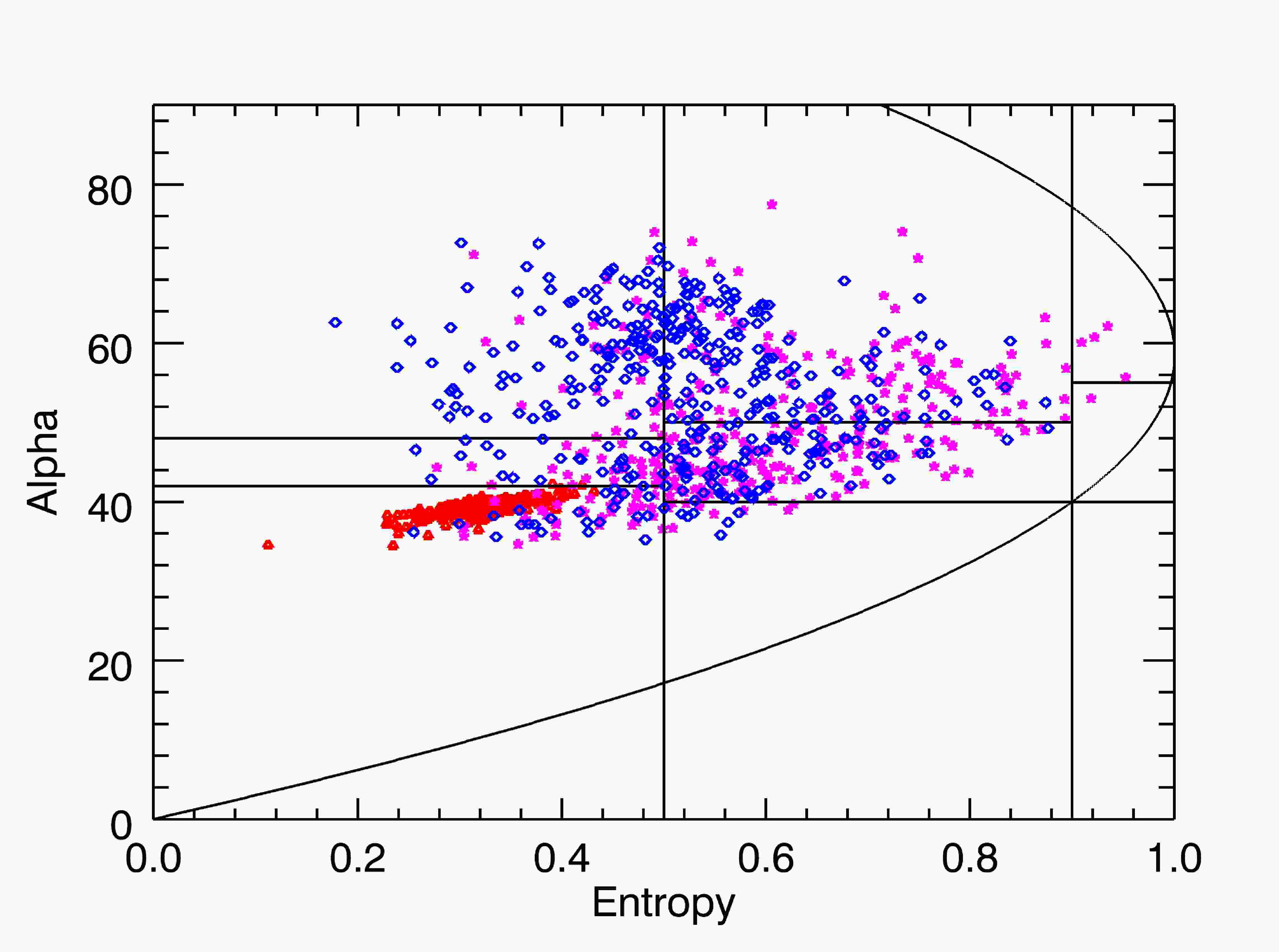}}\\
  \subfigure[Boxcar filtered data\label{fig:plotHalpha_BOX}]{\includegraphics[width=.48\linewidth]{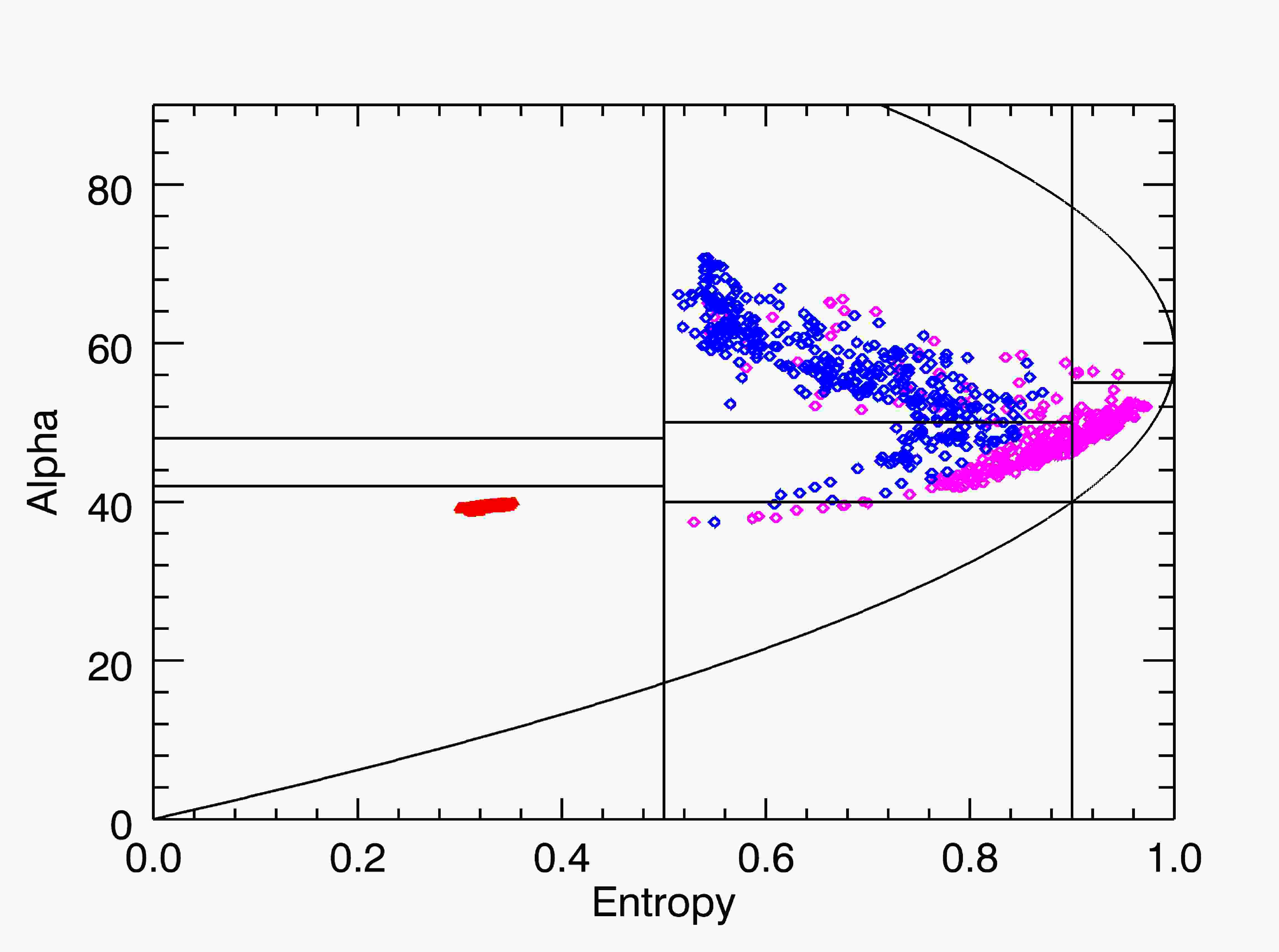}}
  \subfigure[Refined Lee filtered data\label{fig:plotHalpha_LEE}]{\includegraphics[width=.48\linewidth]{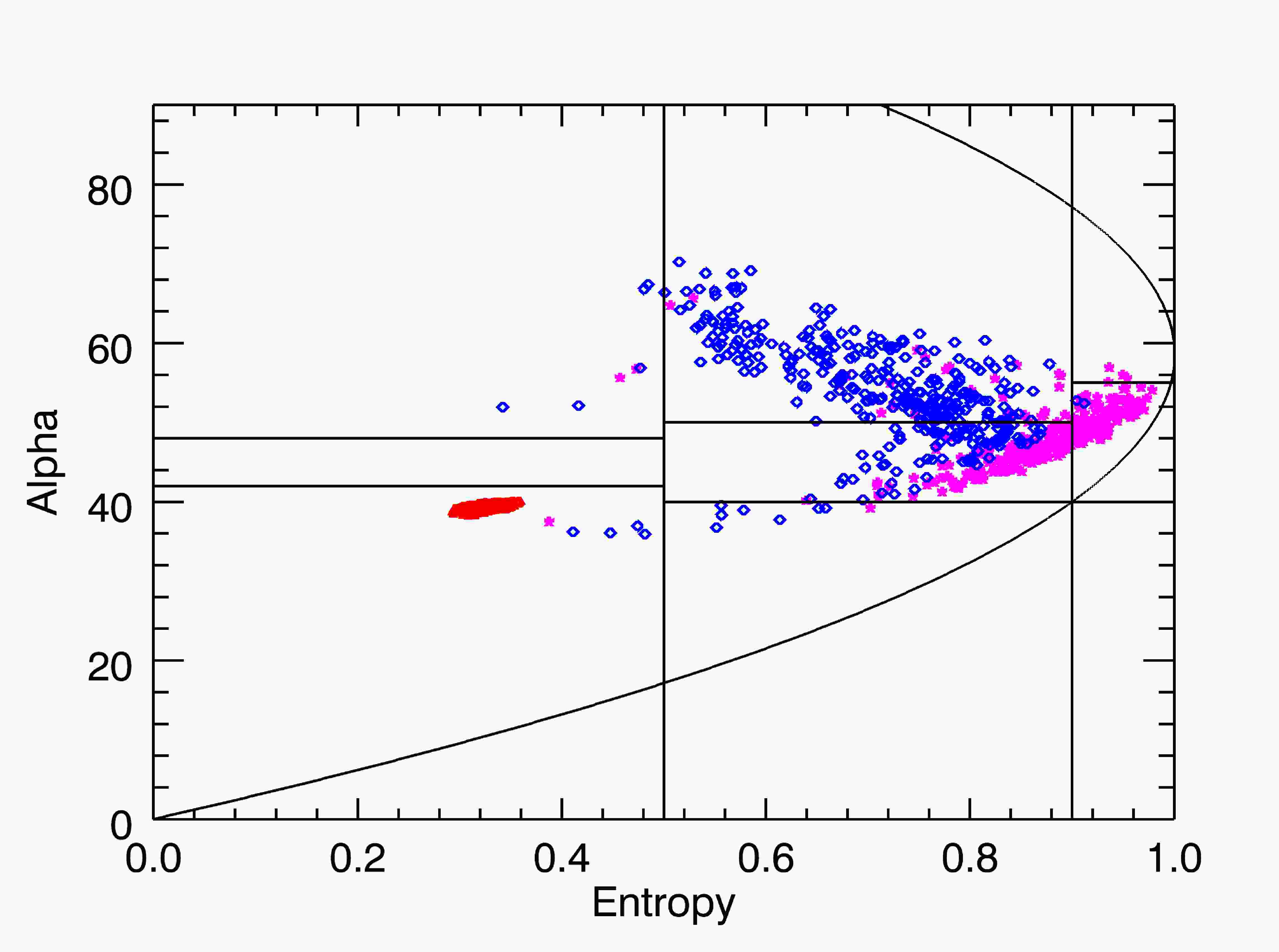}}
  \subfigure[IDAN filtered data\label{fig:plotHalpha_IDAN}]{\includegraphics[width=.48\linewidth]{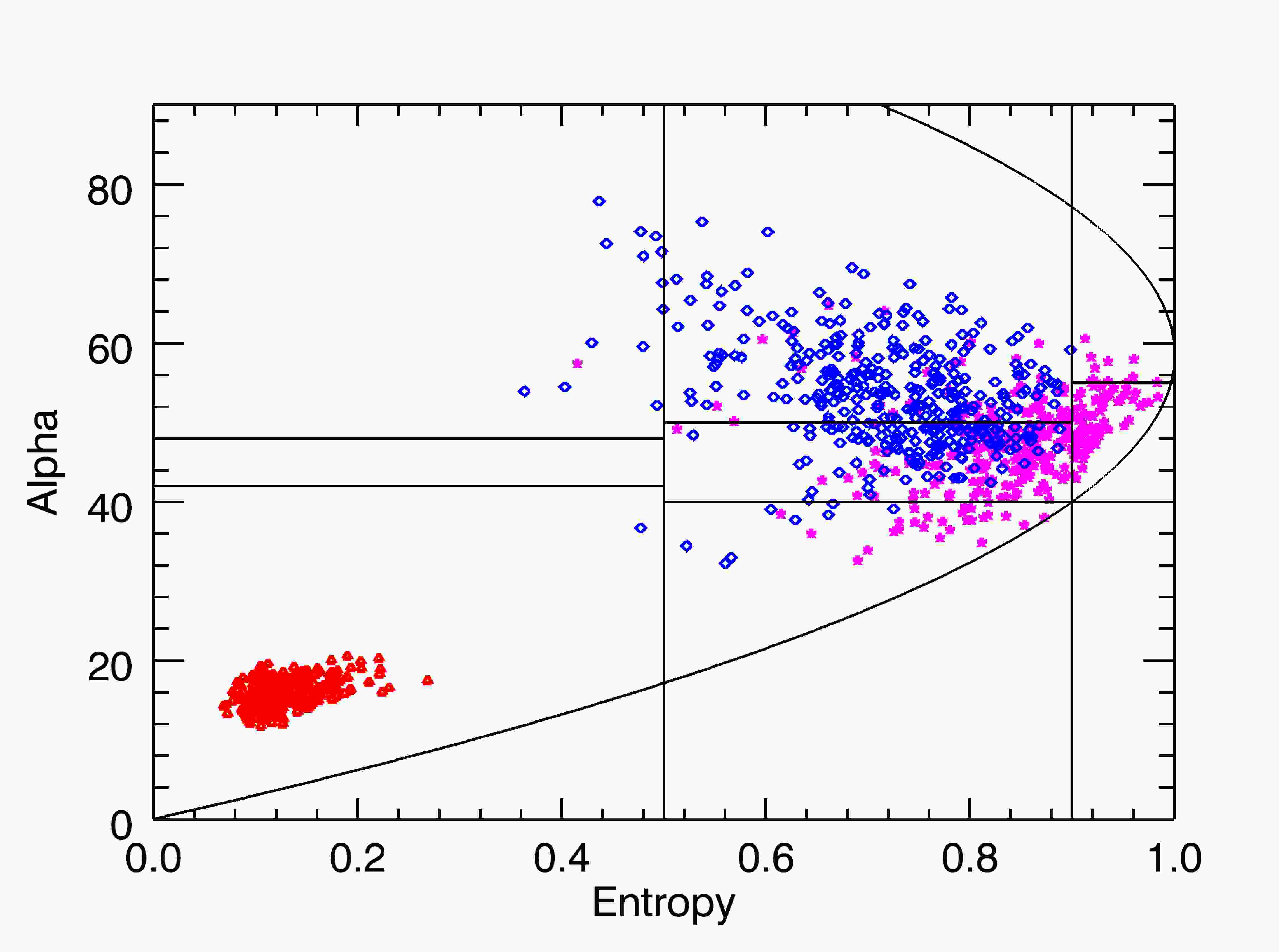}}
  \subfigure[SDNLM filtered data\label{fig:plotHalpha_SDNLM_90}]{\includegraphics[width=.48\linewidth]{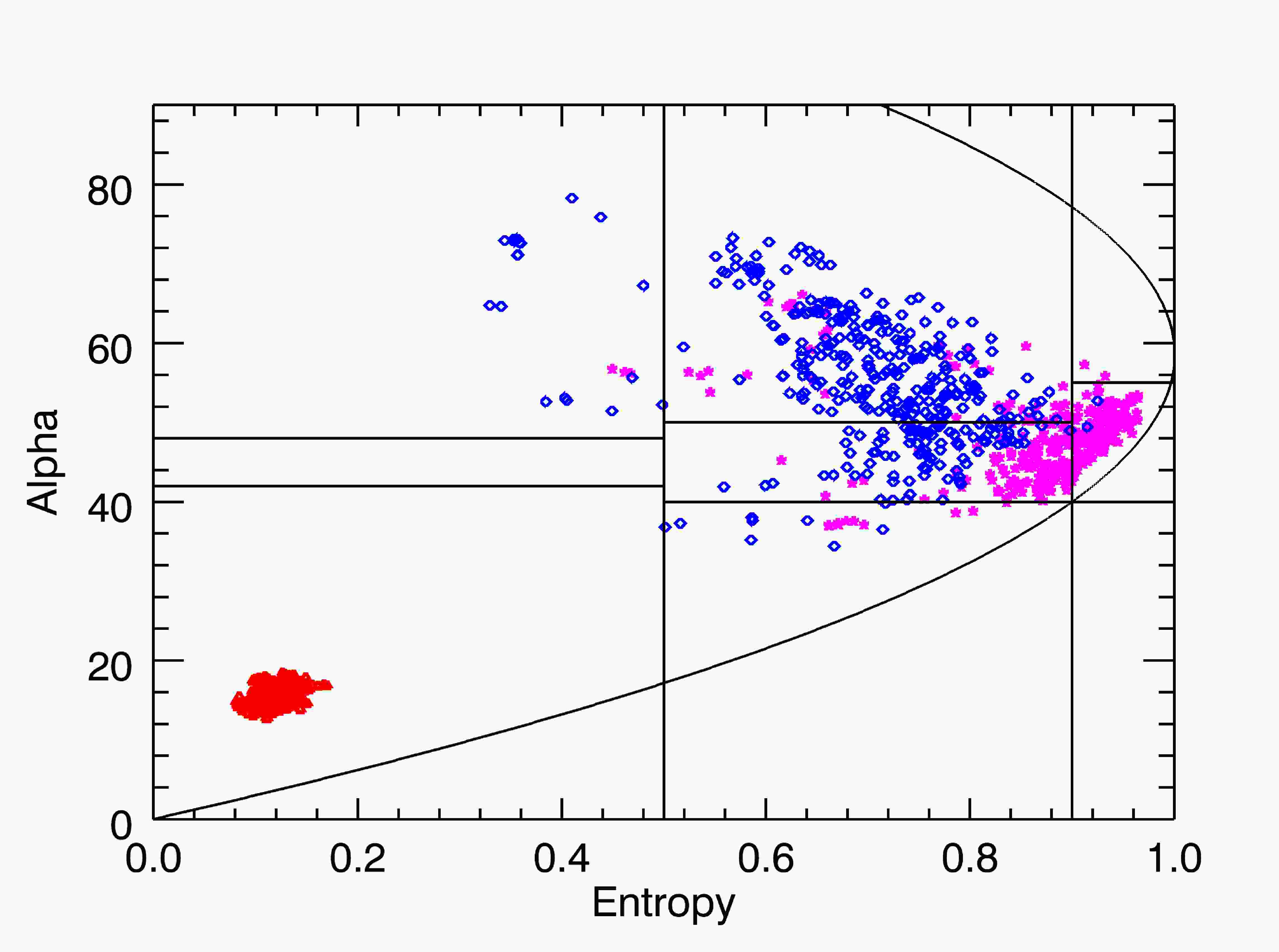}}
\caption{Scatter plot in the ($H,\overline{\alpha}$) plane of samples from the AIRSAR \textsf{L}-band image.}
\label{fig:Halpha_SanFran}
\end{figure}

Samples from the sea, the urban area, and the forest, identified in red, blue and magenta, respectively, in Figure~\ref{fig:Pauli_SanFran}, were taken.
The $(H,\overline{\alpha})$ value of each point from these samples is presented in Figure~\ref{fig:Halpha_SanFran} before and after applying the filters.
The improvement of using filters is notorious when comparing Figure~\ref{fig:plotHalpha_SanFran} with figures~\ref{fig:plotHalpha_BOX}, \ref{fig:plotHalpha_LEE}, \ref{fig:plotHalpha_IDAN} and~\ref{fig:plotHalpha_SDNLM_90}.
While the original data is mixed, specially the samples from urban and forest areas, after applying the filters the data tend to group in clusters.

The sea samples are confined to Zone~9 in all datasets, and all filters have the effect of reducing their variability.
While the Boxcar and Refined Lee filters produce very similar clusters of data, the IDAN and SDNLM filters reduce both the entropy and the $\overline{\alpha}$ coefficient, but still within the zone of low entropy surface scatter, making the sample much more distinguishable from the rest of the data.

The samples from urban (in blue) and forest (in magenta) areas have different mean values of entropy and $\overline{\alpha}$.
The former occupy mostly zones~4 (medium entropy multiple scattering) and~5 (medium entropy vegetation scattering), while the latter span mostly zones~2 (high entropy vegetation scattering) and~5.
While both are present in Zone~5, they seldom overlap; the forest samples have higher values of $\overline{\alpha}$.
Comparing these two samples in the images filtered by the Refined Lee and SDNLM techniques, one notices that they would produce very similar classifications.
The SDNLM produces clusters with more spread than the Refined Lee, but not at the expense of mixing different classes.

In this manner, the filters preserve the scattering properties of the samples, a central feature of every speckle smoothing technique for PolSAR data, according to Lee and Pottier~\cite{LeePottier2009Book}.

\subsection{Effect of iterations number in filtering}

\begin{figure}[hbt]
\centering
  \subfigure[Boxcar filter $I=1$\label{fig:Pauli_SanFran_BOX_1i}]{\includegraphics[width=.31\linewidth,viewport= 320 320 670 670,clip=TRUE]{Pauli_SanFran_BOX}}
  \subfigure[Boxcar filter $I=3$\label{fig:Pauli_SanFran_BOX_3i}]{\includegraphics[width=.31\linewidth,viewport= 320 320 670 670,clip=TRUE]{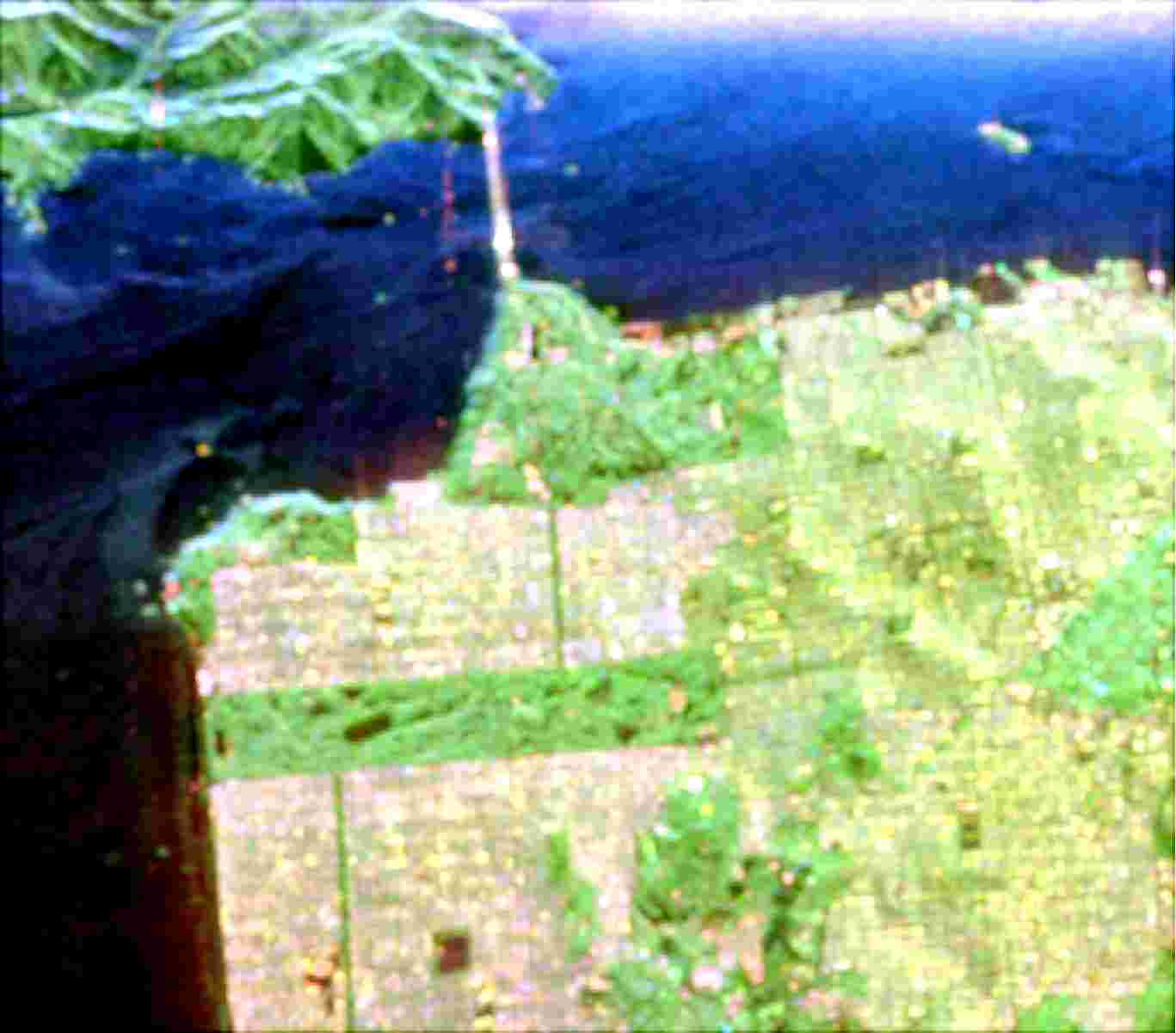}}
  \subfigure[Boxcar filter $I=5$\label{fig:Pauli_SanFran_BOX_5i}]{\includegraphics[width=.31\linewidth,viewport= 320 320 670 670,clip=TRUE]{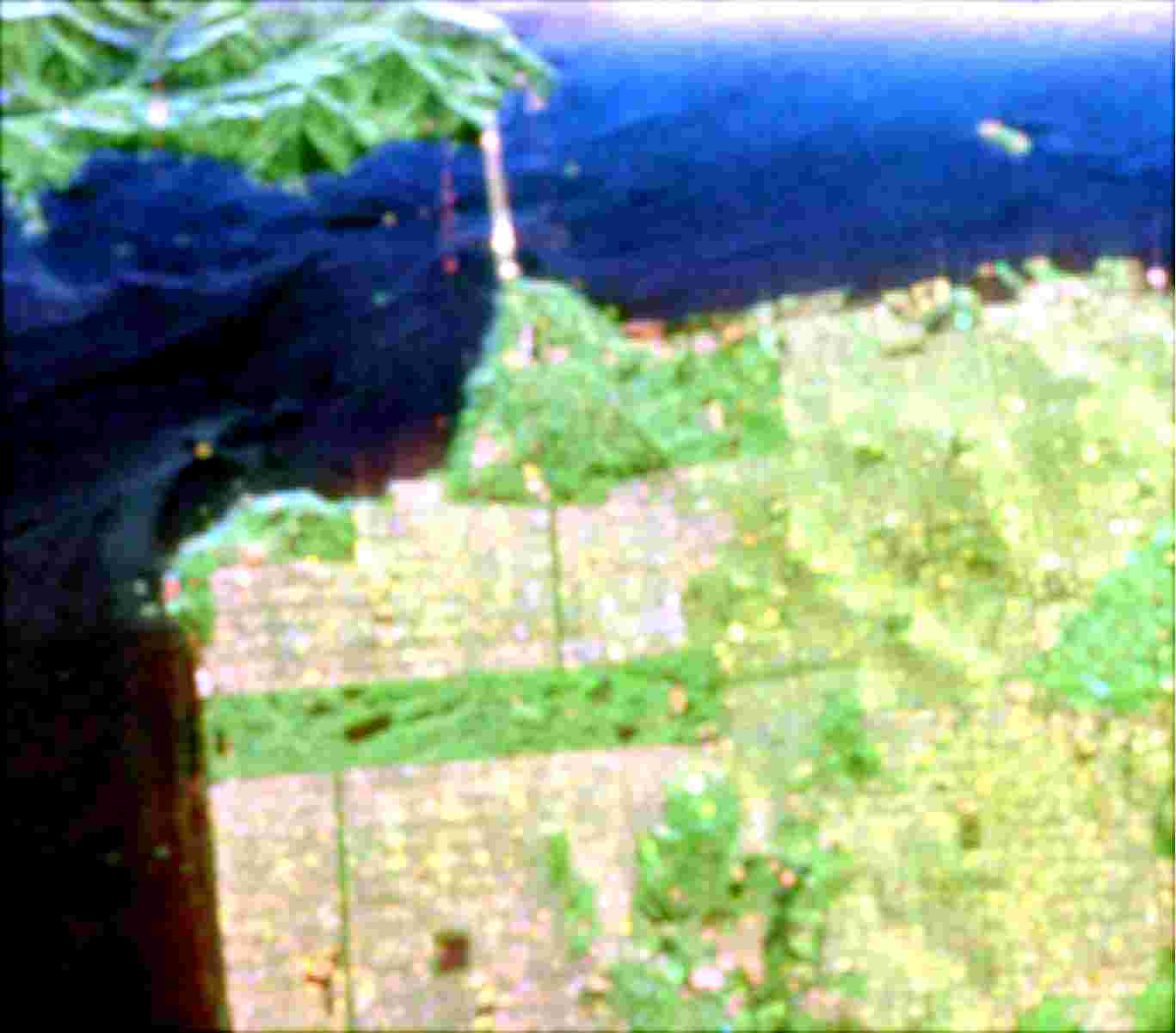}}\\
  \subfigure[Refined Lee filter $I=1$\label{fig:Pauli_SanFran_LEE_Refined_1i}]{\includegraphics[width=.31\linewidth,viewport= 320 320 670 670,clip=TRUE]{Pauli_SanFran_LEE_Refined}}
  \subfigure[Refined Lee filter $I=3$\label{fig:Pauli_SanFran_LEE_Refined_3i}]{\includegraphics[width=.31\linewidth,viewport= 320 320 670 670,clip=TRUE]{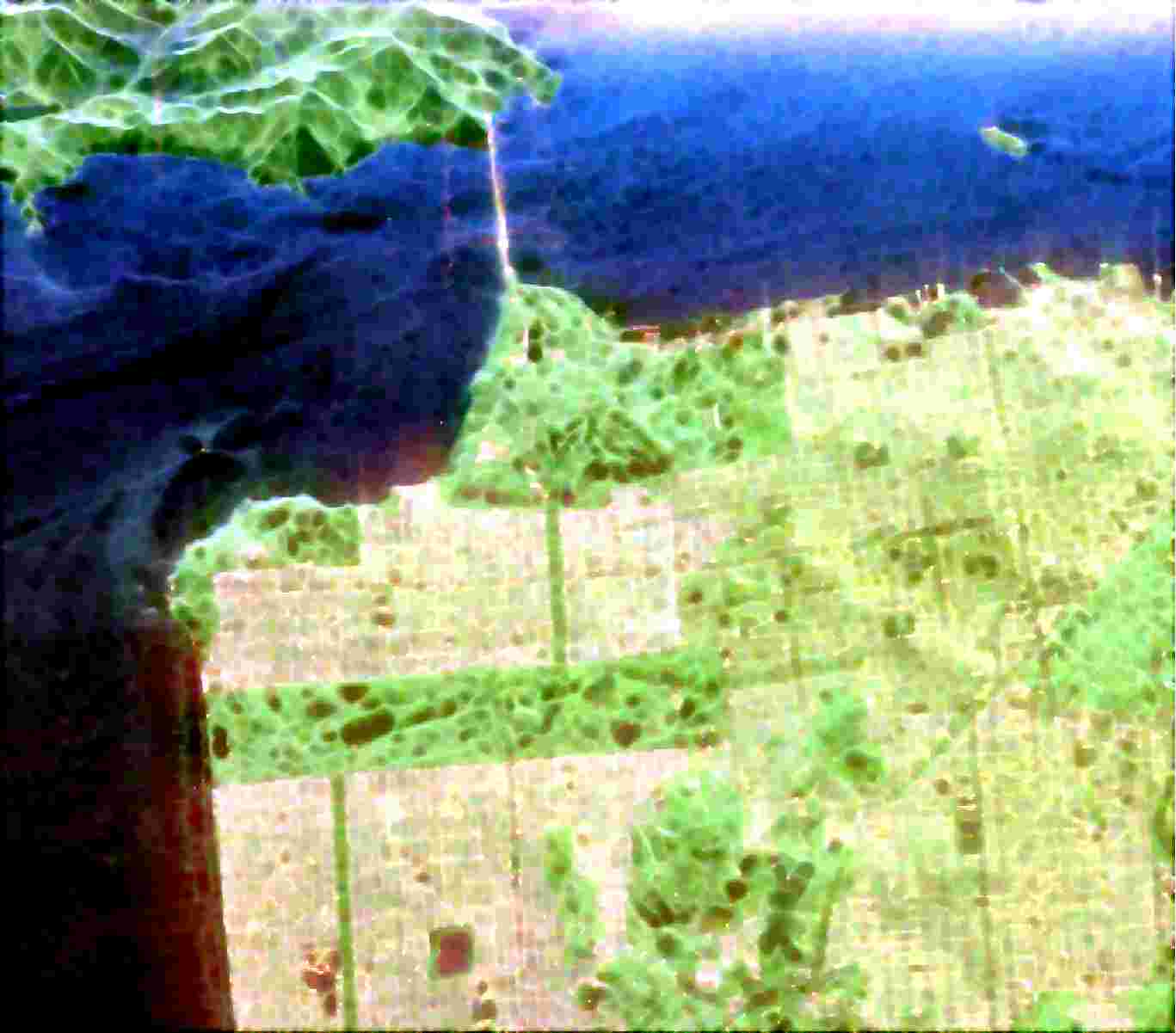}}
  \subfigure[Refined Lee filter $I=5$\label{fig:Pauli_SanFran_LEE_Refined_5i}]{\includegraphics[width=.31\linewidth,viewport= 320 320 670 670,clip=TRUE]{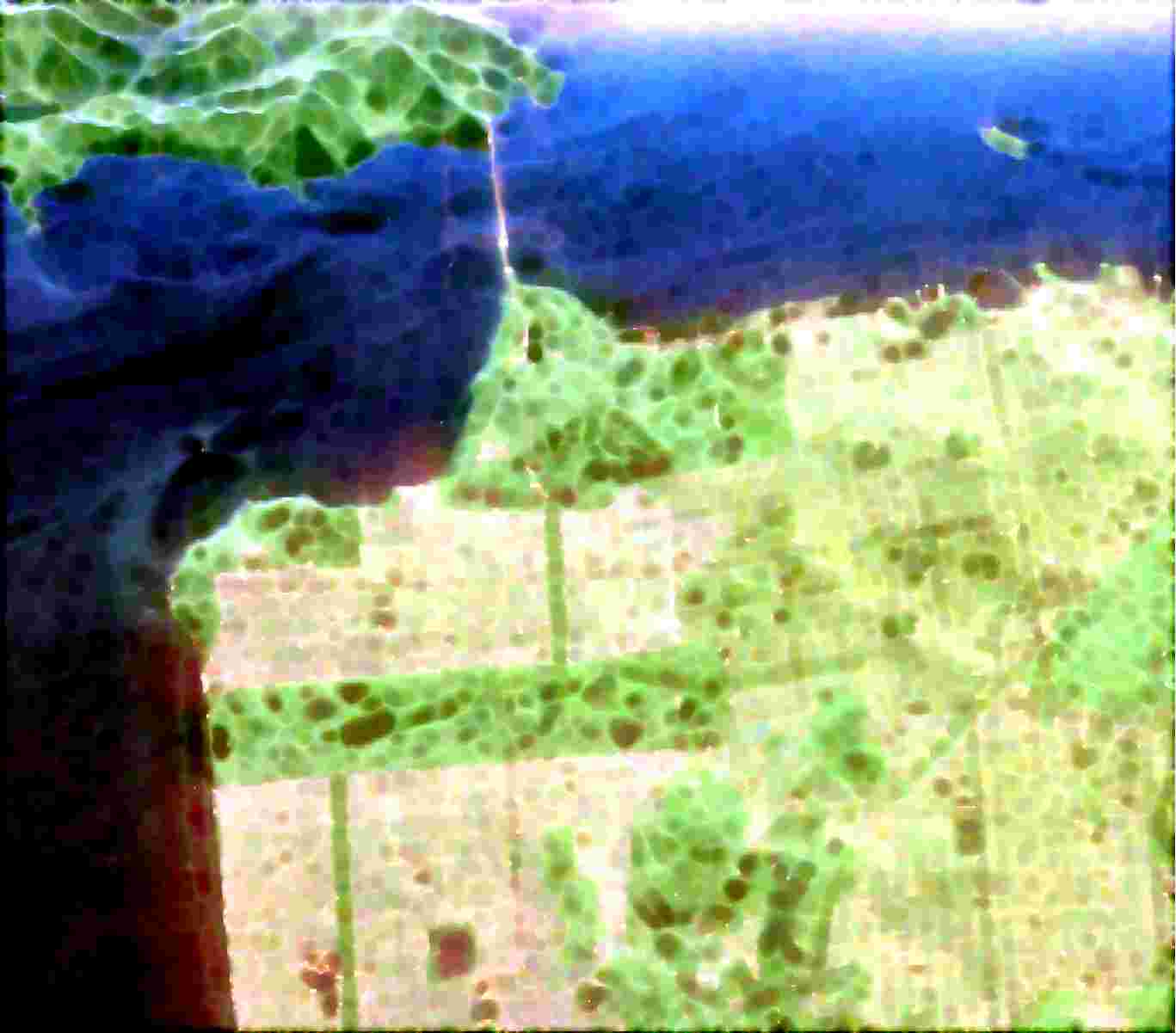}}\\  
  \subfigure[IDAN filter $I=1$\label{fig:Pauli_SanFran_IDAN_1i}]{\includegraphics[width=.31\linewidth,viewport= 320 320 670 670,clip=TRUE]{Pauli_SanFran_IDAN}}
  \subfigure[IDAN filter $I=1$\label{fig:Pauli_SanFran_IDAN_3i}]{\includegraphics[width=.31\linewidth,viewport= 320 320 670 670,clip=TRUE]{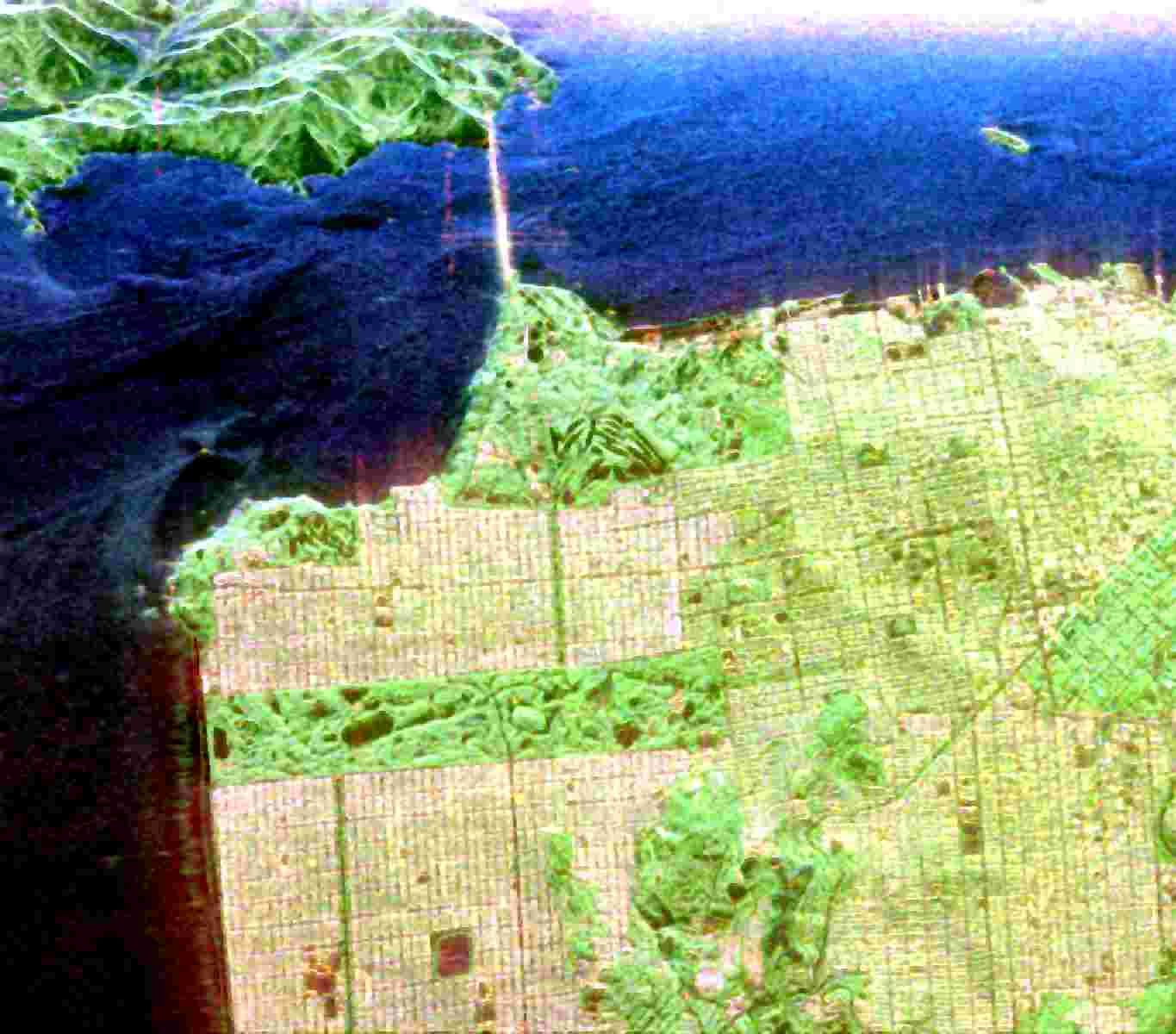}}
  \subfigure[IDAN filter $I=1$\label{fig:Pauli_SanFran_IDAN_5i}]{\includegraphics[width=.31\linewidth,viewport= 320 320 670 670,clip=TRUE]{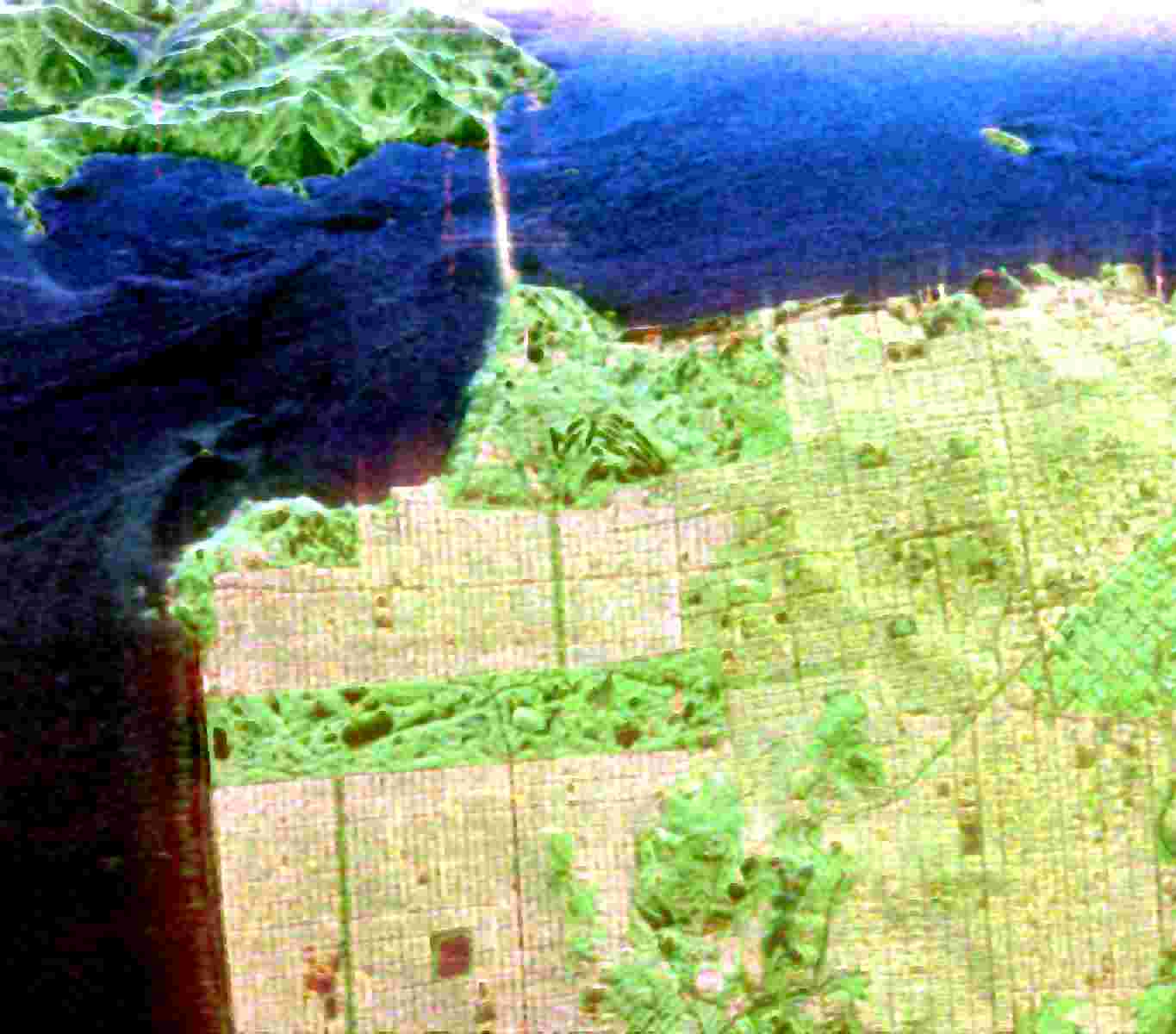}}
  \subfigure[SDNLM filter $I=1$\label{fig:Pauli_SanFran_SDNLM_80_1i}]{\includegraphics[width=.31\linewidth,viewport= 320 320 670 670,clip=TRUE]{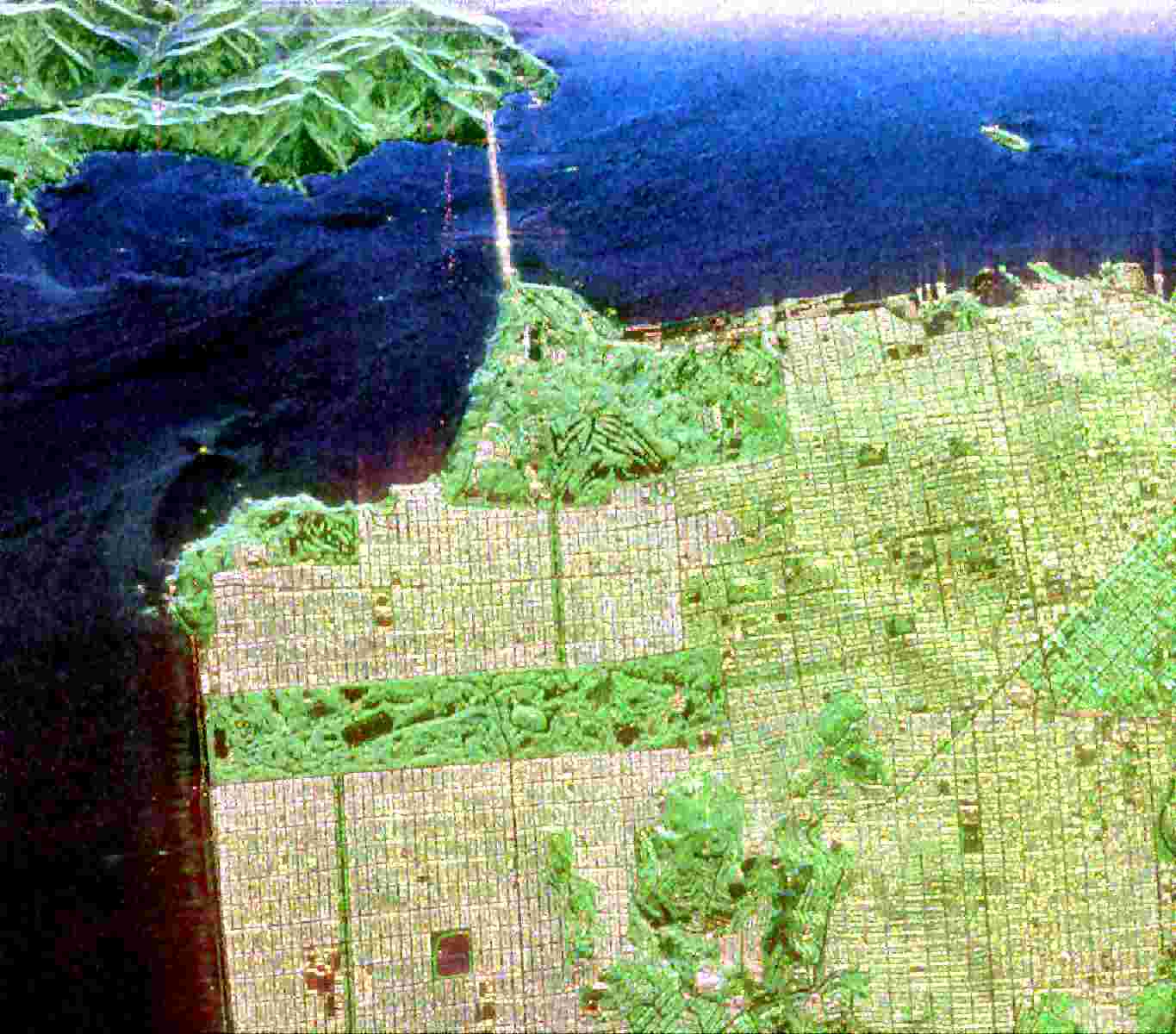}}
  \subfigure[SDNLM filter $I=3$\label{fig:Pauli_SanFran_SDNLM_80_3i}]{\includegraphics[width=.31\linewidth,viewport= 320 320 670 670,clip=TRUE]{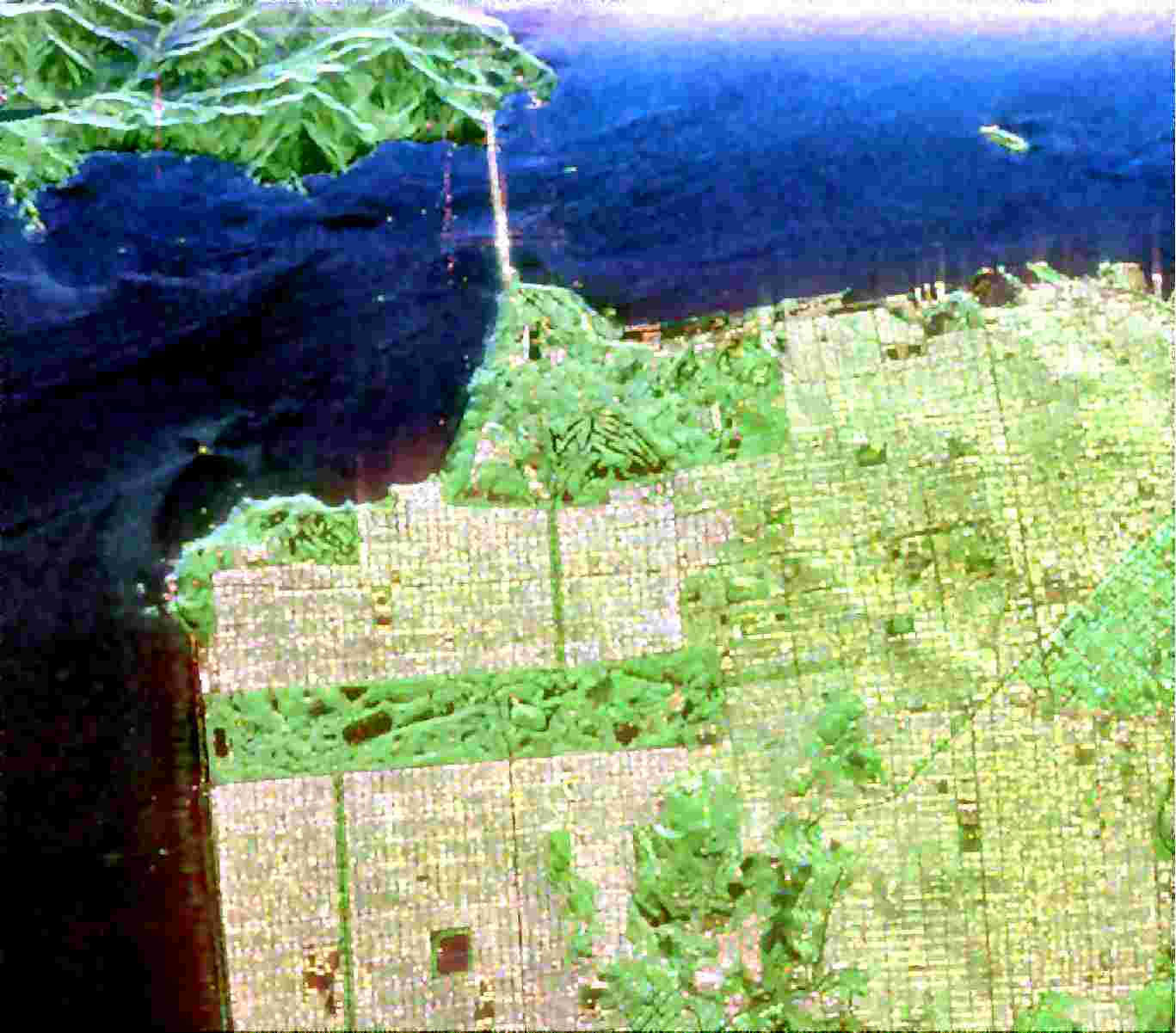}}
  \subfigure[SDNLM filter $I=5$\label{fig:Pauli_SanFran_SDNLM_80_5i}]{\includegraphics[width=.31\linewidth,viewport= 320 320 670 670,clip=TRUE]{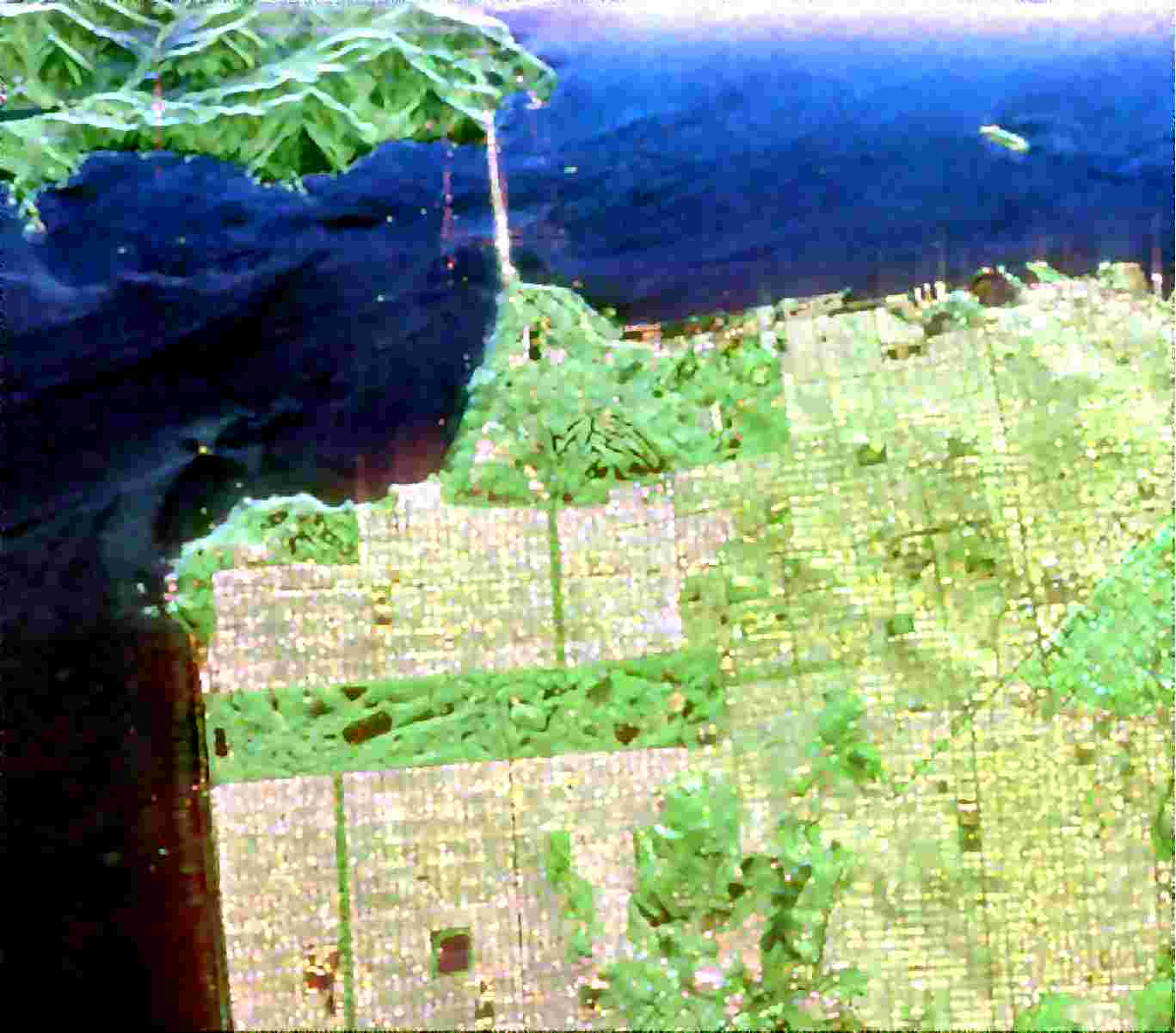}}  
\caption{Pauli decomposition of images filtered one, three and five times with each technique.}
\label{fig:Iterations}
\end{figure}

As previously discussed, SDNLM can be iterated since the properties upon which it is based are preserved by convolutions.
Figure~\ref{fig:Iterations} presents the original image for reference (Figure~\ref{fig:Pauli_SanFran}), and the result of applying each technique (Boxcar, Refined Lee, IDAN and SDNLM with $\eta=80\%$ in each row) one, three and five times (first, second and third column, respectively).

The most notorious new result stems from comparing the IDAN and SDNLM filters.
Three iterations are enough for the former to smudge the original data, and with five iterations the blurring it produces is comparable with that of the Refined Lee and Boxcar filters.
The SDNLM filter, even after five iterations, still preserves most of the spatial information.

\begin{figure}[hbt]
\centering
  \subfigure[AIRSAR \textsf{L}-band data\label{fig:Plot_HAlpha_SanFran}]{\includegraphics[width=.31\linewidth]{Plot_HAlpha_SanFran}}\\
  \subfigure[Boxcar filter $I=1$\label{fig:Plot_HAlpha_BOX_1i}]{\includegraphics[width=.31\linewidth]{Plot_HAlpha_BOX}}
  \subfigure[Boxcar filter $I=3$\label{fig:Plot_HAlpha_BOX_3i}]{\includegraphics[width=.31\linewidth]{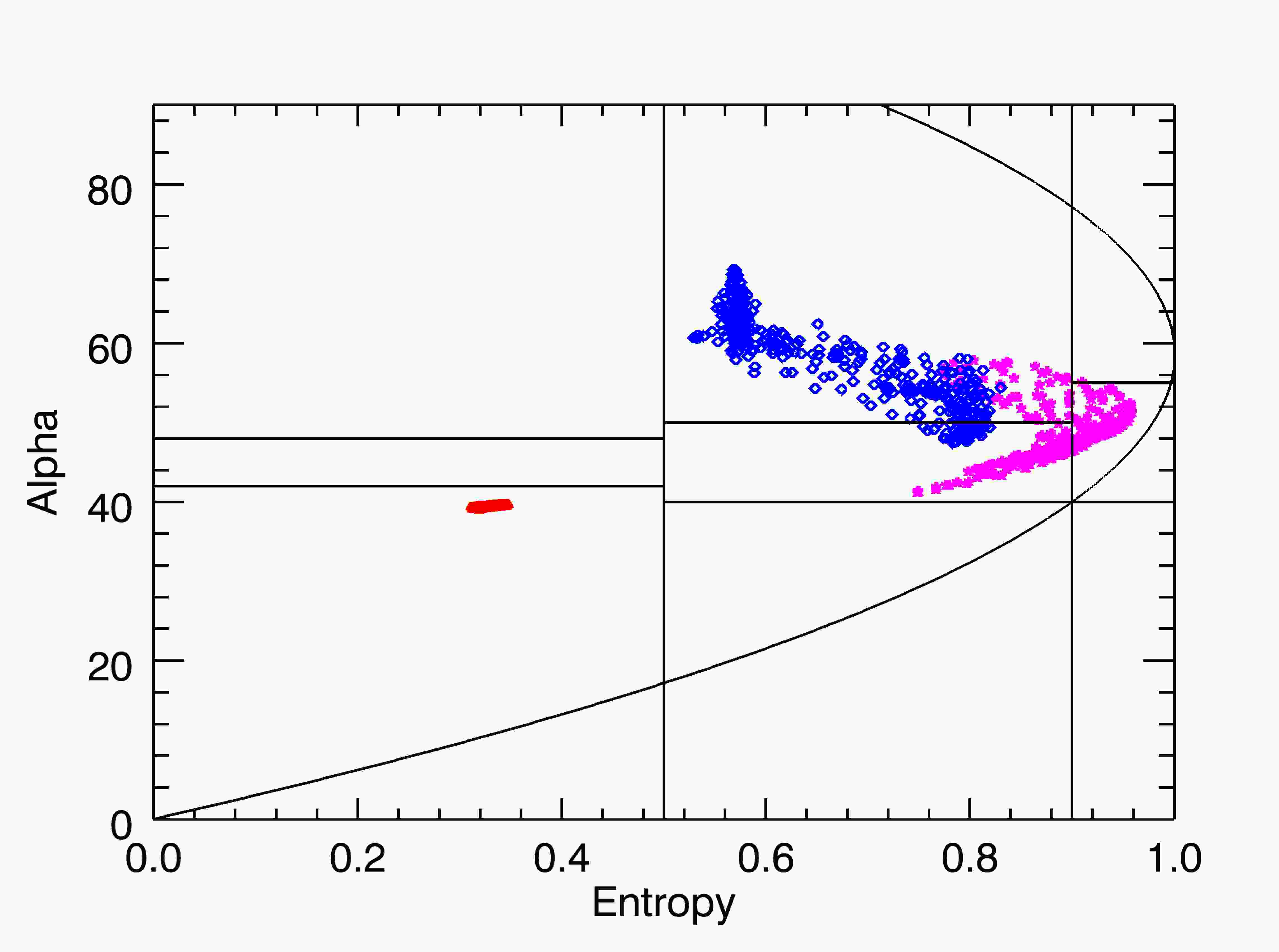}}
  \subfigure[Boxcar filter $I=5$\label{fig:Plot_HAlpha_BOX_5i}]{\includegraphics[width=.31\linewidth]{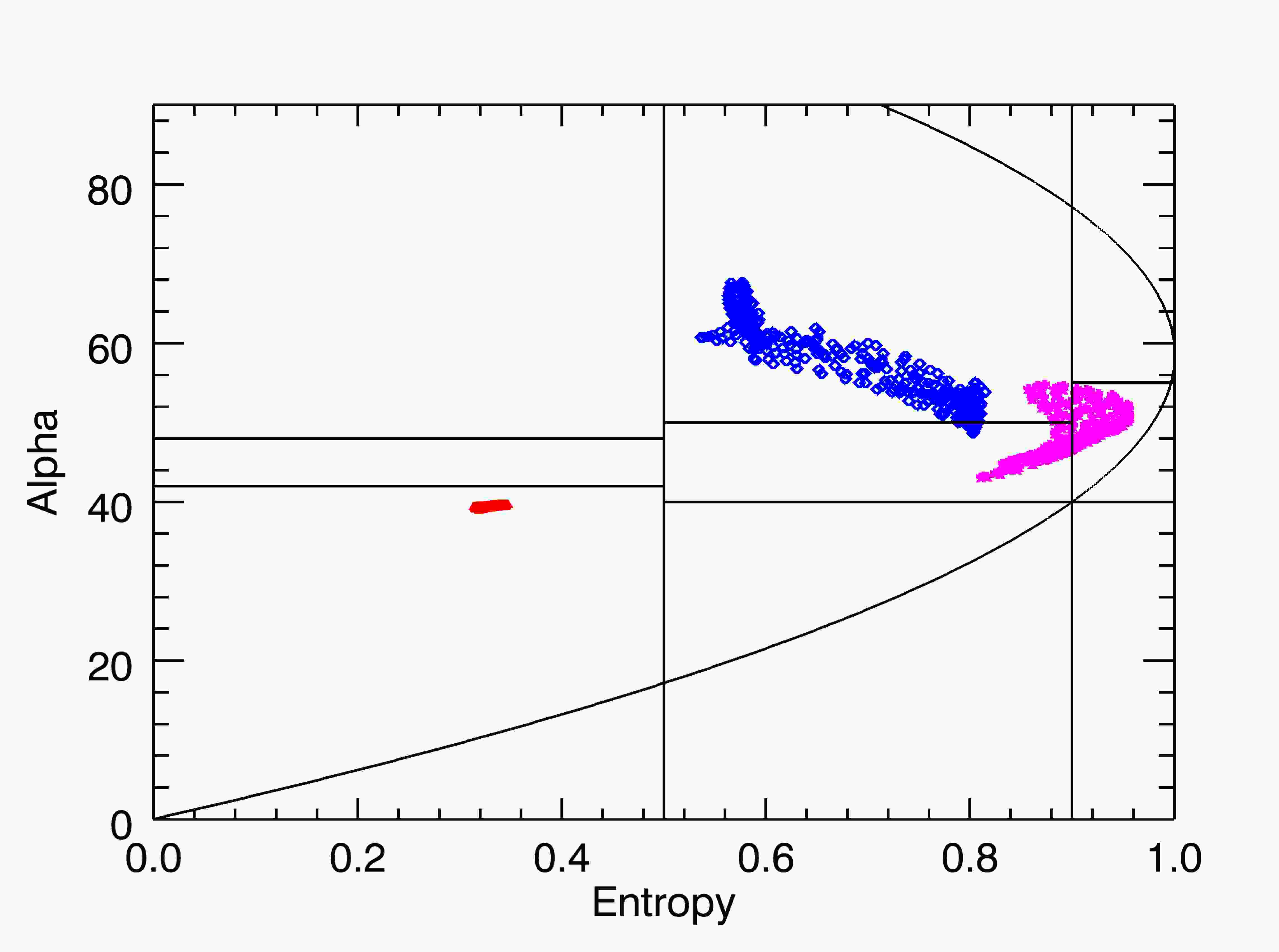}}\\
  \subfigure[Refined Lee filter $I=1$\label{fig:Plot_HAlpha_LEE_1i}]{\includegraphics[width=.31\linewidth]{Plot_HAlpha_LEE}}
  \subfigure[Refined Lee filter $I=3$\label{fig:Plot_HAlpha_LEE_3i}]{\includegraphics[width=.31\linewidth]{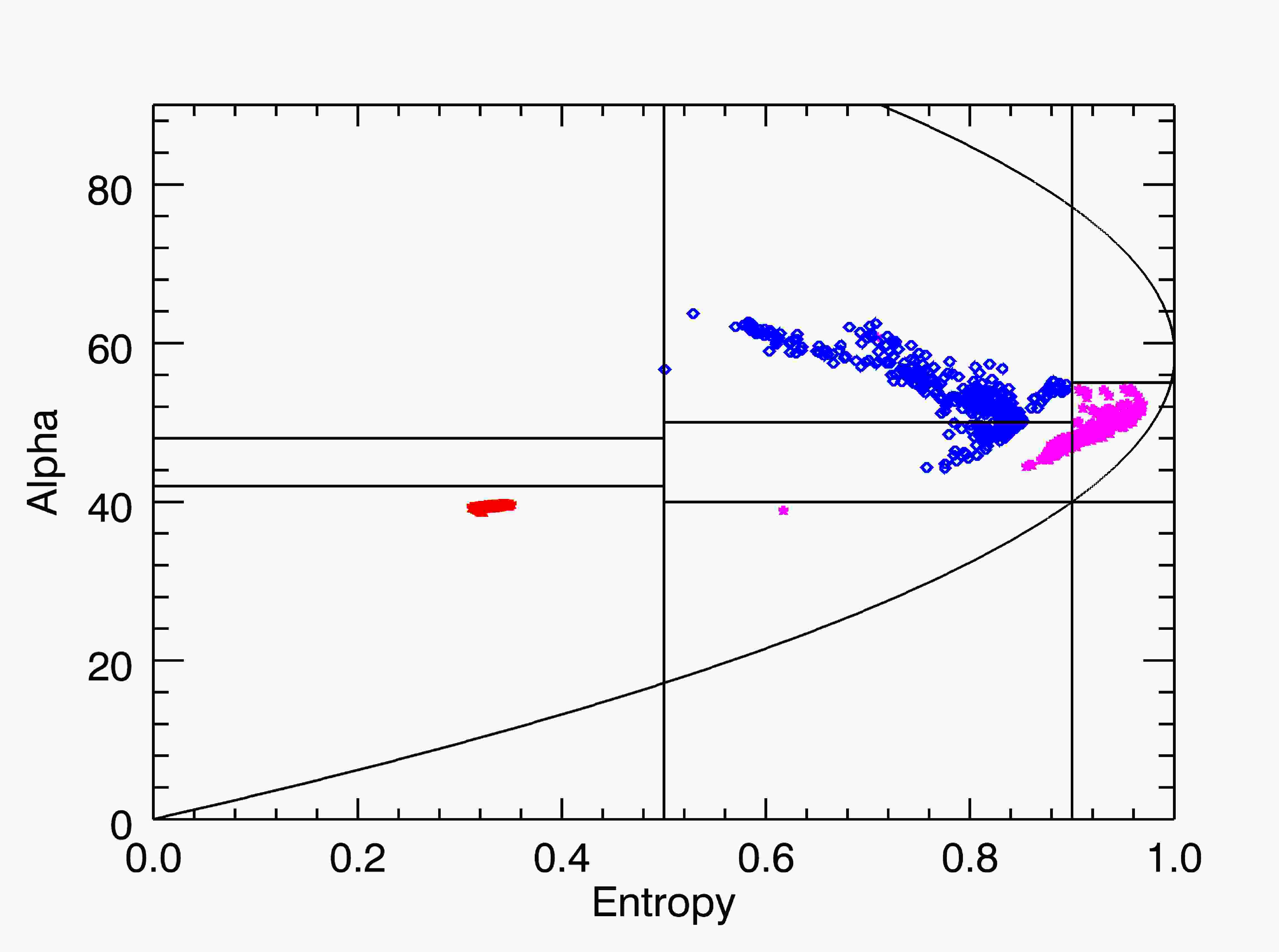}}
  \subfigure[Refined Lee filter $I=5$\label{fig:Plot_HAlpha_LEE_5i}]{\includegraphics[width=.31\linewidth]{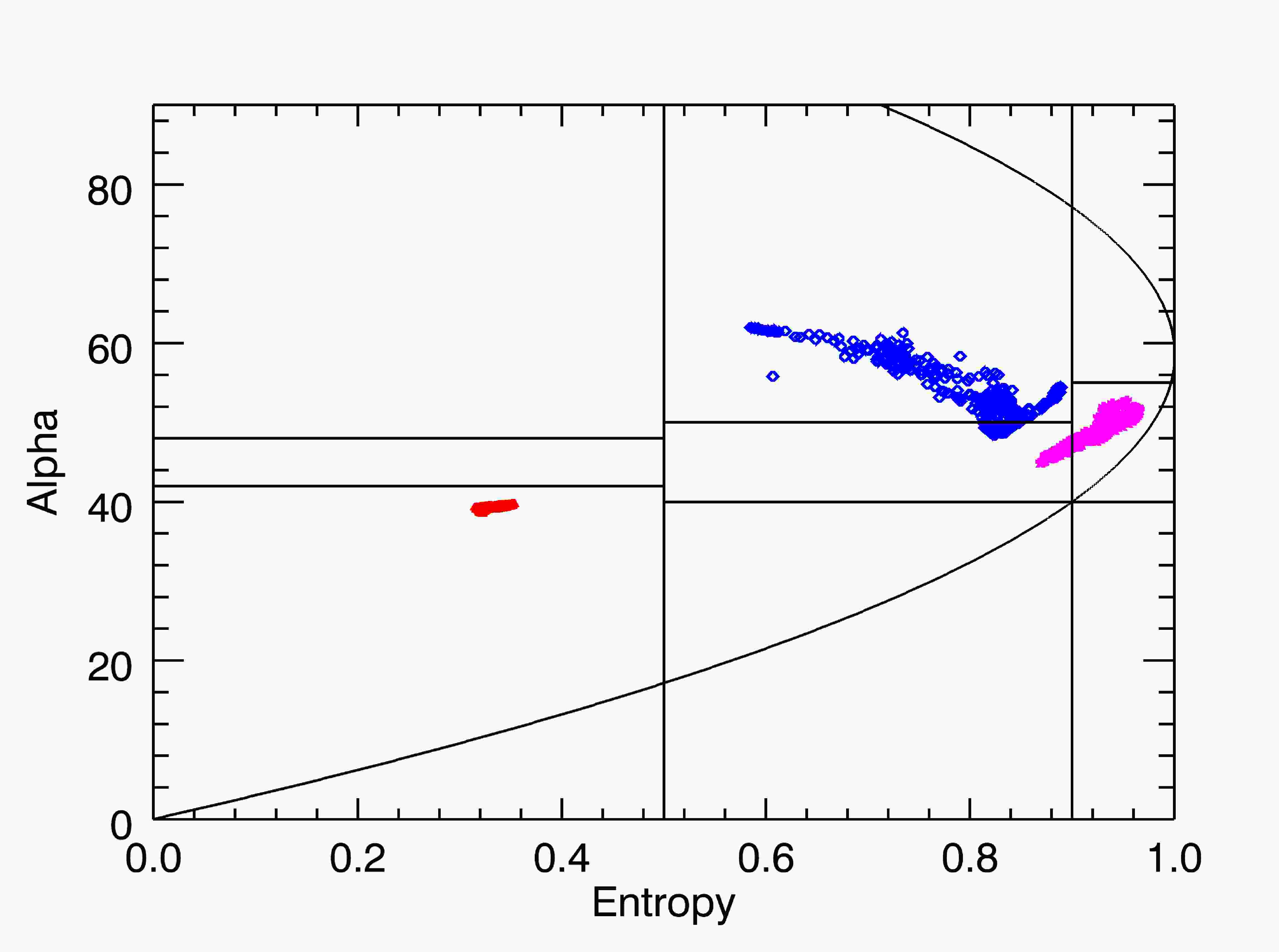}}\\
  \subfigure[IDAN filter $I=1$\label{fig:Plot_HAlpha_IDAN_1i}]{\includegraphics[width=.31\linewidth]{Plot_HAlpha_IDAN}}
  \subfigure[IDAN filter $I=3$\label{fig:Plot_HAlpha_IDAN_3i}]{\includegraphics[width=.31\linewidth]{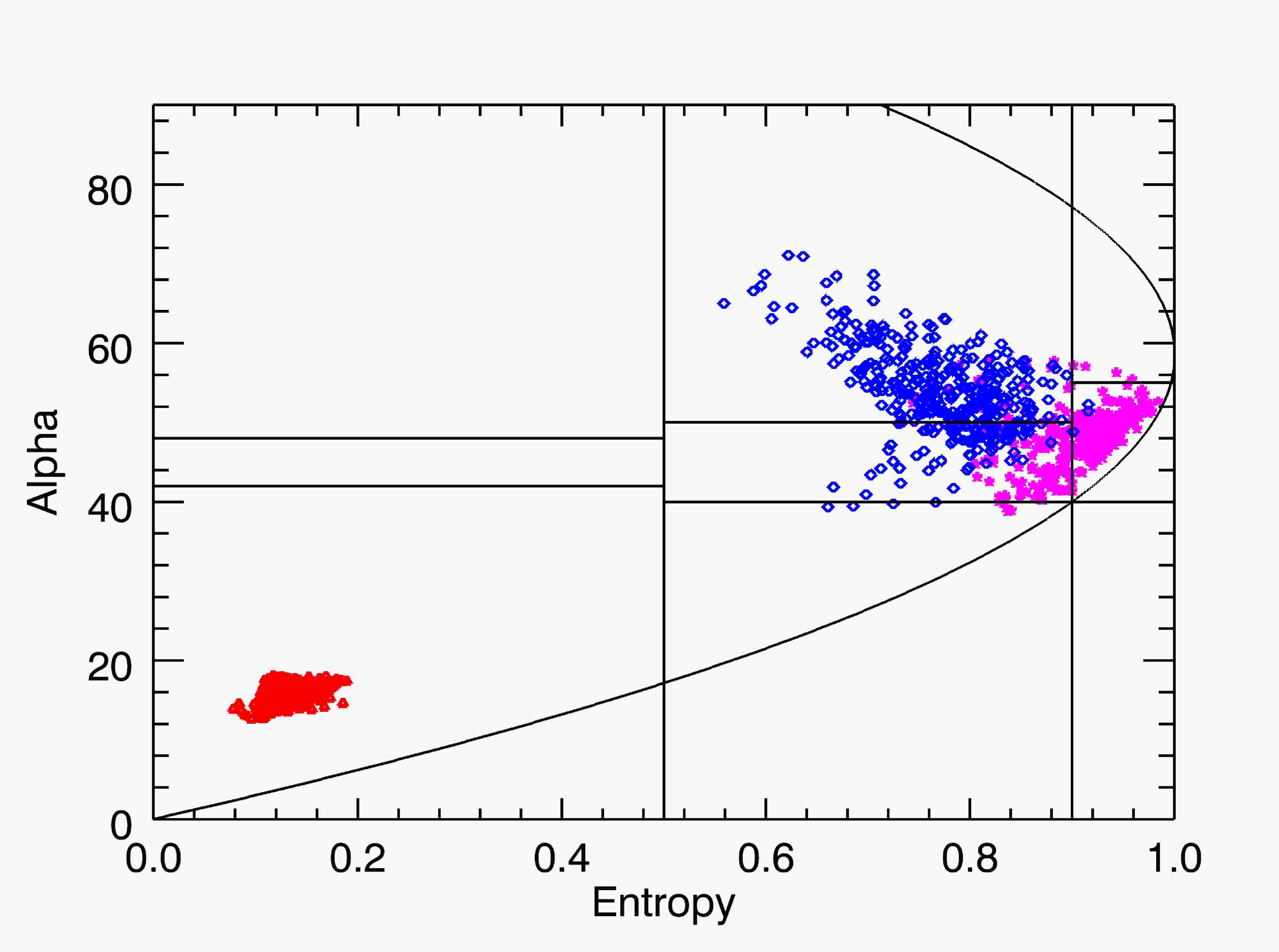}}
  \subfigure[IDAN filter $I=5$\label{fig:Plot_HAlpha_IDAN_5i}]{\includegraphics[width=.31\linewidth]{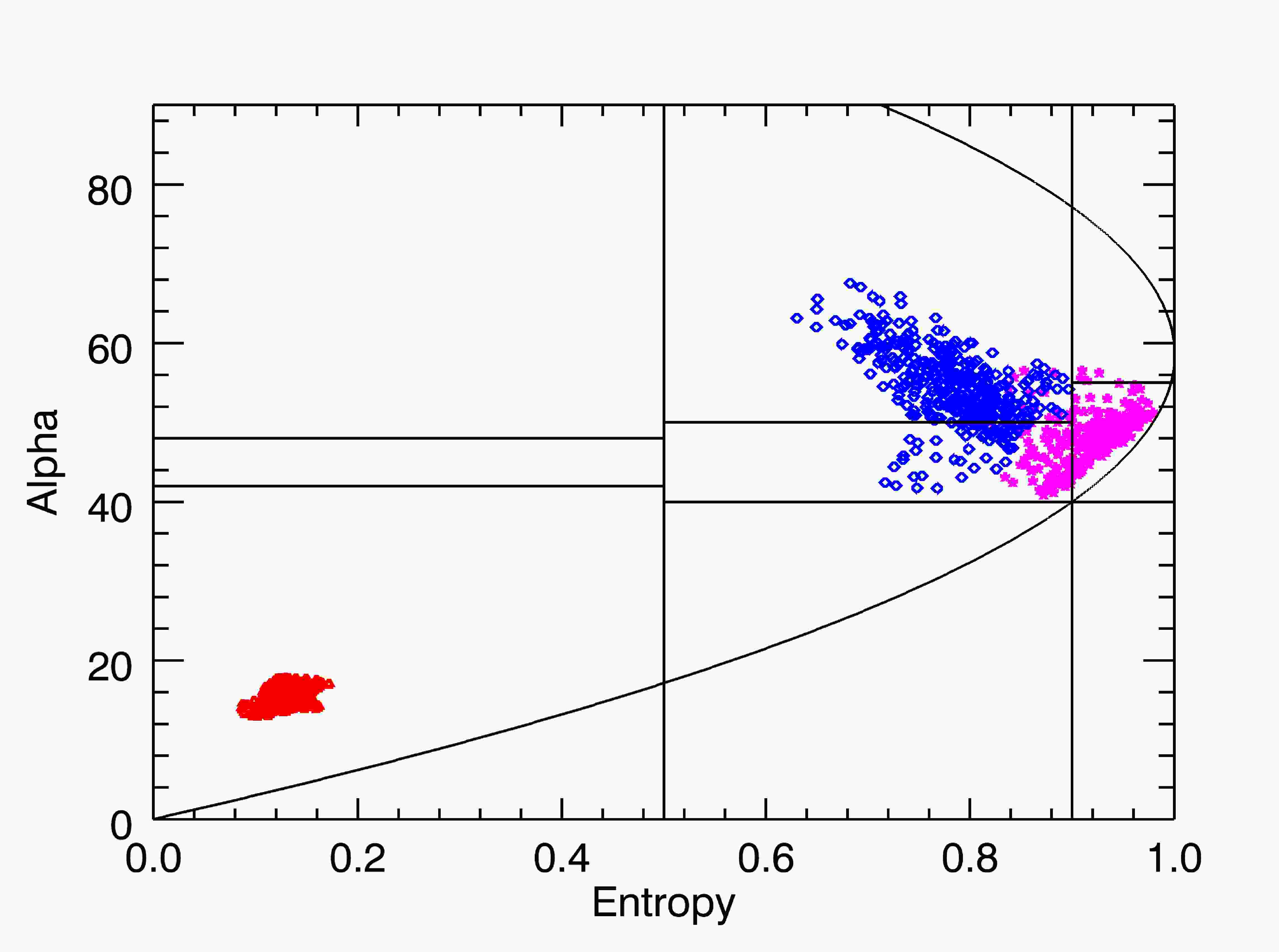}}
  \subfigure[SDNLM filter $I=1$\label{fig:Plot_HAlpha_SDNLM_80_1i}]{\includegraphics[width=.31\linewidth]{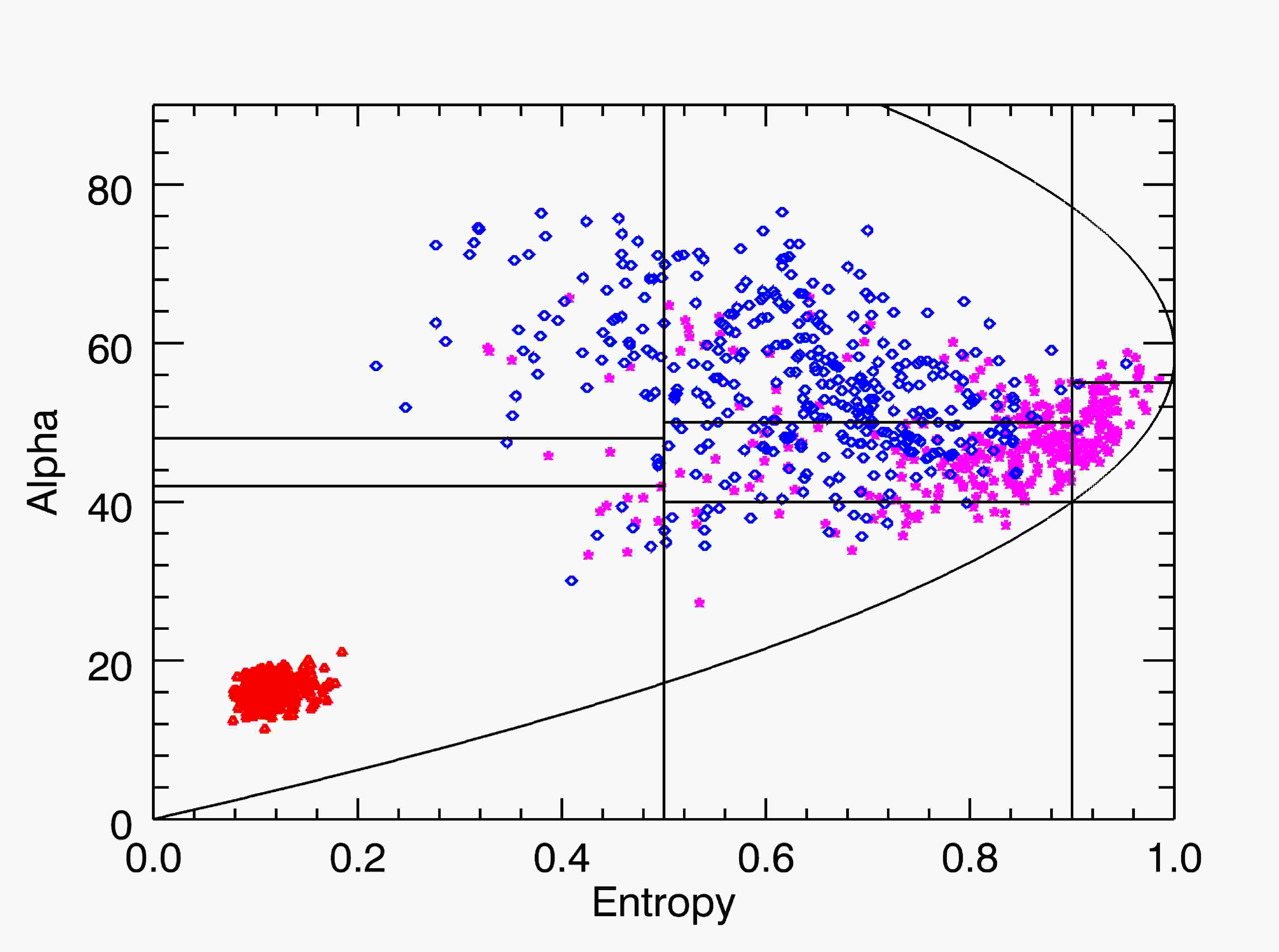}}
  \subfigure[SDNLM filter $I=3$\label{fig:Plot_HAlpha_SDNLM_80_3i}]{\includegraphics[width=.31\linewidth]{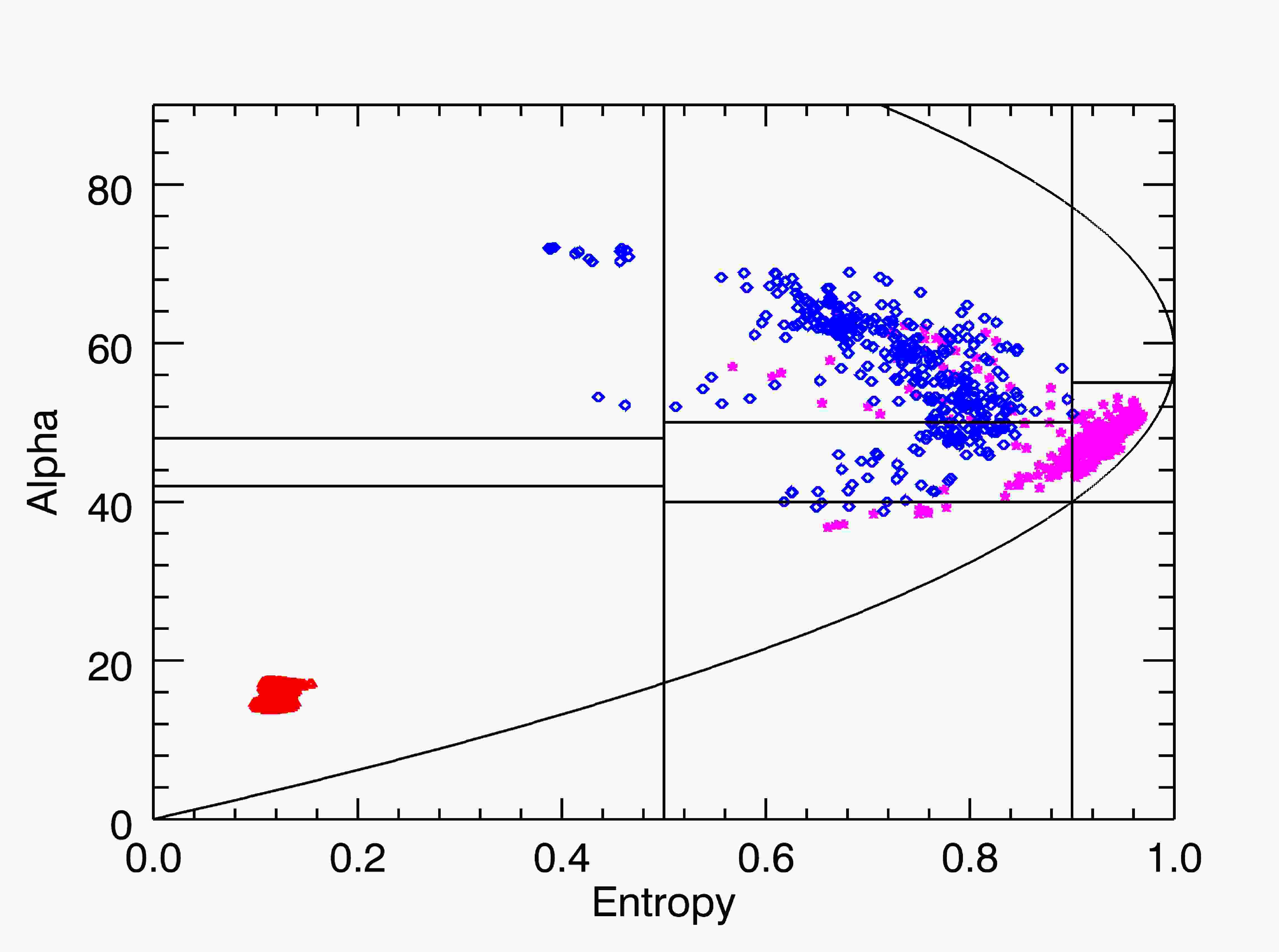}}
  \subfigure[SDNLM filter $I=5$\label{fig:Plot_HAlpha_SDNLM_80_5i}]{\includegraphics[width=.31\linewidth]{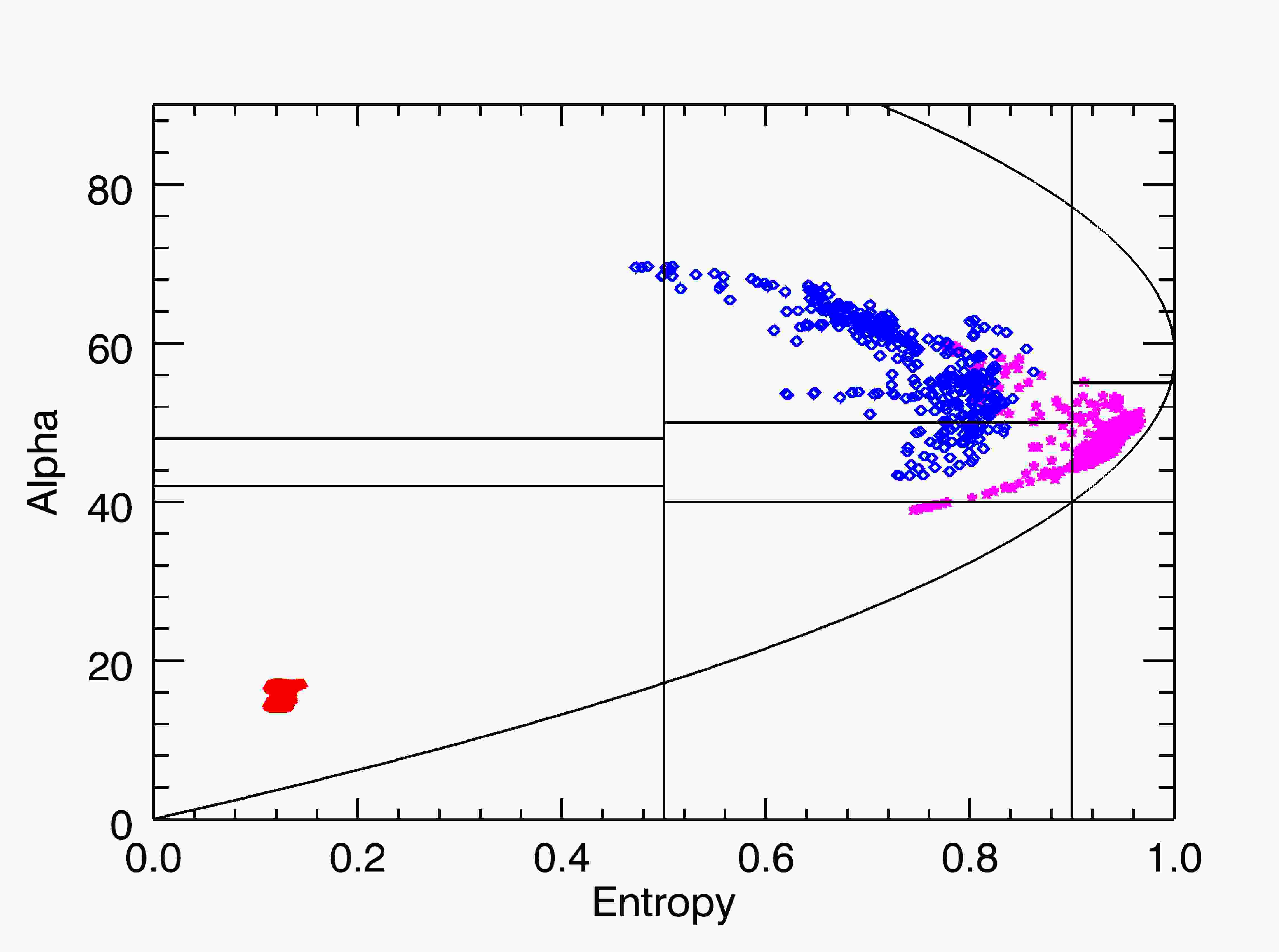}} 
\caption{Scatter plot in the ($H,\overline{\alpha}$) plane of samples from the AIRSAR \textsf{L}-band image after one, three and five iterations $I$}
\label{fig:SanFranPlotHalphaIter}
\end{figure}

Figure~\ref{fig:SanFranPlotHalphaIter} presents the $(H,\overline{\alpha})$ scatter plot of the samples before and after iterating the filters one, three and five times.
Each new iteration adds cohesion to the clusters, whatever the filter employed.
Regarding the SDNLM, the difference between one and three iterations is noticeable.

Table~\ref{tab:QIndexesRealImagesIterated} presents the quantitative analysis of the resulting images.
Again, the Boxcar procedure yields the best noise reduction in smudge-free areas in most of the situations, followed by the Refined Lee filter.
The SDNLM filter tuned somewhere between $80\%$ and $90\%$ of significance produces the best $\operatorname{BRISQUE}$ indexes.

\begin{table}[hbt]
\centering\footnotesize
\caption{Image quality indexes in the real PolSAR image.}\label{tab:QIndexesRealImagesIterated}
\begin{tabular}{c c r@{.}l r@{.}l r@{.}l r@{.}l r@{.}l r@{.}l}
\toprule
  & \multirow{2}{*}{\textbf{Filter}} & \multicolumn{6}{c}{\textbf{ENL}} & \multicolumn{6}{c}{\textbf{$\operatorname{BRISQUE}$ Index}} \\ \cmidrule(lr{.5em}){3-8} \cmidrule(lr{.5em}){9-14} 
  &  & \multicolumn{2}{c}{$H\!H$} & \multicolumn{2}{c}{$H\!V$} & \multicolumn{2}{c}{$V\!V$} & \multicolumn{2}{c}{$H\!H$} & \multicolumn{2}{c}{$H\!V$} & \multicolumn{2}{c}{$V\!V$} \\
\midrule
  & Real data   &  3&867 &  4&227 &  4&494 & 58&258 & 70&845 & 61&593 \\ 
\midrule
\multirow{6}{*}{\rotatebox{90}{\textbf{$1$-iteration}}} 
  & Boxcar      & \textbf{14}&\textbf{564} & \textbf{25}&\textbf{611} & \textbf{18}&\textbf{946} & 36&498 & 37&714 & 36&792 \\ 
  & Refined Lee & 11&491 & 20&415 & 15&407 & 44&997 & 51&547 & 49&412 \\ 
  & IDAN        &  2&994 &  3&732 &  3&923 & 28&823 & 34&853 & 34&691 \\ 
  & SDNLM 80\%  &  7&263 & 11&532 &  8&299 & \textbf{27}&\textbf{841} & \textbf{27}&\textbf{256} & \textbf{33}&\textbf{541} \\ 
  & SDNLM 90\%  &  8&177 & 12&404 &  9&013 & 28&622 & 35&622 & 35&016 \\ 
  & SDNLM 99\%  & 10&828 & 18&379 & 13&075 & 31&026 & 33&881 & 36&881 \\ 
\midrule
\multirow{6}{*}{\rotatebox{90}{\textbf{$3$-iterations}}} 
  & Boxcar      & 24&107 & \textbf{54}&\textbf{348} & 37&501 & 78&379 & 63&823 & 79&307 \\ 
  & Refined Lee & \textbf{24}&\textbf{451} & 45&088 & \textbf{47}&\textbf{423} & 62&421 & 42&882 & 67&016 \\ 
  & IDAN        & 30&880 & 40&220 & 41&640 & 55&811 & 52&470 & 67&016 \\ 
  & SDNLM 80\%  & 15&502 & 32&595 & 21&686 & \textbf{39}&\textbf{500} & \textbf{39}&\textbf{801} & \textbf{42}&\textbf{866} \\ 
  & SDNLM 90\%  & 17&279 & 35&896 & 23&478 & 41&800 & 42&371 & 45&177 \\ 
  & SDNLM 99\%  & 21&892 & 50&069 & 33&347 & 55&824 & 52&824 & 56&844 \\ 
\midrule
\multirow{6}{*}{\rotatebox{90}{\textbf{$5$-iterations}}} 
  & Boxcar      & 29&950 & \textbf{73}&\textbf{504} & 50&811 & 81&585 & 71&628 & 82&365 \\ 
  & Refined Lee & \textbf{36}&\textbf{142} & 72&948 & \textbf{91}&\textbf{634}& 70&908 & 51&728 & 77&271 \\ 
  & IDAN        & 31&310 & 41&050 & 42&340 & 65&906 & 64&803 & 66&400 \\ 
  & SDNLM 80\%  & 20&069 & 52&620 & 31&233 & 49&357 & \textbf{46}&\textbf{318} & 50&872 \\ 
  & SDNLM 90\%  & 23&793 & 61&817 & 36&548 & \textbf{48}&\textbf{803} & 47&177 & \textbf{50}&\textbf{430} \\ 
  & SDNLM 99\%  & 27&798 & 72&209 & 46&450 & 66&251 & 58&663 & 65&944 \\ 
\bottomrule 
\end{tabular}
\end{table}

Table~\ref{tab:statistics} presents the values of mean, variance and ENL estimator on the samples from the three regions of interest (sea, urban and forest), in the cross-polarized band ($HV$).
The fist line in Table~\ref{tab:statistics} presents the values observed in the original (unfiltered) image.
The best values are highlighted in bold; being the best mean the closest to the original value, and the best standard deviation the smallest one.

\begin{sidewaystable}[hbt]
\centering\footnotesize
\caption{The values of mean, variance and ENL estimator on different regions of interest in $H\!V$ polarization}\label{tab:statistics}
\begin{tabular}{c c r@{.}l r@{.}l r@{.}l r@{.}l r@{.}l r@{.}l r@{.}l r@{.}l r@{.}l} 
\toprule
 & \textbf{Filtered}  & \multicolumn{6}{c}{\textbf{Sea}} & \multicolumn{6}{c}{\textbf{Urban}} & \multicolumn{6}{c}{\textbf{Forest}} \\ \cmidrule(lr{.5em}){3-8} \cmidrule(lr{.5em}){9-14} \cmidrule(lr{.5em}){15-20} 
 & \textbf{Versions}  & \multicolumn{2}{c}{$\widehat{\mu}$} & \multicolumn{2}{c}{$\widehat{\sigma}^2$} & \multicolumn{2}{c}{ENL}  & \multicolumn{2}{c}{$\widehat{\mu}$} & \multicolumn{2}{c}{$\widehat{\sigma}^2$} & \multicolumn{2}{c}{ENL} & \multicolumn{2}{c}{$\widehat{\mu}$} & \multicolumn{2}{c}{$\widehat{\sigma}^2$} & \multicolumn{2}{c}{ENL} \\
\midrule
 & Real data  & 90&206 & 33&918 & 7&073 & 200&442 & 50&929 & 15&490 & 161&019 & 53&336 & 9&114 \\ 
\midrule 
\multirow{6}{*}{\rotatebox{90}{\textbf{$1$-iteration}}~} 
 & Boxcar      & 108&820 &  \textbf{8}&\textbf{627} & \textbf{158}&\textbf{408} & 233&589 & \textbf{18}&\textbf{342} & \textbf{162}&\textbf{192} & 156&474 & \textbf{33}&\textbf{925} & \textbf{21}&\textbf{274} \\ 
 & Refined Lee & 113&101 & 12&838 &  77&618 & 232&879 & 18&931 & 151&322 & 164&650 & 40&189 & 16&784 \\ 
 & IDAN        & 109&408 & 22&930 &  22&766 & 220&916 & 32&563 &  46&026 & 171&703 & 38&077 & 20&334 \\ 
 & SDNLM 80\%  & \textbf{105}&\textbf{013} & 15&401 &  46&495 & \textbf{215}&\textbf{326} & 38&632 &  31&068 & \textbf{162}&\textbf{737} & 37&397 & 18&936 \\ 
 & SDNLM 90\%  & 105&386 & 14&050 &  56&261 & 216&555 & 36&698 &  34&822 & 158&836 & 42&379 & 14&047 \\ 
 & SDNLM 99\%  & 107&814 &  9&622 & 125&548 & 221&442 & 32&516 &  46&381 & 156&288 & 40&129 & 15&168 \\ 
\midrule
\multirow{6}{*}{\rotatebox{90}{\textbf{$3$-iterations}}~} 
 & Boxcar      & 109&314 &  \textbf{4}&\textbf{930} & \textbf{491}&\textbf{591} & 242&176 & \textbf{12}&\textbf{331} & \textbf{385}&\textbf{696} & 150&560 & 37&035 & 16&527 \\ 
 & Refined Lee & 121&183 &  8&237 & 216&422 & 245&534 & 10&511 & 545&675 & 172&347 & \textbf{29}&\textbf{176} & \textbf{34}&\textbf{895} \\ 
 & IDAN        & 119&340 & 11&558 & 106&616 & 235&508 & 17&974 & 171&684 & 169&539 & 39&354 & 18&559 \\ 
 & SDNLM 80\%  & 108&954 &  5&798 & 353&120 & \textbf{229}&\textbf{476} & 21&345 & 115&584 & 154&418 & 34&461 & 20&079 \\ 
 & SDNLM 90\%  & 109&039 &  5&154 & 447&608 & 229&900 & 22&228 & 106&976 & \textbf{155}&\textbf{870} & 33&690 & 21&406 \\ 
 & SDNLM 99\%  & \textbf{108}&\textbf{369} &  4&943 & 480&580 & 234&861 & 16&855 & 194&159 & 151&152 & 36&871 & 16&806 \\ 
\midrule
\multirow{6}{*}{\rotatebox{90}{\textbf{$5$-iterations}}~} 
 & Boxcar      & 108&369 &  3&703 & 856&241 & 244&805 & 10&522 & 541&263 & 145&121 & 39&278 & 13&651 \\  
 & Refined Lee & 124&033 &  7&368 & 283&382 & 249&903 &  \textbf{8}&\textbf{389} & \textbf{887}&\textbf{365} & 175&836 & \textbf{25}&\textbf{191} & \textbf{48}&\textbf{723} \\ 
 & IDAN        & 121&134 &  9&301 & 169&603 & 240&984 & 12&865 & 350&900 & 174&783 & 40&405 & 18&712 \\ 
 & SDNLM 80\%  & 107&843 &  3&898 & 765&393 & \textbf{234}&\textbf{505} & 17&174 & 186&439 & \textbf{150}&\textbf{619} & 35&711 & 17&789 \\ 
 & SDNLM 90\%  & 107&614 &  \textbf{3}&\textbf{602} & \textbf{892}&\textbf{750} & 234&600 & 17&160 & 186&911 & 148&879 & 37&514 & 15&750 \\ 
 & SDNLM 99\%  & \textbf{107}&\textbf{199} &  3&693 & 842&776 & 239&963 & 13&178 & 331&599 & 146&762 & 36&236 & 16&404 \\ 
\bottomrule
\end{tabular} 
\end{sidewaystable}

The SDNLM filter is the best at preserving the original mean values, and with reduced standard deviation.
Our proposal does not provide the best variability reduction, a behavior that may be associated with the preservation of fine structures and a smaller loss of spatial resolution, as can be noted in Figure~\ref{fig:Iterations}.
The equivalent number of looks behaves consistently with what was observed in previous examples.

\section{Conclusions}\label{sec:conclu}

The use $h$-$\phi$ divergences, a tool in Information Theory, led to test statistics (with a known and tractable asymptotic distribution) able to check if two samples cannot be described by the same complex Wishart distribution, the classical model for PolSAR data.
Using one of these test statistics, namely the one based on the Hellinger distance, we devised a convolution filter whose weights are function of the $p$-value of tests which compare two patches of size $3\times3$ in a search window of size $5\times5$ pixels.

The filter obtained in this manner (SDNLM -- Stochastic Distances Nonlocal Means) was compared with the Boxcar, Refined Lee and IDAN filters in a variety of PolSAR imagery: data simulated from the complex Wishart law over a realistic phantom using parameters observed in practice, data simulated from the electromagnetic properties of the scattering over a simplified cartoon model, and a real PolSAR image over San Francisco, CA.
The quantitative assessment verified the equivalent number of looks (a measure of noise reduction) over smudge-free samples, the structural $\operatorname{SSIM}$ index, and the $\operatorname{BRISQUE}$ index used appropriately on no-references images (real images or blind).
The Boxcar filter promotes the strongest noise reduction in these conditions, but at the expense of obliterating small details.
The Refined Lee and IDAN filters are competitive, but produce a pixellated effect and their $\operatorname{SSIM}$ index is worse than the produced by the SDNLM filter in all instances.
We noted this same feature with the $\operatorname{BRISQUE}$ index applied to real data and the index values ​​remain stable even during iteration of the SDNLM filter, which does not happen with other filters assessed.

A qualitative assessment was also made checking how the polarimetric entropy is affected by the filters.
We noticed that all the filters enhance it but, in particular, our proposal performs the most refined enhancement since it preserves very small details which are characteristic of complex urban areas.

The effect of the filters and of applying them iteratively was also verified in the $(H,\overline{\alpha})$ plane.
All filters produced more and more compact clusters of observations in this plane as more iterations were applied.
The SDNLM filter yielded the best separation of the sea sample, while the other two were treated at least as well as they were by the other filters.

The SDNLM filter has three tuning parameter: (i)~filtering window size, (ii)~the size of the patches, and (iii)~the significance of the test.
We provide a range of suggested values for the latter, and show good results with an economic choice for the two former.
The filter can also be applied iteratively if more smoothing is requested and, provided an adequate statistical model, it can also be applied to other types of data.

We conclude that our proposal is a good candidate for smoothing PolSAR imagery without compromising either small details or the scattering characteristics of the targets.

The test statistics are invariant with respect to permutations of the sample; directional features will be considered in forthcoming works.
Future research includes the proposal of quality measures for PolSAR imagery, and the use of tests based on entropies~\cite{EntropyBasedStatisticalAnalysisPolSAR}.

\section*{Acknowledgements}

The authors are grateful to CNPq, Capes and FAPESP for the funding of this research, and to Professor Jos\'e Claudio Mura (Divis\~ao de Sensoriamento Remoto, Instituto Nacional de Pesquisas Espaciais, Brazil) for enlightening discussions about the PolSARpro toolbox and PolSAR image decomposition.

\bibliographystyle{splncs03}
\bibliography{Torres_PR2013}

\appendix

\section{Computational information}\label{app:ComputationalInformation}

Computing the ML estimator of the equivalent number of looks $\widehat{L}$ given in equation~\eqref{eq:MLENL} and the test statistic based on the Hellinger distance, c.f.\ equation~\eqref{eq:HellingerTestStatistic}, are part of the computational core of this proposal.
Each weight requires computing these quantities.
It is noteworthy that they involve only two operations on complex matrices: the determinant and the inverse.
Using the fact that the matrices are Hermitian and positive definite, it is possible to reduce drastically the number of operations required to calculate these two quantities.
Specialized accelerated function in the \texttt R programming language \cite{R} were developed with the speed of the filters in mind.

The time required to filter a $128\times 128$ pixels image with one iteration is of about \unit[$75$]{s} in an Intel\textregistered Core\texttrademark\ i7-3632QM CPU \unit[$2.20$]{GHz}, with software developed in \texttt R version~$2.14.1$ running on Ubuntu~$12.04$.
\texttt R was the choice because of its excellent accuracy with respect to similar platforms~\cite{OctaveScilabMatlabCAM}.

\section{Observed covariance matrices}\label{app:CovarianceMatrices}

In the following we present the observed covariance matrices that were used to simulate the data presented in Figure~\ref{fig:R99B}.

\begin{eqnarray*}
 \widehat{\bm{\Sigma}}_1 &=& \left[
 \begin{array}{ccc}
  7.60830 & -0.74901-2.29165\bm i & 1.38157+8.39200\bm i \\
   & 24.8580 & -5.90346-0.45011\bm i \\
   &  & 32.2771 \\
 \end{array}\right]\cdot 10^{-4}\\
  \widehat{\bm{\Sigma}}_2 &=& \left[
 \begin{array}{ccc}
 128.592 & 12.1941-7.12246\bm i & 39.1107+18.7954\bm i \\
  & 336.959 & -8.49716-11.8210\bm i \\
  &  & 154.343 \\
 \end{array}\right]\cdot 10^{-4}\\
  \widehat{\bm{\Sigma}}_3 &=& \left[
 \begin{array}{ccc}
 29.6303 & 4.86985+1.55848\bm i & 3.41851+1.43502\bm i \\
  & 86.8985 & -2.03628-8.24319\bm i \\
  &  & 43.3504 \\
 \end{array}\right]\cdot 10^{-4}\\
  \widehat{\bm{\Sigma}}_4 &=& \left[
 \begin{array}{ccc}
 14.0576 & -0.25731-1.48967\bm i & 4.36926+9.41493\bm i \\
  & 60.5614 & -4.92951-2.16850\bm i \\
  &  & 42.3767 \\
 \end{array}\right]\cdot 10^{-4}\\
 \widehat{\bm{\Sigma}}_5 &=& \left[
 \begin{array}{ccc}
 4.89301 & -0.52225-0.62765\bm i & 1.38866+5.29889\bm i \\
  & 12.1149 & -3.30897-0.85846\bm i \\
  &  & 25.6761 \\
 \end{array}\right]\cdot 10^{-4}\\
 \widehat{\bm{\Sigma}}_6 &=& \left[
 \begin{array}{ccc}
 18.7013 & 0.81235-1.72513\bm i & 1.26677+6.08878\bm i \\
  & 32.8094 & -3.01618-1.67916\bm i \\
  &  & 25.8651 \\
 \end{array}\right]\cdot 10^{-4}.  
\end{eqnarray*}

\end{document}